\documentclass[a4paper,usenames,dvipsnames,11pt]{article}
\pdfoutput=1

\usepackage{jheppub}
\usepackage{slashed}
\usepackage{mathrsfs,booktabs,multirow,tabularx}
\usepackage{stmaryrd}
\usepackage{xspace}
\usepackage{fancyvrb}
\usepackage[makeroom]{cancel}
\usepackage{amsmath}    % need for subequations
\usepackage{amssymb}    % 
\usepackage{graphicx}   % need for figures
\usepackage{verbatim}   % useful for program listings
\usepackage{lscape}
\usepackage{subfig}
\usepackage{listings}
\usepackage{mathtools}
\def\beq{\begin{equation}}
\def\beqn{\begin{eqnarray}}
\def\eeq{\end{equation}}
\def\eeqn{\end{eqnarray}}
\def\abs#1{\left|#1\right|}

\def\pplus#1{\left[#1\right]_+}

\def\PDF#1#2{\Gamma_{\!#1/#2}}
\def\ePDF#1{\Gamma_{\!#1}}
\def\ehPDF#1{\hat{\Gamma}_{\!#1}}

\def\binomial#1#2{
\left(\!\!
\begin{array}{c}
#1\\
#2
\end{array}
\!\!\right)
}
\def\stirlingSo#1#2{
\left[\!\!
\begin{array}{c}
#1\\
#2
\end{array}
\!\!\right]
}

\def\plusdz{\left(\frac{1}{1-z}\right)_+}
\def\plusdo{\left(\frac{\log(1-z)}{1-z}\right)_+}
\def\plusd#1{\left(\frac{\log^#1(1-z)}{1-z}\right)_+}
\def\harm1#1#2{S_#1}

\newcommand\sss{\scriptscriptstyle}
\newcommand\mydot{\!\cdot\!}
\newcommand\ep{\epsilon}
\newcommand\half{\frac{1}{2}}

\newcommand\aem{\alpha}
\newcommand\aemotpi{\frac{\aem}{2\pi}}

\newcommand\gE{\gamma_{\sss\rm E}}

\newcommand{\bN}{\bar{N}}

\newcommand{\epem}{e^+e^-}
\newcommand{\lp}{e^+}
\newcommand{\lm}{e^-}

\newcommand{\ord}{{\cal O}}

\newcommand\NF{n_{\sss F}}

\newcommand\hsig{\hat{\sigma}}
\newcommand\tsig{\tilde{\sigma}}

\newcommand\hS{\hat{S}}

\newcommand\APmat{{\mathbb P}}
\newcommand\Eop{{\mathbb E}}
\newcommand\Rop{{\mathbb R}}

\newcommand\Kmat{{\mathbb K}}
\newcommand\Cmat{{\mathbb C}}

\newcommand\MSb{\overline{\rm MS}}

\newcommand\ePDFs{\ePDF{\rm\sss S}}
\newcommand\ePDFns{\ePDF{\rm\sss NS}}

\newcommand\muz{\mu_0}

\newcommand{\Melleq}{\stackrel{\infty}{=}}
\newcommand\stepf{\Theta}

\newcommand\Domz{\delta(1-z)}
\newcommand\dilog{{\rm Li}}

\newcommand{\Lz}{L_z}

\newcommand{\invM}{M^{-1}}

\allowdisplaybreaks[4]

\title{The electron parton distribution functions at the NNLO in QED}

\author[a,b]{S. Frixione,}
\affiliation[a]{INFN, Sezione di Genova, Via Dodecaneso 33, I-16146, 
Genoa, Italy}
\affiliation[b]{ PH Department, TH Division, CERN, CH-1211 Geneva 23, 
Switzerland}

\author[b]{K. Sch\"onwald}

\emailAdd{Stefano.Frixione@cern.ch}
\emailAdd{Kay.Schonwald@cern.ch}

\abstract{
The electron Parton Distribution Functions (PDFs) are essential
components in the calculations of cross sections within a
collinear-factorisation framework at lepton colliders. We compute them 
at the next-to-next-to-leading order (NNLO) in QED by working in
two different factorisation schemes, namely the standard $\MSb$ 
one and the so-called $\Delta$ scheme, which was originally defined 
at the next-to-leading order, and which we generalise in this work
by extending its definition to all orders. We present analytical
results relevant to the large-$z$ behaviour of the PDFs in both
factorisation schemes, and thus show how the soft-logarithm
enhancement of the $\MSb$ electron PDF is completely absent
in the $\Delta$ scheme. The latter therefore constitutes a natural 
choice that helps significantly reduce the complexity not only of  
numerical simulations in phenomenological applications, but also that
of analytical computations necessary to achieve the soft resummation
of physical observables. We argue that NNLO PDFs are necessary to
attain relative-precision targets of $10^{-4}$ or better, for
colliders whose centre-of-mass energies are in the hundred-GeV range.
}


\preprint{
\begin{flushright}
CERN-TH-2026-196\\
\end{flushright}
}

\begin{document}
\maketitle
\flushbottom

\section{Introduction\label{sec:intro}}
By assuming that the strategy for the foreseeable future of high-energy
particle physics will continue to employ the model of big experiments
at large colliders, it is highly probable that the first post-LHC
practical realisation of that model will entail the construction 
of an $\epem$ machine. While the FCC-ee is at present the foremost
candidate, other options (circular or otherwise) are still viable.
From a theoretical viewpoint, the details of which collider will
eventually be built are mostly irrelevant, since all of the projects
share a common trait, namely an expected enormous amount of data
taken, and a subsequent very significant increase of the statistical 
power of experimental results w.r.t.~those of the LEP era. This wealth
of information can be properly exploited only if the corresponding
theoretical predictions have comparably small systematic uncertainties,
which at present is most certainly not the case.

An ubiquitous effect in $\epem$ physics is that of initial-state
radiation (ISR henceforth), defined as the cumulative result of
the branchings stemming from incoming leptons,
that can be meaningfully separated from both those which are 
associated with beam dynamics (including beam-beam interactions),
and those which are specific of the production of a given final state
(e.g.~a lepton or a $W^+W^-$ pair). Such a separation does not have
to be necessarily physical (in fact, generally it is not), but it must
be unambiguous given some external constraints (e.g.~the choice of a
scale and/or a scheme). ISR constitutes a major source of uncertainty
in theoretical predictions owing to its being enhanced, order by order,
by two classes of large logarithms, which we schematically denote as
follows:
\beq
L=\log\frac{Q^2}{m^2}\,,\;\;\;\;\;\;\;\;
\ell=\log\frac{Q^2}{\langle E_\gamma\rangle^2}\,,
\label{logdef}
\eeq
and refer to them as collinear and soft logarithms, respectively.
In eq.~(\ref{logdef}) $Q$ is a typical hard scale of the problem 
(e.g.~the centre of mass energy, or the invariant mass of the 
system produced in the collision), $m$ is the electron mass, and
$\langle E_\gamma\rangle$ is the average photon energy available
given the collider operating conditions and the constraints implicit
in the observable definition. Since typically
\beq
m^2\ll Q^2\,,\;\;\;\;\;\;\;\;
\langle E_\gamma\rangle^2\ll Q^2\,,
\eeq
both $L$ and $\ell$ grow large, potentially compensating the
coupling-constant suppression of perturbative computations, and
thus spoiling the ``convergence'' of the perturbative series.
A re-arrangement of such a series by means of resummation is
therefore to be systematically envisaged. Lacking a general framework
for the simultaneous resummation of both $L$ and $\ell$, roughly speaking
one resorts to resumming one of the logarithms, and takes into account
the other logarithm order-by-order in perturbation theory. If the
logarithm to be resummed is $L$, one employs a factorisation theorem,
where the resummation is embedded in lepton PDFs, and is achieved by
means of solving the RGE evolution equations for the latter. In this
context, soft logarithms may appear in both the PDFs and the short-distance
cross sections. Conversely, if the logarithm to be resummed is $\ell$,
one employs the YFS formulation~\cite{Yennie:1961ad}, where collinear 
logarithms are accounted for by order-by-order non-soft residues. Within 
these two frameworks, one can further consider the resummation of soft
and of collinear logarithms, respectively; in order for this to happen,
it is clearly desirable that the expressions of the PDFs and of the
short-distance cross sections in terms of $\ell$ in the former approach,
and of the soft residues in terms of $L$ in the latter approach, be
as simple as possible.

In this paper, we focus on the collinear-factorisation strategy.
We remind the reader that, while formally collinear factorisation
formulae in QED are identical to their QCD counterparts, routinely
used in LHC physics, there is a fundamental difference between the two,
in that QED PDFs are entirely calculable in perturbation theory. In
particular, one typically computes them at a certain order in 
the fine structure constant, $\ord(\aem^k)$,
at a scale $\muz$ of the order of the electron mass, and then evolves
them to the desired hard scale $Q$ by means of the Altarelli-Parisi
equations~\cite{Gribov:1972ri,Lipatov:1974qm,Altarelli:1977zs,
Dokshitzer:1977sg}, where splitting kernels of an order matching that
of the initial conditions (namely, $\ord(\aem^{k+1})$) must be
employed. Strictly speaking, the resulting PDFs should be referred
to as having N$^k$LO-N$^k$LL accuracy, but we shall often understand
the N$^k$LL bit in the following. As is the case in QCD, QED PDFs are
unphysical objects, since they depend on an unphysical factorisation
scale $\mu$, and on an unphysical factorisation scheme $K$, which
basically controls the finite part of the subtraction of the collinear
singularity in the evolution equations for the bare PDFs. While
physically meaningful only when convoluted with short-distance cross
sections, with the result of such a convolution being scale- and
factorisation-scheme-independent at the perturbative order at which
one is working, mathematically PDFs can be handled as perfectly
well-defined standalone objects.

In order to be definite, we consider the case of an incoming electron (the 
case of an incoming positron can be trivially obtained by charge-conjugating
the former). Denoting by $\ePDF{i}$ the PDF of parton $i$ inside the
electron, the dominant contribution to a cross section is generally due
to $i$ also being an electron. Its PDF in the large-$z$ region 
reads~\cite{Gribov:1972ri,Lipatov:1974qm,Skrzypek:1990qs,Skrzypek:1992vk,
Cacciari:1992pz}:
\beq
\ePDF{e^-}(z,\mu^2)\stackrel{z\to 1}{\longrightarrow}
\frac{e^{-\gE\xi}e^{\hat{\xi}}}{\Gamma(1+\xi)}\,
\xi(1-z)^{-1+\xi}\,\ehPDF{e^-}(z,\mu^2,\xi)\,.
\label{Gammae}
\eeq
Here, $\xi$ and $\hat{\xi}$ are two $\mu$-dependent but scheme-independent
parameters
\beq
\xi=\frac{\aem L}{\pi}+\ldots\,,\;\;\;\;\;\;\;\;
\hat{\xi}=\frac{3\aem L}{4\pi}+\ldots\,,
\eeq
which include up to N$^k$LL collinear-logarithm contributions
(as well as effects due to the running of $\aem$)
in N$^k$LO+N$^k$LL PDFs, while $\ehPDF{e^-}(z,\mu^2,\xi)$ is a scale-
and scheme-dependent function beyond the LO -- at the LO, this function
is identically equal to one. 
The essence of the issues stemming from adopting $\MSb$ as the factorisation
scheme for electron PDFs become clear thanks to the following result
(see e.g.~refs.~\cite{Blumlein:2011mi,Bertone:2019hks,
Ablinger:2020qvo}\footnote{While the explicit form of the r.h.s.~of 
eq.~(\ref{MSblogs}) is known only up to NNLO, it is straightforward 
to argue that it is indeed valid to all orders.})
\beq
\ehPDF{e^-}^{(\MSb)}(z,\mu^2,\xi)\stackrel{z\to 1}{\longrightarrow}
1+\sum_{k=1}^\infty\aem^k\sum_{i=0}^{2k}c_{k,i}\log^i(1-z)\,,
\label{MSblogs}
\eeq
where the $c_{k,i}$ coefficients are $z$-independent. Therefore,
by increasing the perturbative order of the calculations, one finds
an increasing number of soft logarithms, whose leading power is twice
as large as that of the coupling constant. Some of these logarithms
are spurious, i.e.~they will cancel against analogous contributions in
the short-distance cross sections. However, such a cancellation is never
achieved analytically\footnote{Unless one re-expands the lepton PDFs,
which one is not supposed to do in the context of high-precision
phenonomenological applications, but as a possible way towards a 
deeper understanding of the structure of the cross section.}, 
which significantly complicates both numerical
computations and the potential resummation of soft logarithms.

What we have just schematically discussed represents a serious problem
in the context of the large-precision landscape relevant to the next
lepton collider. This observation, as well as the necessity of a
rigorous way to assess the factorisation-scheme systematics, has
been the main motivation for the definition of the $\Delta$
scheme~\cite{Frixione:2021wzh}. In that scheme, at the NLO the
counterpart of eq.~(\ref{MSblogs}) reads thus:
\beqn
\ehPDF{e^-}^{(\Delta,{\rm NLO})}(z,\mu^2,\xi)
&\stackrel{z\to 1}{\longrightarrow}&
\frac{\aem(\mu)}{\aem(m)}\left(
1+\sum_{k=1}^\infty d_{k}\big(\log(1-z)\big)\right)\,,
\label{DelNLOlogs}
\\*
d_{k}\big(\log(1-z)\big)
&\stackrel{z\to 1}{\longrightarrow}&
a_k/\log^{k+1}(1-z)\,,
\label{Delcoeff}
\eeqn
for some $a_k$ $z$-independent coefficient of order one,
with a well-behaved perturbative expansion in $\aem$.
In other words, the NLO $\Delta$-scheme electron PDF differs from
its LO counterpart by terms which are {\em not} enhanced by soft 
logarithms. Therefore, by construction the $\Delta$ scheme gets rid 
of all of the spurious soft logarithms which plague the $\MSb$ 
scheme\footnote{A recent paper also discusses the impact of factorisation 
schemes different from $\MSb$ in QCD~\cite{Delorme:2026vln}.}. This immediately 
renders any numerical simulation easier than its analogous $\MSb$ one, as well
as giving one the possibility of addressing soft resummation solely at the
level of short-distance cross sections.

It is therefore a crucial question whether these properties of the
$\Delta$ scheme can be generalised beyond the NLO. This is a question
that we address in this work, showing that the answer is indeed in
the affirmative. Having said that, a proper factorisation-scheme
systematics, which is by nature non-parametric, implies that one still
needs at least two schemes to compare to one another. We thus also 
consider the NNLO electron PDFs in the $\MSb$ scheme, in order to find 
their large-$z$ behaviour which is essential for their reliable integration.

This paper is organised as follows. Our notation and conventions
are established in sect.~\ref{sec:conv}. In sect.~\ref{sec:ini} we 
introduce the $\MSb$ NNLO initial conditions which we subsequently employ 
in the PDF evolution. The all-order generalisation of the $\Delta$ scheme 
is presented in sect.~\ref{sec:DS}. The PDF evolution equations 
are discussed in sect.~\ref{sec:eopK}, where they are handled by
means of an evolution operator, whose RGE in a generic scheme
is derived there. The actual $\Delta$-scheme PDFs emerging from 
such an evolution operator are studied in sect.~\ref{sec:asy}, 
with particular attention paid to the large-$z$ behaviour. 
Alternative definitions of the $\Delta$ scheme are considered in 
sect.~\ref{sec:others}, which clarifies that $\Delta$ should be more 
properly referred to as a class of factorisation schemes. The impact 
of the factorisation-scheme choice on short-distance cross sections is 
then briefly considered in sect.~\ref{sec:DYxsec}, among other things
by means of Drell-Yan as a representative example. Finally, we give our 
conclusions in sect.~\ref{sec:concl}. Technical material, which is essential 
for a complete understanding of the manipulations carried out in the main
text, is collected in appendix~\ref{sec:Mell}, and further expanded
in appendix~\ref{sec:proof}. All of the analytical results derived here
for the first time, or needed in the context of such derivations,
are given in appendix~\ref{sec:res}, in both Mellin and configuration
spaces. An off-topic mathematical curiosity is reported in 
appendix~\ref{sec:gE}.

\section{Notation and conventions\label{sec:conv}}
In general, a subscript $N$ indicates that the corresponding
quantity is defined in Mellin space -- see appendix~\ref{sec:Mell}
for our conventions for the Mellin transform. As is customary,
the variable
\beq
\bN=N\,e^{\gE}\,,
\eeq
will often appear in the manipulations of Mellin-space functions.
Following ref.~\cite{Frixione:2021wzh}, we denote by $\Gamma_N$ 
a column vector that collects all of the PDFs:
\beq
\Gamma_N=\left(
\begin{array}{c}
\ePDF{\alpha_1,N}\\
\ePDF{\alpha_2,N}\\
\vdots\\
\ePDF{\alpha_n,N}\\
\end{array}
\right)\,,
\label{FN}
\eeq
with $\ePDF{\alpha_i}$ the PDF of parton $\alpha_i$ inside the
particle of interest. Conventionally, we shall label the partons
so that $\alpha_1$ coincides with the identity of the particle (thus,
for an electron particle, $\alpha_1=\lm$). Throughout this paper we
shall limit ourselves to considering a single fermion family, so
that $\NF=1$\footnote{Nevertheless, we shall keep a functional dependence
on $\NF$ in view of the fact that, as has been shown in
ref.~\cite{Bertone:2022ktl}, this is largely sufficient to generalise
single-fermion-family results to a variable flavour number scheme.}.
This implies \mbox{$(\alpha_1,\alpha_2,\alpha_3)=(e^-,\gamma,e^+)$}
for electron PDFs, in the so-called physical basis. As is well known,
evolution equations partially decouple when expressed in terms of what
is referred to as the evolution basis, for which
\mbox{$(\alpha_1,\alpha_2,\alpha_3)=(\Sigma,\gamma,NS)$}, and:
\beq
\left(
\begin{array}{c}
\ePDF{\Sigma}\\
\ePDF{\gamma}\\
\ePDF{NS}\\
\end{array}
\right)=
T
\left(
\begin{array}{c}
\ePDF{e^-}\\
\ePDF{\gamma}\\
\ePDF{e^+}\\
\end{array}
\right)\,,
\;\;\;\;\;\;\;\;
T=\left(
\begin{array}{rrr}
1 & 0 & 1\\
0 & 1 & 0\\
1 & 0 &-1
\end{array}
\right)\,.
\label{Tmatdef}
\eeq
If $M_P$ is an operator that acts on the physical basis (i.e.~a $3\times 3$
matrix in the representation above), then its evolution-basis counterpart is:
\beq
M_E=T\,M_P\,T^{-1}\,.
\eeq
Henceforth, we shall often omit to indicate whether a vector or an
operator is defined in the physical or the evolution basis, since this
is typically clear from the context.

The quantity defined in eq.~(\ref{FN}) may feature an argument, which 
indicates the value of the evolution variable at which the PDFs are 
evaluated. A special notation, $\Gamma_{0,N}$, may also be used 
for the initial conditions. These, we impose at a scale $\muz\sim m$
in Mellin space; in other words:
\beq
\Gamma_{0,N}\equiv \Gamma_{N}(\muz)\,.
\eeq
We also denote by
\beq
\Gamma_{N}^{(K)}
\eeq
the PDFs relevant to a given factorisation scheme $K$; conventionally,
the absence of that superscript implies that one works in $\MSb$.

In Mellin space we shall frequently consider the large-$N$ limit
of some quantity $F_N$, which is dual to the large-$z$ behaviour
of its inverse Mellin transform $\invM[F_N]$. We shall typically
denote such a large-$N$ quantity by $\widetilde{F}_N$, so that:
\beq
F_N\;\stackrel{N\to\infty}{=}\;
\widetilde{F}_N+\ord\left(\frac{1}{N^{\rho+2}}\right)\,,
\label{wtFNdef}
\eeq
with $\rho\ge -1$ a given integer. Equation~(\ref{wtFNdef}) implies
that $\widetilde{F}_N$ is a polynomial in $\log\bN$, whose coefficients
are polynomials in $1/N$ of degree \mbox{$\rho+1$}. In other words,
$\widetilde{F}_N$ captures the leading behaviour of $F_N$, up to and
including terms suppressed by $1/N^{\rho+1}$. The dual of eq.~(\ref{wtFNdef}) 
reads:
\beq
\invM\big[F_N\big](z)\;\stackrel{z\to 1}{=}\;
\invM\big[\widetilde{F}_N\big](z)+\ord\left((1-z)^{\rho+1}\right)\,,
\label{wtFNdefz}
\eeq
that is, $\invM\big[\widetilde{F}_N\big](z)$ gives the leading behaviour
of $\invM\big[F_N\big](z)$ up to and including terms suppressed by
$(1-z)^\rho$ at $z\to 1$.

We also remind the reader that, in keeping with
ref.~\cite{Bertone:2019hks}, it is convenient to express the PDF 
evolution in terms of the variable $t$, defined thus:
\beq
t=\frac{1}{2\pi b_0}\log\frac{\aem(\mu)}{\aem(\mu_0)}\,.
\label{tdef}
\eeq
We use the following definition of the QED $\beta$ function:
\beq
\frac{\partial\aem(\mu)}{\partial\log\mu^2}=\beta(\aem)=
b_0\aem^2+b_1\aem^3+b_2\aem^4+\ldots\,,
\label{betaQED}
\eeq
with~\cite{Tarasov:1980au,Larin:1993tp}
\beq
b_0=\frac{\NF}{3\pi}\,,\;\;\;\;\;\;\;\;
b_1=\frac{\NF}{4\pi^2}\,,\;\;\;\;\;\;\;\;
b_2=-\frac{\NF}{32\pi^3}-\frac{11\NF^2}{144\pi^3}\,,
\label{b0b1}
\eeq
and $\NF$ the number of active charged fermion families. Equation~(\ref{tdef}) 
implies that:
\beq
\frac{\partial}{\partial\log\mu^2}=\frac{1}{2\pi b_0}\,
\frac{\beta(\aem(\mu))}{\aem(\mu)}\,\frac{\partial}{\partial t}\,,
\label{dmudt}
\eeq
and thus:
\beq
\frac{\partial\aem(\mu)}{\partial t}=2\pi b_0\aem(\mu)
\;\;\;\;\Longrightarrow\;\;\;\;
\aem(\mu)=\aem(\mu_0)e^{2\pi b_0 t}\,.
\label{aemvst}
\eeq
By defining
\beq
L=\log\frac{\mu^2}{\muz^2}\,,\;\;\;\;\;\;\;\;
L_0=\log\frac{\muz^2}{m^2}\,,
\label{LL0def}
\eeq
whereby $L$ essentially coincides with that introduced in
eq.~(\ref{logdef}), and gives a practical meaning to that generic
definition, then
\beq
t=\frac{\aem L}{2\pi}+
\frac{(2b_1L-b_0^2L^2)\aem^2}{4b_0\pi}+
\frac{(6b_2L-9b_0b_1L^2+2b_0^3L^3)\aem^3}{12b_0\pi}+\ldots
\label{texp}
\eeq
Used as an argument of functions or operators, $t$ corresponds to $\mu$,
and $t=0$ to $\mu=\muz$. This creates a potential notation ambiguity;
specifically, in the case of $\aem$, one has
\beq
\mu\;:\longrightarrow\;\aem(\mu)\,,\aem(\muz)\;\;\;\;\;\;\;\;\;
t\;:\longrightarrow\;\aem(t)\,,\aem(0)\,.
\label{aemaem}
\eeq
Since in this paper we solely use $\MSb$-renormalised quantities
(i.e., we never employ the so-called $\aem(0)$ scheme), the possibility
of confusion is nil, and thus we may adopt either of the two notations in
eq.~(\ref{aemaem}).

Finally, we shall use the following notation for the coefficients 
in a series expansion in~$\aem$:
\beq
f(\aem)=\sum_{i=0}^\infty f^{[i]}\left(\aemotpi\right)^i\,.
\label{fcoeff}
\eeq

\section{$\MSb$ initial conditions\label{sec:ini}}
The initial conditions at the NNLO in $\MSb$ have been first calculated in
refs.~\cite{Blumlein:2011mi,Ablinger:2020qvo}. In those papers, they were
extracted by computing matrix elements of twist-2 operators 
sandwiched between massive external electrons. Recently, the results
of these calculations have been found to be in agreement with those
obtained by means of SCET-based approaches in refs.~\cite{Stahlhofen:2025hqd,
Schnubel:2025ejl}.  
In ref.~\cite{Stahlhofen:2025hqd} effects stemming from a second heavy fermion 
have also been considered. Since we limit our discussion to a single
massive fermion in this paper, we do not take these additional results 
into account.

In order to be self-contained, we give the explicit formulae for the $\MSb$
initial conditions in both Mellin and configuration space, as well as their
asymptotic $N\to\infty$ and $z\to 1$ limits, which will be of importance 
in defining the $\Delta$ scheme in section~\ref{sec:DS}. However,
these explicit results are very long, and therefore we report them in the
subsections of appendix~\ref{sec:resMSbb}; in particular, we use
the physical basis \mbox{($\alpha\in\{e^-,\gamma,e^+\}$)}, and give the 
complete and asymptotic Mellin-space predictions here:
\beq
\Gamma_{\alpha,0,N}\,,\;\;\;
\widetilde{\Gamma}_{\alpha,0,N}
\;\;\longrightarrow\;\;{\rm appendix}~{\protect\ref{sec:iniN}},
\eeq
and their $z$-space counterparts here:
\beq
\Gamma_{\alpha,0}(z)\,,\;\;\;
\widetilde{\Gamma}_{\alpha,0}(z)
\;\;\longrightarrow\;\;{\rm appendix}~{\protect\ref{sec:iniZ}}.
\eeq
In keeping with our general strategy for asymptotic quantities
introduced in eqs.~(\ref{wtFNdef}) and~(\ref{wtFNdefz}),
$\widetilde{\Gamma}_{\alpha,0,N}$ and $\widetilde{\Gamma}_{\alpha,0}(z)$
include up to terms of \mbox{$\ord(1/N^2)$} and \mbox{$\ord((1-z))$},
respectively.

\section{The $\Delta$ scheme\label{sec:DS}}

\subsection{All-order definition\label{sec:DSall}}
We generalise the definition given in ref.~\cite{Frixione:2021wzh}
by constructing an operator $\Rop(t)$ which
transforms, for any given value of the evolution variable, 
the $\MSb$ PDFs into their $\Delta$-scheme counterparts, namely:
\beq
\Gamma_{N}^{(\Delta)}(t)=\Rop_N(t)\Gamma_{N}(t)
\;\;\;\;\Longrightarrow\;\;\;\;
\Gamma_{0,N}^{(\Delta)}=\Rop_N(0)\Gamma_{0,N}\,.
\label{Rrot}
\eeq
Furthermore, we assume that $\Rop_N$ be invertible, and  
defined by means of a characteristic function
\beq
r(x)=I+r_1x+\sum_{i=2}^\infty r_ix^i\,,\;\;\;\;\;\;\;\;
r_1\ne 0\,,
\label{rfun}
\eeq
through the following expressions, at order N$^k$LO:
\beqn
\Rop_N(t)&=&r\left(\Kmat_N^{(\Delta)}(t)\right)\,,
\label{Ropdef}
\\
\Kmat_N^{(\Delta)}(t)&=&\sum_{i=1}^k\left(\frac{\aem(t)}{2\pi}\right)^i
\Kmat_{N}^{(\Delta)[i]}\,.
\label{Kmatdef}
\eeqn
Thus, the evolution-parameter dependence of $\Rop_N(t)$ is only 
due to the running coupling. Furthermore, the function $r(x)$
admits a series expansion around $x=0$ whose zero-order term is
equal to the identity\footnote{In a one-dimensional flavour space,
such as that relevant to non-singlet evolution, $I\equiv 1$.} $I$,
and whose first-order term is non-null; these conditions guarantee
a smooth perturbative behaviour.

It is easy to see that the definition above leads one to that
of ref.~\cite{Frixione:2021wzh}, simply by choosing (with $k=1$)
\beq
r(x)=I+x\,.
\label{rfundef}
\eeq
Equation~(\ref{rfundef}) constitutes the simplest possible form
of $r(x)$, and will be our default in the following. Alternative
options will be discussed in sect.~\ref{sec:others}. Apart from
the choice of $r(x)$, we underline again how definitions at various
perturbative orders differ by the presence of additional terms
on the r.h.s.~of eq.~(\ref{Kmatdef}). On top of this, there is
the central issue of the very definition of the matrices 
$\Kmat_{N}^{(\Delta)[i]}$, which we now turn to discussing.

The basic idea is that of determining $\Kmat_{N}^{(\Delta)[i]}$ by imposing 
the initial conditions in the $\Delta$ scheme to have a specific form.
In view of the fact that the initial conditions are defined at
the perturbative order one is working at, this implies that
the constraint above must be imposed up to the same perturbative
order\footnote{Owing to this, only the coefficients \mbox{$r_1,\ldots r_k$}
of eq.~(\ref{rfun}) will be relevant in this determination. However,
the unexpanded form of $r(x)$ will be used within a kernel of the evolution 
equation.}. Thus, at N$^k$LO, eqs.~(\ref{Rrot})--(\ref{Kmatdef}) imply:
\beqn
\left.\Gamma_{0,N}^{(\Delta)}\right|_{\aem^k}&=&
\left.r\left(\sum_{i=1}^k\left(\frac{\aem(0)}{2\pi}\right)^i
\Kmat_{N}^{(\Delta)[i]}\right)\Gamma_{0,N}\right|_{\aem^k}
\label{Ksol1}
\\*
&\stackrel{r(x)=I+x}{=}&
\left.\left(I+\sum_{i=1}^k\left(\frac{\aem(0)}{2\pi}\right)^i
\Kmat_{N}^{(\Delta)[i]}\right)\Gamma_{0,N}\right|_{\aem^k}\,,
\label{Ksol2}
\eeqn
where we have denoted by a subscript $\aem^k$ an expansion to
$\ord(\aem^k)$ of what is on the left of that symbol. 
Equations~(\ref{Ksol1}) or~(\ref{Ksol2}) have to be solved for 
$\Kmat_{N}^{(\Delta)[i]}$ given the $\MSb$ initial conditions $\Gamma_{0,N}$,
and their required $\Delta$-scheme counterparts $\Gamma_{0,N}^{(\Delta)}$.
Although there is in principle ample freedom in the choice of the latter,
in practice one wants to deal with as simple forms as possible, 
which does happen if one imposes the $\Delta$-scheme initial conditions
to be equal to the LO ones, namely (in the physical basis):
\beq
\Gamma_{0,N}^{(\Delta)}=\left(
\begin{array}{c}
1\\
0\\
0\\
\end{array}
\right)\,.
\label{iniDelNLO}
\eeq
As a matter of fact, the strict equality of the $\Delta$-scheme initial 
conditions with eq.~(\ref{iniDelNLO}) is technically complicated and 
unnecessary, and that equality will actually be imposed in an asymptotic 
sense; fuller details are given below.

As all perturbative-compliant constraints, eqs.~(\ref{Ksol1}) 
or~(\ref{Ksol2}) must be solved order by order in $\aem$. In the $\NF=1$ 
case, at order N$^k$LO this gives one $3k$ equations to solve in $5k$ variables
(with $5$ being the number of independent elements in each of the matrices
$\Kmat_{N}^{(\Delta)[i]}$, taking into account the symmetries of QED). The 
resulting system of equations is therefore underconstrained, and one is 
entitled to set some of these elements equal to zero before solving the system. 
A convenient and non-restrictive choice is the one made\footnote{By working 
at the NLO, in that paper the element $K_{e\bar{e}}^{(\Delta)}$ could be set
equal to zero, being of relative $\ord(\aem^2)$.} in 
ref.~\cite{Frixione:2021wzh} (both in the $N$ and the $z$ space)
\beq
\Kmat_P^{(\Delta)}=\left(
\begin{array}{ccc}
K_{ee}^{(\Delta)} & 0 & K_{e\bar{e}}^{(\Delta)}  \\
K_{\gamma e}^{(\Delta)} & 0 & K_{\gamma e}^{(\Delta)} \\
K_{e\bar{e}}^{(\Delta)} & 0 & K_{ee}^{(\Delta)} \\
\end{array}
\right)
\;\;\;\;\longrightarrow\;\;\;\;
\Kmat_E^{(\Delta)}=\left(
\begin{array}{ccc}
K_{\Sigma\Sigma}^{(\Delta)} & 0 & 0 \\
K_{\gamma\Sigma}^{(\Delta)} & 0 & 0 \\
0 & 0 & K_{NS}^{(\Delta)} \\
\end{array}
\right)=
T\,\Kmat_P^{(\Delta)}\,T^{-1}
\,,
\label{KmatDel}
\eeq
with the two expressions given in the physical and evolution basis,
respectively, as the subscripts indicate (since no confusion is possible,
such subscripts will soon be removed, in the interest of a simpler
notation). 

Equation~(\ref{KmatDel}) features $3$ independent 
matrix elements per perturbative order, which
implies that now the system of constraints can be solved in a
deterministic manner. In particular, by imposing the initial
conditions in the $\Delta$ scheme to be identical to their LO
counterparts, eq.~(\ref{iniDelNLO}), and by employing eqs.~(\ref{Ksol2}) 
and~(\ref{KmatDel}), one finds the NLO solution 
of ref.~\cite{Frixione:2021wzh}, namely\footnote{In the $K$
matrices there cannot be a dependence upon $\muz$, which conversely
is present in the initial conditions if $\muz\ne m$, whence the
choice of scale in eqs.~(\ref{KeeDel}) and~(\ref{KgeDel}).
We shall further discuss this point later.}:
\beqn
K_{ee}^{(\Delta)}(z)\equiv 
\frac{\aem(0)}{2\pi}K_{ee}^{(\Delta)[1]}(z)&=&
-\frac{\aem(0)}{2\pi}\Gamma_e^{[1]}(z,\muz=m)\,,
\label{KeeDel}
\\
K_{\gamma e}^{(\Delta)}(z)\equiv 
\frac{\aem(0)}{2\pi}K_{\gamma e}^{(\Delta)[1]}(z)&=&
-\frac{\aem(0)}{2\pi}\Gamma_\gamma^{[1]}(z,\muz=m)\,.
\label{KgeDel}
\eeqn
However, in general eq.~(\ref{Ksol1}) is an overkill for the problem at hand.
One should bear in mind that the primary goal of the $\Delta$ scheme is
that of removing the soft logarithms $\log(1-z)$ from the large-$z$ 
expression of the electron PDF. While eq.~(\ref{Ksol1}) with 
eq.~(\ref{iniDelNLO}) indeed achieves that goal, so does the usage of
any expression whose {\em asymptotic} large-$z$ behaviour is the same as 
that of eq.~(\ref{iniDelNLO}), without being equal to it elsewhere in 
the $z$ range. Furthermore, by strictly imposing the equality to
eq.~(\ref{iniDelNLO}), one may actually make the small-$z$ behaviour 
of the PDFs, and specifically of the PDF of the photon, worse than 
is necessary. While this is an issue of negligible 
overall relevance at any planned $\epem$ collider,
it is still disturbing from a theoretical viewpoint, because it
shows how what is essentially a large-$z$ procedure affects the
opposite end of the $z$ spectrum as an undesirable byproduct.

The discussion above suggests that a more conservative procedure
which allows one to not affect the small-$z$ behaviour of the PDFs
stems from turning eq.~(\ref{Ksol1}) into what follows:
\beqn
\left.\Gamma_{0,N}^{(\Delta)}\right|_{\aem^k}&=&
\left.r\left(\sum_{i=1}^k\left(\frac{\aem(0)}{2\pi}\right)^i
\Kmat_{N}^{(\Delta)[i]}\right)\widetilde{\Gamma}_{0,N}\right|_{\aem^k}\,;
\label{Ksol1b}
\eeqn
we recall that by $\widetilde{\Gamma}_{0,N}$ we denote the large-$N$
asymptotic form of the $\MSb$ initial conditions, up to and including
terms of order $1/N^{\rho+1}$ for a given $\rho\ge -1$. In addition to
giving one a better separate control on the small- and large-$z$
regions, this also solves the problem associated with the fact that,
at variance with what happens at the NLO, the complete expressions
of the initial conditions at the NNLO (and thus, presumably, at yet
higher orders) are extremely involved, and thus potentially troublesome
when exponentiated (as it happens in the $\Delta$ scheme).

We point out that adding to $\widetilde{\Gamma}_{0,N}$ terms suppressed
by higher powers of $1/N$, i.e.~of $\ord(1/N^{\rho+2})$ and beyond,
is also legit. This helps one solve a potential further issue with 
eq.~(\ref{Ksol1b}). Namely, by using $\widetilde{\Gamma}_{0,N}$ instead
of the complete $\MSb$ initial conditions one is not guaranteed any 
longer of an automatic compliance of the $\Delta$-scheme initial
conditions with conservation laws -- specifically, flavour and momentum
conservation. In order to restore those, the solution anticipated above
amounts to imposing what follows, rather than eq.~(\ref{Ksol1b}):
\beqn
\left.\Gamma_{0,N}^{(\Delta)}\right|_{\aem^k}&=&
\left.r\left(\sum_{i=1}^k\left(\frac{\aem(0)}{2\pi}\right)^i
\left(\widetilde{\Kmat}_{N}^{(\Delta)[i]}+
\Cmat_{N}^{[i]}\right)\right)\widetilde{\Gamma}_{0,N}\right|_{\aem^k}\,.
\label{Ksol1c}
\eeqn
Here, we have denoted by $\widetilde{\Kmat}_{N}^{(\Delta)[i]}$ the solutions
of eq.~(\ref{Ksol1c}) obtained by setting \mbox{$\Cmat_{N}^{[i]}=0$} (thus, 
these are potentially flavour and momentum-violating solutions); then, 
$\Cmat_{N}^{[i]}$ are determined by imposing that the $\Delta$-scheme 
initial conditions given by the l.h.s.~of eq.~(\ref{Ksol1c}) do conserve 
flavour and momentum. As was discussed before, this two-step procedure is 
self-consistent only if each element of $\Cmat_{N}^{[i]}$ is suppressed 
by a power of $N$ at least one unity larger than the one which gives the 
largest suppression in the adopted asymptotic forms of the $\MSb$ initial 
conditions; that is, the elements of $\Cmat_{N}^{[i]}$ must be of
$\ord(1/N^{\rho+2})$ and higher.

The possibility of choosing the parameter $\rho$ implies that, once
the initial conditions in the $\Delta$ scheme are defined in the Mellin
space, their $z$-space counterparts are unambiguously defined only up to
and including terms of $\ord((1-z)^\rho)$. In other words, the Mellin 
inversion of those initial conditions is itself ambiguous, and the resolution
of such an ambiguity must become part of the prescription adopted.
One can, for example, choose to employ the {\em exact} inverse Mellin
transforms of eqs.~(\ref{invMlgzplus}) and~(\ref{invMlgz}). While
natural, this choice has two drawbacks: {\em i)} the use of $\log z$
as a core variable in the resulting functions and distributions is
highly non-standard, and can render it difficult to manipulate 
subsequent expressions at the same time as the complete $\MSb$ initial
conditions; {\em ii)} quantities which contain powers of $\log z$
are again liable to worsen the small-$z$ behaviour of the resulting
PDFs. In order to address both problems at once, we introduce what
we call the:\\

\begin{center}
\begin{minipage}{0.95\textwidth}
$\bullet$ {\em $\log z\to 1-z$ prescription}. Loosely speaking, we define
it as follows: given a polynomial in $\log\bN$, with coefficients
themselves polynomial in $1/N$:
\beq
\widetilde{P}_N\big(\log\bN,N\big)=\sum_k\sum_i p_{ki}\frac{\log^k\bN}{N^i}\,,
\label{lzomz1}
\eeq 
we turn it into a $z$-space expression 
by means of eqs.~(\ref{invMlgzplus}) and~(\ref{invMlgz}):
\beqn
&&M^{-1}\left[\widetilde{P}_N(\log\bN,N)\right]=
\sum_k\sum_i p_{ki}M^{-1}\left[\frac{\log^k\bN}{N^i}\right]
\label{lzomz2}
\\*&&\phantom{aaaaaaaaa}
=\sum_m\sum_q r_{mq}(1-z)^q\frac{(\log(-\log z))^m}{\log z}=
P\left(\log(-\log z),\frac{1}{\log z}\right)\,,
\nonumber
\eeqn
where for simplicity we have omitted any plus-distribution notations,
and understood contributions proportional to $\delta(1-z)$.
We next convert all of the $\log z$ terms into terms that 
feature powers of $\log(1-z)$, by means of eqs.~(\ref{lgzvslgom3}) 
and~(\ref{lgzvslgom4}):
\beqn
&&\overline{P}^{({\rm mod})}\left(\log(1-z),\frac{1}{1-z}\right)
\label{lzomz3}
\\*&&\phantom{aaa}=
P\left(\log(-\log z)\to \log(1-z)\sum_{s=0}^\rho(1-z)^s,
\frac{1}{\log z}\to\frac{1}{1-z}\sum_{r=0}^\rho(1-z)^r\right).
\nonumber
\eeqn
For consistency with the function thus obtained, we have therefore 
to modify the original $N$-space expression by using the exact Mellin 
transforms of eqs.~(\ref{Mlgomzplus}) and~(\ref{Mlgomz}):
\beqn
M\left[\overline{P}^{({\rm mod})}\left(\log(1-z),\frac{1}{1-z}\right)\right]=
\overline{P}_N^{({\rm mod})}\Big(\psi_k(N),\Gamma(N+q)\Big).
\label{lzomz4}
\eeqn
By construction:
\beq
\widetilde{P}_N\big(\log\bN,N\big)=
\overline{P}_N^{({\rm mod})}\Big(\psi_k(N),\Gamma(N+q)\Big)+
\ord\left(\frac{1}{N^{\rho+2}}\right).
\eeq
The whole prescription amounts to applying 
eqs.~(\ref{GNfun})--(\ref{czMlgNappr}).
\end{minipage}
\end{center}

\noindent
This prescription is convenient since subsequent operations will require the 
computations of products of asymptotic expressions and of their inverse Mellin 
transforms, which are series not necessarily summable in a closed analytical
form. After the conversions to expressions in terms of $\log(1-z)$ in 
configuration space, and of polygamma's and rational functions of $N$
in Mellin space, the former are standard variables, and the inverse Mellin 
transforms of the latter are finite sums which can be efficiently handled 
e.g.~by means of the \texttt{Sigma} and \texttt{HarmonicSums} packages.

In summary, the definition of initial conditions in the $\Delta$
scheme is achieved by means of the following steps:
\begin{enumerate}
\item Choose a function $r(x)$ according to eq.~(\ref{rfun}).
\item After choosing an integer $\rho\ge -1$, define the elements of
$\widetilde{\Gamma}_{0,N}$ to be the large-$N$ asymptotic forms
of the $\MSb$ initial conditions, i.e.~some polynomials in $\log\bN$
with coefficients polynomial in $1/N$ of degree \mbox{$\rho+1$}.
Any function of $\log\bN$ suppressed by at least \mbox{$1/N^{\rho+2}$}
may be included as well.
\item Solve eq.~(\ref{Ksol1c}) for the elements of 
$\widetilde{\Kmat}_{N}^{(\Delta)[i]}$, with these matrices in the
form of eq.~(\ref{KmatDel}), after setting $\Cmat_{N}^{[i]}=0$ and
the l.h.s.~of eq.~(\ref{Ksol1c}) equal to eq.~(\ref{iniDelNLO}). Order
by order in $\aem$, these solutions will have polynomial structures
in $\log\bN$ and $1/N$ of degrees not higher than those of the
asymptotic $\MSb$ initial conditions they stem from.
\item $\log z\to 1-z$ prescription: 
transform each element of $\widetilde{\Kmat}_{N}^{(\Delta)[i]}$ into 
$\overline{\Kmat}_{N}^{(\Delta)[i]}$, according to eq.~(\ref{cNMlgNappr}); 
their exact $z$-space counterparts\footnote{Owing to the fact that
we use exact formulae for Mellin and inverse-Mellin transforms, these
operations could equivalently be achieved by starting from a series
expansion in the configuration space.} are given by eq.~(\ref{czMlgNappr}).
\item Choose a functional form for each element of the momentum- and 
flavour-conservation restoring terms $\Cmat_{N}^{[i]}$, suppressed by 
at least \mbox{$1/N^{\rho+2}$}, and let their normalisations float.
\item Plug the $\overline{\Kmat}_{N}^{(\Delta)[i]}$ matrices so obtained
back into the definition of $\Rop$ per eqs.~(\ref{Ropdef}) 
and~(\ref{Kmatdef}), and also include there the matrices $\Cmat_{N}^{[i]}$ 
chosen in the previous step, thus arriving at:
\beq
\Kmat_{N}^{(\Delta)[i]}:=
\overline{\Kmat}_{N}^{(\Delta)[i]}+\Cmat_{N}^{[i]}
\;\;\;\;\Longrightarrow\;\;\;\;
\Rop_N(t)=r\left(
\sum_{i=1}^k\left(\frac{\aem(t)}{2\pi}\right)^i
\left(\overline{\Kmat}_{N}^{(\Delta)[i]}+
\Cmat_{N}^{[i]}\right)
\right).
\label{RopdefDelf}
\eeq
Equation~(\ref{RopdefDelf}) is finally what is employed in the definition
of the initial conditions in the $\Delta$ scheme according to the rightmost
side of eq.~(\ref{Rrot}), expanded to $\ord(\aem^k)$; the so-far floating 
normalisations of the $\Cmat_{N}^{[i]}$ matrices are determined by imposing 
the conservation of flavour and momentum.
\end{enumerate}
We stress that by construction (see appendix~\ref{sec:Mell}) we have:
\beq
\widetilde{\Kmat}_{N}^{(\Delta)[i]}-
\left(\overline{\Kmat}_{N}^{(\Delta)[i]}+
\Cmat_{N}^{[i]}\right)=\ord\left(\frac{1}{N^{\rho+2}}\right)\,.
\label{wKvsoK}
\eeq

\subsection{A simple example\label{sec:DSex}}
In order to give an explicit demonstration of the procedure outlined in
sect.~\ref{sec:DSall}, let us consider an academic case, and let us 
follow step by step the items of the list above. For further simplicity, 
this example will be relevant to the non-singlet component at $\ord(\aem)$,
and therefore all of the matrices and column vectors will become 
c-numbers. Let us assume the $\MSb$ initial conditions to be
\beqn
\Gamma_{0,N}&=&1+\frac{\aem(0)}{2\pi}\Gamma_{0,N}^{[1]}
\label{exa0}
\\
\Gamma_{0,N}^{[1]}&=&\psi_0(N)+\frac{\psi_1(N+1)}{N}+1-\frac{\pi^2}{6}+\gE\,,
\label{exa1}
\\*
\invM\Big[\Gamma_{0,N}^{[1]}\Big]&=&
\left(1-\frac{\pi^2}{6}\right)\delta(1-z)-\pplus{\frac{1}{1-z}}+
\frac{\pi^2}{6}-\log(1-z)\log z-{\rm Li}_2(z)\,,\phantom{aaa}
\label{exa2}
\eeqn
whose first moment is equal to one (non-singlet flavour conservation).
We can now follow the procedure outlined before.
\begin{enumerate}
\item We choose $r(x)=1+x$ as in eq.~(\ref{rfundef}).
\item We choose $\rho=1$. Therefore, we shall need to retain up to 
terms of $\ord(1/N^2)$ in the asymptotic forms of the $\MSb$ initial
conditions. By direct calculation from eq.~(\ref{exa1}):
\beqn
\widetilde{\Gamma}_{0,N}^{[1]}=
\log\bN+1-\frac{\pi^2}{6}-\frac{1}{2N}+ \frac{11}{12 N^2}\,.
\label{exa3}
\eeqn
We do not include any extra function suppressed by $1/N^3$ or 
higher powers.
\item The solution of eq.~(\ref{Ksol1c}) with the l.h.s.~equal
to one (i.e.~with the coefficient of the term of $\ord(\aem)$ equal 
to zero), and with $\Cmat_{N}^{[i]}=0$ is then
\beq
\widetilde{K}_{N}^{(\Delta)[1]}=-\widetilde{\Gamma}_{0,N}^{[1]}\,.
\label{exaKsolc1}
\eeq
\item $\log z\to 1-z$ prescription: by using eq.~(\ref{cNMlgNappr}) 
in eq.~(\ref{exa3}) we obtain the sought transformed $K$-matrix 
element in Mellin space:
\beqn
\widetilde{K}_{N}^{(\Delta)[1]}\;\longrightarrow\;
\overline{K}_{N}^{(\Delta)[1]}=-\left(
\gE+\psi_0(N)+1-\frac{\pi^2}{6}+\frac{1}{N(1+N)}
\right),
\label{exa4}
\eeqn
whence
\beqn
\invM\left[\overline{K}_{N}^{(\Delta)[1]}\right]=-\left[
\left(1-\frac{\pi^2}{6}\right)\delta(1-z)-\pplus{\frac{1}{1-z}}
+(1-z)\right].
\label{exa5}
\eeqn
\item We choose
\beq
C^{[1]}=\beta(1-z)^2\;\;\;\;\Longrightarrow\;\;\;\;
C_N^{[1]}=\frac{2\beta}{N(N+1)(N+2)}\,,
\label{exaCchoice}
\eeq
which is of $\ord(1/N^3)$ in Mellin space, in keeping with 
our choice of $\rho$.
\item With these choices, eq.~(\ref{RopdefDelf}) leads to:
\beq
\Gamma_{0,N}^{(\Delta)}=
\Rop_N(t)\Gamma_{0,N}=1+\frac{\aem(0)}{2\pi}\left(\Gamma_{0,N}^{[1]}+
\overline{K}_{N}^{(\Delta)[1]}+C_N^{[1]}\right)+\ord(\aem^2)\,,
\label{exafin}
\eeq
with the quantities in round brackets taken from eqs.~(\ref{exa1}),
(\ref{exa4}), and~(\ref{exaCchoice}). In this expression, there
is still the parameter $\beta$ which is undetermined. We can compute
its value by imposing flavour conservation, i.e.~the requirement that
first moment ($N=1$) of the r.h.s.~of eq.~(\ref{exafin}) be equal
to one, which leads one to:
\beq
\beta=\frac{\pi^2}{2}-\frac{9}{2}\,.
\eeq
\end{enumerate}
In summary, eq.~(\ref{exafin}) implies
\beqn
\Gamma_{0,N}^{(\Delta)}&=&1+\frac{\aem(0)}{2\pi}\Gamma_{0,N}^{(\Delta)[1]}\,,
\label{exaf1}
\\*
\Gamma_{0,N}^{(\Delta)[1]}&=&
\frac{\psi_1(N+1)}{N}-\frac{1}{N(1+N)}+
\frac{9-\pi^2}{N(N+1)(N+2)}
\label{exaf2}
\\*&=&
\frac{19-2\pi^2}{2N^3}+\ord\left(\frac{1}{N^4}\right)\,,
\label{exaf3}
\eeqn
and therefore
\beqn
\Gamma_{0,N}^{(\Delta)}(z)&=&\delta(1-z)+
\frac{\aem(0)}{2\pi}\Gamma_{0}^{(\Delta)[1]}(z)\,,
\label{exaf4}
\\*
\Gamma_{0}^{(\Delta)[1]}(z)&=&
\frac{\pi^2}{6}-1+z+\half(9-\pi^2)(1-z)^2
-\log(1-z)\log z-{\rm Li}_2(z)
\label{exaf5}
\\*&=&
\frac{1}{4}\left(19-2\pi^2\right)(1-z)^2+\ord((1-z)^3)\,.
\label{exaf6}
\eeqn
We observe that had we literally followed ref.~\cite{Frixione:2021wzh} 
the $\Delta$-scheme initial conditions would not have any contribution
of $\ord(\aem)$. However, as is shown in eqs.~(\ref{exaf3})
and~(\ref{exaf6}), the prescription proposed here leads to non-null
terms only beyond the chosen accuracy $\rho$, the results being
of order \mbox{$1/N^3\equiv 1/N^{\rho+2}$} in Mellin space,
and of order \mbox{$(1-z)^2\equiv (1-z)^{\rho+1}$} in configuration
space. We also point out that the small-$z$ behaviour of
eq.~(\ref{exaf5}) is the same as that of eq.~(\ref{exaf2}),
i.e.~the leading term is proportional to $z\log z$. For this to
happen, it is crucial that the transformation advocated in item~4
be applied. In fact, without that one would need to employ
\beqn
\invM\left[\widetilde{K}_{N}^{(\Delta)[1]}\right]=-\left\{
\left(1-\frac{\pi^2}{6}+\gE\right)\delta(1-z)-\pplus{\frac{1}{\log z}}
-\half-\frac{11\log z}{12}\right\},
\label{exa6d}
\eeqn
which diverges at $z\to 0$, thus presenting an evidence of the
small-$z$ issue we have repeatedly mentioned. While eqs.~(\ref{cNMlgNappr})
and~(\ref{czMlgNappr}) constitute our default solution to this problem,
another rather straightforward possibility is that of defining
\beq
\overline{\Kmat}_{N}^{(\Delta)[i]}:=
\widetilde{\Kmat}_{N+\delta N}^{(\Delta)[i]}\,,\;\;\;\;
\delta N=1~{\rm or}~2\,,
\eeq
which corresponds to multiplying the Mellin transform of
$\widetilde{\Kmat}_{N}^{(\Delta)[i]}$ by $z$ or $z^2$ in
configuration space. Such an option will not be pursued
any further in this paper.

This discussion, which emphasises the freedom inherent to the
definition of the $\Delta$ scheme through various ambiguities
in the steps that lead to its actual construction, should ultimately 
render it clear that such a scheme is more appropriately thought
as a {\em class} of schemes, all characterised by the absence of
soft logarithms at large $z$'s, but otherwise flexible enough to
accommodate possibly different needs associated with different perturbative
orders and collider configurations. Having said that, it is paramount
to keep in mind that all of these differences will be compensated
by their counterparts at the short-distance cross section level,
up to terms beyond the perturbative accuracy one is working at.

\subsection{$\Delta$ at the NNLO\label{sec:DSnnlo}}
We now finally consider the actual construction of the $\Delta$ scheme at
the NNLO. As was done in the example of sect.~\ref{sec:DSex}, we choose for 
simplicity $r(x)$ according to eq.~(\ref{rfundef}), and set $\rho=1$. 
With this, the results reported in appendix~\ref{sec:res} (see in 
particular eq.~(\ref{iniMSb}) and the coefficients\footnote{In these
coefficients, and as a consequence of that in the whole of the
$\Delta$ scheme, we set $L_0=0$. While in principle this constitutes
a small loss in generality, in practice one never chooses $L_0\ne 0$,
since that logarithm cannot be resummed by the evolution of the PDFs,
and is thus best eliminated\label{ft:L0}.} which appear there) are
all is needed to construct the large-$N$ $\ord(\aem^2)$ $\MSb$
initial conditions that enter eq.~(\ref{Ksol1c}) (steps~1 and 2). 
The matrix elements which solve that equation (step~3) are then 
the following ones (here given in the evolution basis):
\beqn
\widetilde{K}_{NS,N}^{(\Delta)[1]}&=&
-\widetilde{\Gamma}_{NS,0,N}^{[1]}\,,
\label{wKaswG1}
\\
\widetilde{K}_{NS,N}^{(\Delta)[2]}&=&
\left(\widetilde{\Gamma}_{NS,0,N}^{[1]}\right)^2
-\widetilde{\Gamma}_{NS,0,N}^{[2]}\,,
\label{wKaswG2}
\\
\widetilde{K}_{\Sigma\Sigma,N}^{(\Delta)[1]}&=&
-\widetilde{\Gamma}_{\Sigma,0,N}^{[1]}\,,
\label{wKaswG3}
\\
\widetilde{K}_{\Sigma\Sigma,N}^{(\Delta)[2]}&=&
\left(\widetilde{\Gamma}_{\Sigma,0,N}^{[1]}\right)^2
-\widetilde{\Gamma}_{\Sigma,0,N}^{[2]}\,,
\label{wKaswG4}
\\
\widetilde{K}_{\gamma\Sigma,N}^{(\Delta)[1]}&=&
-\widetilde{\Gamma}_{\gamma,0,N}^{[1]}\,,
\label{wKaswG5}
\\
\widetilde{K}_{\gamma\Sigma,N}^{(\Delta)[2]}&=&
\widetilde{\Gamma}_{\gamma,0,N}^{[1]}\widetilde{\Gamma}_{\Sigma,0,N}^{[1]}
-\widetilde{\Gamma}_{\gamma,0,N}^{[2]}\,.
\label{wKaswG6}
\eeqn
The subsequent transformation 
$\widetilde{\Kmat}_{N}^{(\Delta)[i]}\to 
\overline{\Kmat}_{N}^{(\Delta)[i]}$
of step~4, done according to eq.~(\ref{cNMlgNappr}), implies the usage
of the transformed asymptotic $\MSb$ initial conditions, 
i.e.~eqs~(\ref{wKaswG1}) and~(\ref{wKaswG2}) lead to:
\beqn
\widetilde{K}_{NS,N}^{(\Delta)[1]}\;\longrightarrow\;
\overline{K}_{NS,N}^{(\Delta)[1]}&=&
-\overline{\Gamma}_{NS,0,N}^{[1]}\,,
\label{oKasoG1}
\\
\widetilde{K}_{NS,N}^{(\Delta)[2]}\;\longrightarrow\;
\overline{K}_{NS,N}^{(\Delta)[2]}&=&
\left(\overline{\Gamma}_{NS,0,N}^{[1]}\right)^2
-\overline{\Gamma}_{NS,0,N}^{[2]}\,,
\label{oKasoG2}
\eeqn
and so forth. Note that in eq.~(\ref{oKasoG2}) we have implicitly
resolved an ambiguity, since squaring the transform of
$\widetilde{\Gamma}_{NS,0,N}^{[1]}$ is in general not identical 
to transforming the square of that quantity. However:
\beq
\left(\overline{\Gamma}_{NS,0,N}^{[1]}\right)^2=
\overline{\left(\widetilde{\Gamma}_{NS,0,N}^{[1]}\right)^2}+
\ord\left(\frac{1}{N^{\rho+2}}\right)\,,
\eeq
i.e.~the two operations are indeed equivalent at the order
at which we are working. Finally, after adding the flavour- and
momentum-restoring matrices and fixing their normalisation we
arrive at the sought initial conditions:
\beqn
\Gamma_{NS,0,N}^{(\Delta)[1]}&=&
\Gamma_{NS,0,N}^{[1]}-
\overline{\Gamma}_{NS,0,N}^{[1]}
+C_{NS,N}^{[1]}\,,
\\
\Gamma_{NS,0,N}^{(\Delta)[2]}&=&
\Gamma_{NS,0,N}^{[2]}-
\Gamma_{NS,0,N}^{[1]}\overline{\Gamma}_{NS,0,N}^{[1]}+
\left(\overline{\Gamma}_{NS,0,N}^{[1]}\right)^2
\nonumber\\*&&
-\overline{\Gamma}_{NS,0,N}^{[2]}+
\Gamma_{NS,0,N}^{[1]}C_{NS,N}^{[1]}
+C_{NS,N}^{[2]}\,,
\\
\Gamma_{\Sigma,0,N}^{(\Delta)[1]}&=&
\Gamma_{\Sigma,0,N}^{[1]}-
\overline{\Gamma}_{\Sigma,0,N}^{[1]}
+C_{NS,N}^{[1]}\,,
\\
\Gamma_{\Sigma,0,N}^{(\Delta)[2]}&=&
\Gamma_{\Sigma,0,N}^{[2]}-
\Gamma_{\Sigma,0,N}^{[1]}\overline{\Gamma}_{\Sigma,0,N}^{[1]}+
\left(\overline{\Gamma}_{\Sigma,0,N}^{[1]}\right)^2
\nonumber\\*&&
-\overline{\Gamma}_{\Sigma,0,N}^{[2]}+
\Gamma_{\Sigma,0,N}^{[1]}C_{\Sigma,N}^{[1]}
+C_{\Sigma,N}^{[2]}\,,
\\
\Gamma_{\gamma,0,N}^{(\Delta)[1]}&=&
\Gamma_{\gamma,0,N}^{[1]}-
\overline{\Gamma}_{\gamma,0,N}^{[1]}
+C_{\gamma,N}^{[1]}\,,
\\
\Gamma_{\gamma,0,N}^{(\Delta)[2]}&=&
\Gamma_{\gamma,0,N}^{[2]}-
\Gamma_{\Sigma,0,N}^{[1]}\overline{\Gamma}_{\gamma,0,N}^{[1]}+
\overline{\Gamma}_{\Sigma,0,N}^{[1]}\overline{\Gamma}_{\gamma,0,N}^{[1]}+
\nonumber\\*&&
-\overline{\Gamma}_{\gamma,0,N}^{[2]}+
\Gamma_{\Sigma,0,N}^{[1]}C_{\gamma,N}^{[1]}
+C_{\gamma,N}^{[2]}\,.
\eeqn
The flavour- and momentum-restoring matrix elements 
have been chosen thus:
\beqn
\alpha_{i,j}&=&1\,,
\label{aijv}
\\
C_{NS}^{[1]}(z)&=&\alpha_{3,1}f_{3,1}(1-z)^3\,,
\\
C_{NS}^{[2]}(z)&=&\alpha_{3,2}f_{3,2}(1-z)^3\,,
\\
C_{\Sigma}^{[1]}(z)&=&\alpha_{1,1}f_{1,1}(1-z)^3\,,
\\
C_{\Sigma}^{[2]}(z)&=&\alpha_{1,2}f_{1,2}(1-z)^3\,,
\\
C_{\gamma}^{[1]}(z)&=&\alpha_{2,1}f_{2,1}\frac{(1-z)^2}{z}\,,
\\
C_{\gamma}^{[2]}(z)&=&\alpha_{2,2}f_{2,2}(1-z)^2\,,
\eeqn
Momentum and flavour conservation lead to the following values of
the floating normalisations\footnote{Flavour conservation (first Mellin
moment) constraints the normalisation of the non-singlet, $f_{3,1}$ and 
$f_{3,2}$ (the former turns out to be equal to zero since by setting 
$\rho=1$ the $\MSb$ asymptotic initial conditions are identical to the 
exact ones).
Momentum conservation (second Mellin moments) constraints the 
singlet-plus-photon system. Hence, two floating normalisations are
left unconstrained -- we choose them to be $f_{1,1}$ and $f_{1,2}$,
and set their values equal to one. These choices can be freely modified.
\label{ft:params}
}
\beqn
f_{3,1}&=&0\,,
\\
f_{3,2}&=&\frac{3071}{24}+\frac{9971\NF}{810}
-\frac{4\pi^2\NF}{3}
-\frac{\pi^2}{9}-8\pi^2\log (2)
-\frac{5\pi^4}{9}
+\frac{8\NF\zeta_3}{3}-22\zeta_3\,,
\\              
f_{1,1}&=&1\,,
\\              
f_{1,2}&=&1\,,
\\              
f_{2,1}&=&-\frac{409}{60}\,,
\\            
f_{2,2}&=&\frac{724717}{1080}+\frac{6367\NF}{270}
-4\pi^2\NF-19\pi^2-24\pi^2\log (2)
-\frac{5\pi^4}{3}
\nonumber\\*&&\phantom{\Big(}
-162\zeta_3+8\NF\zeta_3\,.
\eeqn
The constants $\alpha_{i,j}$ must be set equal to one in any practical 
applications, as is indicated in eq.~(\ref{aijv}). However, in the results
which follow, we still have kept them as symbols, in order for the reader
to be able to make different choices for the flavour- and momentum-restoring
functional forms w.r.t.~those adopted here. While such choices are arbitrary,
they must reproduce the first and second moments stemming from the functions
above, since those are guaranteed to lead to flavour and momentum 
conservation. In particular, one finds:
\beqn
\int_0^1 dz C_{NS}^{[i]}(z)&=&\frac{1}{4}\alpha_{3,i}f_{3,i}\,,
\\
\int_0^1 dz z\left(C_{\Sigma}^{[1]}(z)+C_{\gamma}^{[1]}(z)\right)&=&
\frac{1}{20}\alpha_{1,1}f_{1,1}+
\frac{1}{3}\alpha_{2,1}f_{2,1}\,,
\\
\int_0^1 dz z\left(C_{\Sigma}^{[2]}(z)+C_{\gamma}^{[2]}(z)\right)&=&
\frac{1}{20}\alpha_{1,2}f_{1,2}+
\frac{1}{12}\alpha_{2,2}f_{2,2}\,.
\eeqn
The explicit expressions of the $K$-matrix elements are reported in the
subsections of appendix~\ref{sec:resK}. We use
the evolution basis \mbox{($\alpha\in\{\Sigma\Sigma,\gamma\Sigma,NS\}$)},
and give the $N$-space results here ($k=1,2$):
\beq
K_{\alpha,N}^{(\Delta)[k]}\,,\;\;\;
K_{\alpha,N}^{(\Delta)[1]}K_{\beta,N}^{(\Delta)[1]}
\;\;\longrightarrow\;\;{\rm appendix}~{\protect\ref{sec:KN}}\,,
\eeq
while their $z$-space counterparts are found here:
\beq
K_{\alpha}^{(\Delta)[k]}(z)\,,\;\;\;
K_{\alpha}^{(\Delta)[1]}(z)K_{\beta}^{(\Delta)[1]}(z)
\;\;\longrightarrow\;\;{\rm appendix}~{\protect\ref{sec:KZ}}\,.
\eeq

\section{Evolution operator\label{sec:eopK}}
In this section we write the evolution equations for the 
PDFs~\cite{Gribov:1972ri,Lipatov:1974qm,Altarelli:1977zs,Dokshitzer:1977sg}
in keeping with what has been done in 
refs.~\cite{Bertone:2019hks,Frixione:2021wzh}, namely
by introducing a PDF evolution operator, and by defining it so as it 
works in a generic factorisation scheme. The present derivation has
significant overlaps with that of those papers; in the interest of
a self-contained presentation we report here the basic steps, emphasising
only those which are relevant to going beyond the NLO accuracy of
refs.~\cite{Bertone:2019hks,Frixione:2021wzh}.

The $\MSb$ evolution equations for a given particle type (i.e.~the electron
in our case) are written in Mellin space as follows:
\beq
\frac{\partial \Gamma_N(\mu)}{\partial\log\mu^2}=
\frac{\aem(\mu)}{2\pi}\,\APmat_N(\mu)\,\Gamma_N(\mu)=
\sum_{i=0}^\infty\left(\frac{\aem(\mu)}{2\pi}\right)^{i+1}
\APmat_N^{[i]}\,\Gamma_N(\mu)\,.
\label{matAPmell}
\eeq
The PDFs are regarded as
evolved from the values they assume at the given scale $\muz\sim m$
by means of an evolution operator (an $n\times n$ matrix, with
$n=2\NF+1$ in QED), thus:
\beq
\Gamma_N(\mu)=\Eop_N(\mu,\muz)\,\Gamma_{0,N}\,,\;\;\;\;\;\;
\Eop_N(\muz,\muz)=I\,.
\label{FvsE}
\eeq
Equation~(\ref{matAPmell}) is then equivalent to:
\beqn
\frac{\partial \Eop_N(\mu,\muz)}{\partial\log\mu^2}&=&
\sum_{i=0}^\infty\left(\frac{\aem(\mu)}{2\pi}\right)^{i+1}
\APmat_N^{[i]}\,\Eop_N(\mu,\muz)\,.
\label{matAPmell3}
\eeqn
As was anticipated in sect.~\ref{sec:conv}, in order to take the running 
of $\aem$ into account in an easier manner, one uses the variable $t$ of 
eq.~(\ref{tdef}) instead of the scale $\mu$, whereby eq.~(\ref{matAPmell3}) 
becomes:
\beqn
\frac{\partial \Eop_N(t)}{\partial t}&=&
\frac{b_0\aem^2(t)}{\beta(\aem(t))}
\sum_{i=0}^\infty\left(\frac{\aem(t)}{2\pi}\right)^i
\APmat_N^{[i]}\,\Eop_N(t)\,.
\label{matAPmell4}
\eeqn
In order to generalise the $\MSb$ result of
eq.~(\ref{matAPmell4}) to a generic factorisation scheme $K$, we need
to introduce fictitious short-distance cross sections. Since this
will be sufficient for achieving our goals, we shall consider only
one incoming particle, and therefore such cross sections can be
collected in a column vector (as opposed to a square matrix), analogously 
to what has been done for the PDFs (eq.~(\ref{FN})):
\beq
\hS=\left(
\begin{array}{c}
\hsig_{\alpha_1}\\
\vdots\\
\hsig_{\alpha_n}\\
\end{array}
\right)\,,
\label{hxsec}
\eeq
each element of which thus corresponds to a partonic channel. The
(physical) particle-level cross section in Mellin space therefore
follows from the factorisation theorem, i.e.~is the incoherent sum of the 
PDFs convoluted with the cross sections\footnote{Equation~(\ref{physxsec})
is the Mellin-space version of the one-leg factorisation theorem. The
$N$ vs $N+1$ structure of that formula stems from assuming that the
short-distance cross sections include the partonic flux.\label{ft:facth}}:
\beq
\sigma_N(\mu)=\hS_N^{\rm T}(\mu)\,\Gamma_{N+1}(\mu)=
\hS_N^{\rm T}(\mu)\,\Eop_{N+1}(\mu,\muz)\,\Gamma_{0,N+1}\,.
\label{physxsec}
\eeq
Note that the RGE invariance of the physical observables implies that:
\beq
\frac{\partial\sigma_N(\mu)}{\partial\log\mu^2}=0+\ord(\aem^{b+k+1})\,,
\eeq
where $b$ is the power of $\alpha$ that factors out at the Born level,
and $k$ is the accuracy of the computation ($k=0$ being the LO, $k=1$ the 
NLO, and so forth, in keeping with the notation established before).

The functional form on the r.h.s.~of eq.~(\ref{physxsec}) must have
a general validity, i.e.~it should apply to any factorisation scheme $K$.
Therefore:
\beq
\sigma_N(\mu)=
\hS_N^{{(K)}^{\rm T}}(\mu)\,\Eop_{N+1}^{(K)}(\mu,\muz)\,
\Gamma_{0,N+1}^{(K)}\,.
\label{physxsecK}
\eeq
Equations~(\ref{physxsec}) and~(\ref{physxsecK}) establish the
factorisation-scheme invariance of observables, at the same time
implicitly implying that the contributions to such observables
(i.e.~the short-distance cross sections, the evolution operator,
and the initial conditions) are all scheme dependent. The cancellation
of such a dependence constitutes a powerful check on the correctness
of the results, and (since it occurs only up to the perturbative order
one is working at) provides one with an estimate of the impact of
factorisation-scheme-independence violations due to missing higher orders.

We equate the r.h.s.~of ~(\ref{physxsec}) and~(\ref{physxsecK}),
trade the $\mu$ dependence for that upon $t$, and suitably insert
the identity in the following representations
\beq
I=\Rop_{N+1}^{-1}(t)\Rop_{N+1}(t)=\Rop_{N+1}^{-1}(0)\Rop_{N+1}(0)\,,
\label{IRR}
\eeq
where $\Rop$ is the analogue of (and coincides with, when $K=\Delta$) 
the operator introduced in sect.~\ref{sec:DS}. We obtain:
\beqn
&&\hS_N^{\rm T}(t)\,\Rop_{N+1}^{-1}(t)\Rop_{N+1}(t)\,
\Eop_{N+1}(t)\,\Rop_{N+1}^{-1}(0)\Rop_{N+1}(0)\,\Gamma_{0,N+1}
\nonumber
\\*&&\phantom{aaa}=
\hS_{N}^{(K)^{\rm T}}(t)\,\Eop_{N+1}^{(K)}(t)\,\Gamma_{0,N+1}^{(K)}\,.
\label{sigeqsig}
\eeqn
In keeping with eq.~(\ref{Rrot}) we must have
\beq
\Gamma_{0,N+1}^{(K)}=\Rop_{N+1}(0)\Gamma_{0,N+1}\,,
\label{RrotK}
\eeq
whence eq.~(\ref{sigeqsig}) leads us to
\beqn
\hS_{N}^{(K)}(t)&=&
\Rop_{N+1}^{-1^{\rm T}}(t)\,\hS_N(t)\,,
\label{xsecrot}
\\*
\Eop_{N}^{(K)}(t)
&=&
\Rop_{N}(t)\,\Eop_{N}(t)\,\Rop_{N}^{-1}(0)\,.
\label{Eoprot}
\eeqn
Equation~(\ref{xsecrot}) defines the $K$-scheme short-distance
cross sections in terms of their $\MSb$ counterparts. Conversely,
eq.~(\ref{Eoprot}) can be used to arrive at the evolution equation
for $\Eop_{N}^{(K)}(t)$. In order to do that, one derives w.r.t.~$t$
the two sides of that equation:
\beq
\frac{\partial\Eop_{N}^{(K)}(t)}{\partial t}
=
\frac{\partial\Rop_{N}(t)}{\partial t}\,\Eop_{N}(t)\,\Rop_{N}^{-1}(0)+
\Rop_{N}(t)\,\frac{\partial\Eop_{N}(t)}{\partial t}\,\Rop_{N}^{-1}(0)\,.
\eeq
We can now use eqs.~(\ref{matAPmell4}), (\ref{IRR}), and~(\ref{Eoprot})
to arrive at the sought result:
\beqn
\frac{\partial\Eop_{N}^{(K)}(t)}{\partial t}&=&
\frac{\partial\Rop_{N}(t)}{\partial t}\,\Rop_{N}^{-1}(t)\,\Eop_{N}^{(K)}(t)
\nonumber\\*&+&
\frac{b_0\aem^2(t)}{\beta(\aem(t))}
\sum_{i=0}^\infty\left(\frac{\aem(t)}{2\pi}\right)^i
\Rop_{N}(t)\,\APmat_N^{[i]}\,\Rop_{N}^{-1}(t)\,\Eop_N^{(K)}(t)\,.
\label{Eopevol}
\eeqn
This equation encompasses the $\MSb$ case of eq.~(\ref{matAPmell4}),
and generalises the result of eq.~(2.25) of ref.~\cite{Frixione:2021wzh}. 
As is shown in that paper, it is crucial that the kernel of the leftmost 
term on the r.h.s.~of this equation not be expanded in $\aem$ and then 
truncated; conversely, the kernel of the rightmost term may be expanded, 
but also kept as is.

\begin{figure}[thb]
  \begin{center}
  \includegraphics[width=0.47\textwidth]{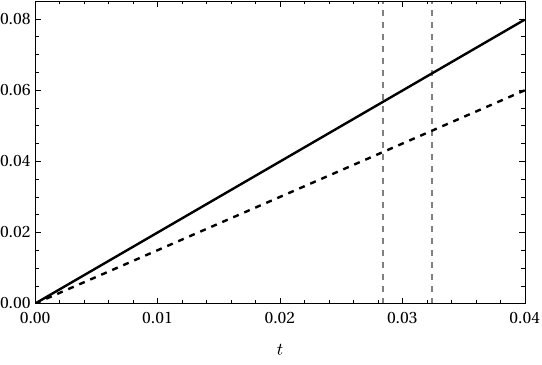}
$\phantom{a}$
  \includegraphics[width=0.47\textwidth]{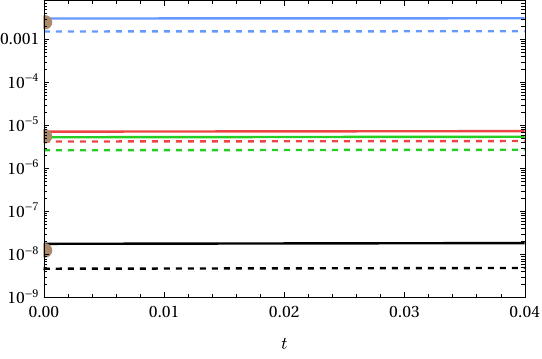}
\caption{\label{fig:xi} 
Left panel: $\xi_{2,r}$ (solid) and $\hat{\xi}_{2,r}$ (dashed), as
a function of $t$. Right panel: relative differences according to
eq.~(\ref{xirat}) (solid lines; dashed lines represent their
$\hat{\xi}$ analogues), as a function of $t$.
}
  \end{center}
\end{figure}

In view of the central role it plays, it is worth writing explicitly
eq.~(\ref{Eopevol}) in the non-singlet subspace, where all matrices
and vectors become c-numbers. Thus:
\beqn
\frac{\partial E_{N}^{(K)}(t)}{\partial t}&=&
\frac{\partial r\!\left(K_{NS,N}^{(K)}(t)\right)}{\partial t}\,
\frac{1}{r\!\left(K_{NS,N}^{(K)}(t)\right)}\,E_{N}^{(K)}(t)
\nonumber\\*&+&
\frac{b_0\aem^2(t)}{\beta(\aem(t))}
\sum_{i=0}^\infty\left(\frac{\aem(t)}{2\pi}\right)^i\,
P_{NS,N}^{[i]}\,E_N^{(K)}(t)\,,
\label{EopevolNS}
\eeqn
with
\beq
K_{NS,N}^{(K)}(t)=\sum_{i=1}^k\left(\frac{\aem(t)}{2\pi}\right)^i
K_{NS,N}^{(K)[i]}\,.
\label{KmatK}
\eeq
Equation~(\ref{EopevolNS}) can immediately be solved by separating
the variables; we obtain:
\beq
\log E_{N}^{(K)}(t)=
\log\frac{r\!\left(K_{NS,N}^{(K)}(t)\right)}
{r\!\left(K_{NS,N}^{(K)}(0)\right)}+
\sum_{i=0}^\infty P_{NS,N}^{[i]}
\int_0^t du
\frac{b_0\aem^2(u)}{\beta(\aem(u))}
\left(\frac{\aem(u)}{2\pi}\right)^i\,.
\label{EopNSsol}
\eeq
This implies that, in $\MSb$, we have
\beq
\log E_{N}(t)=
\sum_{i=0}^\infty P_{NS,N}^{[i]}
\int_0^t du
\frac{b_0\aem^2(u)}{\beta(\aem(u))}
\left(\frac{\aem(u)}{2\pi}\right)^i\,.
\label{EopNSsolMSb}
\eeq
If one does {\em not} expand in $\aem$ the $\beta$-dependent factor
in eq.~(\ref{EopNSsolMSb}), at the NNLO the following integrals are
relevant:
\beq
{\cal B}_i=\int_0^t du
\frac{b_0\aem^2(u)}{\beta(\aem(u))}
\left(\frac{\aem(u)}{2\pi}\right)^i\,,
\;\;\;\;\;\;\;\;
i=0,\,1,\,2\,.
\label{hBi}
\eeq
Explicit computations can be performed by employing the three-loop
expression of the $\beta$ function, whence:
\beqn
{\cal B}_0&=&
t+\frac{1}{4\pi b_0}
\log\frac{b_0+b_1\aem(0)+b_2\aem(0)^2}
{b_0+b_1\aem(t)+b_2\aem(t)^2}
\\*&-&
\frac{1}{4\pi b_0}\frac{b_1}{\sqrt{b_1^2-4b_0b_2}}
\nonumber
\\*&&\phantom{aa}\times
\log\frac{
\Big(
b_1+2b_2\aem(t)-\sqrt{b_1^2-4b_0b_2}
\Big)
\Big(
b_1+2b_2\aem(0)+\sqrt{b_1^2-4b_0b_2}
\Big)
}
{
\Big(
b_1+2b_2\aem(t)+\sqrt{b_1^2-4b_0b_2}
\Big)
\Big(
b_1+2b_2\aem(0)-\sqrt{b_1^2-4b_0b_2}
\Big)
}\,,
\nonumber
%%\\
\eeqn
\beqn
{\cal B}_1&=&
\frac{1}{8\pi^2\sqrt{b_1^2-4b_0b_2}}
\\*&&\phantom{aa}\times
\log\frac{
\Big(
b_1+2b_2\aem(t)-\sqrt{b_1^2-4b_0b_2}
\Big)
\Big(
b_1+2b_2\aem(0)+\sqrt{b_1^2-4b_0b_2}
\Big)
}
{
\Big(
b_1+2b_2\aem(t)+\sqrt{b_1^2-4b_0b_2}
\Big)
\Big(
b_1+2b_2\aem(0)-\sqrt{b_1^2-4b_0b_2}
\Big)
}\,,
\nonumber
\\
{\cal B}_2&=&
-\frac{1}{16\pi^3 b_2}
\log\frac{b_0+b_1\aem(0)+b_2\aem(0)^2}
{b_0+b_1\aem(t)+b_2\aem(t)^2}
\\*&-&
\frac{1}{16\pi^3 b_2}\frac{b_1}{\sqrt{b_1^2-4b_0b_2}}
\nonumber
\\*&&\phantom{aa}\times
\log\frac{
\Big(
b_1+2b_2\aem(t)-\sqrt{b_1^2-4b_0b_2}
\Big)
\Big(
b_1+2b_2\aem(0)+\sqrt{b_1^2-4b_0b_2}
\Big)
}
{
\Big(
b_1+2b_2\aem(t)+\sqrt{b_1^2-4b_0b_2}
\Big)
\Big(
b_1+2b_2\aem(0)-\sqrt{b_1^2-4b_0b_2}
\Big)
}\,,
\nonumber
\eeqn
so that
\beq
\left.\phantom{\Big(}\log E_{N}(t)\right|_{NNLO}=
\sum_{i=0}^2 {\cal B}_i\,P_{NS,N}^{[i]}\,.
\label{EopNSsolMSbNNLOr}
\eeq
Conversely, if one does expand in $\aem$ the $\beta$-dependent factor
in eq.~(\ref{EopNSsolMSb}), at the NNLO one obtains:
\beqn
\left.\phantom{\Big(}\log E_{N}(t)\right|_{NNLO}&=&
P_{NS,N}^{[0]}+
\aemotpi\left(P_{NS,N}^{[1]}-\frac{2\pi b_1}{b_0}P_{NS,N}^{[0]}\right)
\label{EopNSsolMSbNNLO}
\\*&+&
\left(\aemotpi\right)^2\left(
P_{NS,N}^{[2]}-\frac{2\pi b_1}{b_0}P_{NS,N}^{[1]}+
\frac{4\pi^2(b_1^2-b_0b_2)}{b_0^2}P_{NS,N}^{[0]}\right)\,.
\nonumber
\eeqn
The integrals of interest are therefore:
\beq
\hat{\cal B}_i=\int_0^t du
\left(\frac{\aem(u)}{2\pi}\right)^i\,,
\;\;\;\;\;\;\;\;
i=0,\,1,\,2\,,
\eeq
whose explicit calculation gives:
\beqn
\hat{\cal B}_0&=&t\,,
\\
\hat{\cal B}_1&=&\frac{\aem(t)-\aem(0)}{4\pi^2 b_0}\,,
\\
\hat{\cal B}_2&=&\frac{\aem(t)^2-\aem(0)^2}{16\pi^3 b_0}\,.
\eeqn
We now remind the reader that the parameters $\xi$ and $\hat{\xi}$
that enter eq.~(\ref{Gammae}) and their analogues are defined implicitly
as in eq.~(4.6) of ref.~\cite{Frixione:2021wzh}, namely
\beq
\log E_{N}(t)\stackrel{N\to\infty}{\longrightarrow}
-\xi\log\bN+\hat{\xi}\,,
\label{xihxidef}
\eeq
where in the large-$N$ expression of the evolution operator
all terms suppressed by a power of $N$ are discarded. Therefore,
the parameters relevant to NNLO-accurate results can be obtained
from eqs.~(\ref{EopNSsolMSbNNLOr}) and~(\ref{EopNSsolMSbNNLO})
by inserting there the large-$N$ forms of the splitting kernels
up to $\ord(\aem^3)$~\cite{Moch:2004pa,Blumlein:2021enk}. By means 
of explicit computations we obtain what follows:
\beqn
P_{NS,N}^{[0]}&=&
-2\log\bN+\frac{3}{2}\,,
\\
P_{NS,N}^{[1]}&=&
\frac{20\NF}{9}\log\bN+\frac{3}{8}-\frac{\NF}{6}-
\left(\half+\frac{2\NF}{9}\right)\pi^2+6\zeta_3\,,
\\
P_{NS,N}^{[2]}&=&
\left(\frac{55\NF}{6}+\frac{8\NF^2}{27}-8\NF\zeta_3\right)\log\bN+
\frac{29}{16}-\frac{23\NF}{4}-\frac{17\NF^2}{18}
\\*&+&
\left(\frac{3}{8}+\frac{5\NF}{18}+\frac{20\NF^2}{81}\right)\pi^2+
\left(\frac{1}{5}+\frac{29\NF}{270}\right)\pi^4
\\*&+&
\left(\frac{17}{2}-\frac{34\NF}{3}-\frac{8\NF^2}{9}\right)\zeta_3-
\frac{2}{3}\pi^2\zeta_3-30\zeta_5\,.
\nonumber
\eeqn
We now now put everything back together, and write the sought parameters
in a closed form. Denoting by $\xi_k$ and $\hat{\xi}_k$ the results
at N$^k$LO stemming from eq.~(\ref{EopNSsolMSbNNLO}), we obtain:
\beqn
\xi_0&=&2t\,,
\label{xi0res}
\\
\hat{\xi}_0&=&\frac{3}{2}t\,,
\\
\xi_1&=&\xi_0-
\frac{\aem(t)-\aem(0)}{4\pi^2 b_0}
\left(\rho_1+\frac{4\pi b_1}{b_0}\right)\,,
\\
\hat{\xi}_1&=&\hat{\xi}_0+
\frac{\aem(t)-\aem(0)}{4\pi^2 b_0}
\left(\lambda_1-\frac{3\pi b_1}{b_0}\right)\,,
\\
\xi_2&=&\xi_1+
\frac{\aem(t)^2-\aem(0)^2}{8\pi^3 b_0}
\left(\rho_2+\frac{\pi b_1}{b_0}\rho_1+
\frac{4\pi^2(b_1^2-b_0b_2)}{b_0^2}\right)\,,
\\
\hat{\xi}_2&=&\hat{\xi}_1+
\frac{\aem(t)^2-\aem(0)^2}{8\pi^3 b_0}
\left(\lambda_2-\frac{\pi b_1}{b_0}\lambda_1+
\frac{3\pi^2(b_1^2-b_0b_2)}{b_0^2}\right)\,,
\label{hxi2res}
\eeqn
having defined:
\beqn
\rho_1&=&\frac{20\NF}{9}\,,
\\
\lambda_1&=&\frac{3}{8}-\frac{\pi^2}{2}-\frac{\NF}{18}\left(3+4\pi^2\right)
+6\zeta_3\,,
\\
\rho_2&=&-\frac{55\NF}{12}-\frac{4\NF^2}{27}+4\NF\zeta_3\,,
\\
\lambda_2&=&\frac{29}{32}-\frac{23\NF}{8}-\frac{17\NF^2}{36}+
\left(\frac{3}{16}+\frac{5\NF}{36}+\frac{10\NF^2}{81}\right)\pi^2
+\left(\frac{1}{10}+\frac{29\NF}{540}\right)\pi^4
\nonumber\\*
&+&\left(\frac{17}{4}-\frac{34\NF}{6}-\frac{4\NF^2}{9}\right)\zeta_3
-\frac{1}{3}\pi^2\zeta_3-15\zeta_5\,.
\eeqn
The results for $\xi_0$, $\hat{\xi}_0$, $\xi_1$, and $\hat{\xi}_1$
are identical to those originally found in ref.~\cite{Bertone:2019hks}.
If we employ eq.~(\ref{EopNSsolMSbNNLOr}) and denote the corresponding 
results by $\xi_{2,r}$ and~$\hat{\xi}_{2,r}$, we find instead:
\beqn
\xi_{2,r}&=&2t+\frac{1}{2\pi b_0}
\left(1-\frac{b_0\rho_2}{4\pi^2 b_2}\right)
\log\frac{b_0+b_1\aem(0)+b_2\aem(0)^2}
{b_0+b_1\aem(t)+b_2\aem(t)^2}
\label{xi2rres}
\\*&-&
\frac{1}{2\pi}\frac{1}{\sqrt{b_1^2-4b_0b_2}}
\left(\frac{b_1}{b_0}+\frac{\rho_1}{2\pi}+
\frac{b_1\rho_2}{4\pi^2 b_2}\right)
\nonumber
\\*&&\phantom{aa}\times
\log\frac{
\Big(
b_1+2b_2\aem(t)-\sqrt{b_1^2-4b_0b_2}
\Big)
\Big(
b_1+2b_2\aem(0)+\sqrt{b_1^2-4b_0b_2}
\Big)
}
{
\Big(
b_1+2b_2\aem(t)+\sqrt{b_1^2-4b_0b_2}
\Big)
\Big(
b_1+2b_2\aem(0)-\sqrt{b_1^2-4b_0b_2}
\Big)
}\,,
\nonumber
\\
\hat{\xi}_{2,r}&=&\frac{3}{2}t+\frac{1}{8\pi b_0}
\left(3-\frac{b_0\lambda_2}{\pi^2 b_2}\right)
\log\frac{b_0+b_1\aem(0)+b_2\aem(0)^2}
{b_0+b_1\aem(t)+b_2\aem(t)^2}
\label{hxi2rres}
\\*&-&
\frac{1}{2\pi}\frac{1}{\sqrt{b_1^2-4b_0b_2}}
\left(\frac{3b_1}{4b_0}-\frac{\lambda_1}{2\pi}+
\frac{b_1\lambda_2}{4\pi^2 b_2}\right)
\nonumber
\\*&&\phantom{aa}\times
\log\frac{
\Big(
b_1+2b_2\aem(t)-\sqrt{b_1^2-4b_0b_2}
\Big)
\Big(
b_1+2b_2\aem(0)+\sqrt{b_1^2-4b_0b_2}
\Big)
}
{
\Big(
b_1+2b_2\aem(t)+\sqrt{b_1^2-4b_0b_2}
\Big)
\Big(
b_1+2b_2\aem(0)-\sqrt{b_1^2-4b_0b_2}
\Big)
}\,.
\nonumber
\eeqn
The NLO-level counterparts of eqs.~(\ref{xi2rres}) and~(\ref{hxi2rres})
are obtained by computing the integrals of eq.~(\ref{hBi}) with $b_2=0$,
and by ignoring the contribution of ${\cal B}_2$. We thus arrive at
what follows:
\beqn
\xi_{1,r}&=&2t+\frac{1}{4\pi b_0}
\left(4+\frac{b_0\rho_1}{\pi b_1}\right)
\log\frac{b_0+b_1\aem(0)}{b_0+b_1\aem(t)}\,,
\label{xi1rres}
\\
\hat{\xi}_{1,r}&=&\frac{3}{2}t+\frac{1}{4\pi b_0}
\left(3-\frac{b_0\lambda_1}{\pi b_1}\right)
\log\frac{b_0+b_1\aem(0)}{b_0+b_1\aem(t)}\,.
\label{hxi1rres}
\eeqn
The $\xi$ and $\hat{\xi}$ parameters are relatively small in 
absolute value, and the differences among them are extremely tiny, 
in keeping with their being perturbative quantities. Such small
values imply, in view of eq.~(\ref{Gammae}), 
that the electron PDF is barely an integrable function, and therefore 
that for an efficient numerical handling the analytical knowledge of 
its $z\to 1$ behaviour is crucial. In the left panel of fig.~\ref{fig:xi}
we show $\xi_{2,r}$ and $\hat{\xi}_{2,r}$ as a solid and a dashed
line, respectively, as a function of $t$; the two vertical dashed
lines indicate the values of $t$ which correspond to $\mu=m_Z$
\mbox{($t\simeq 0.02840$)} and $\mu=500$~GeV \mbox{($t\simeq 0.03244$)},
having set\footnote{The reader must bear in mind eq.~(\ref{aemaem}), whereby
in this context $\aem(0)$ denotes the coupling constant computed at
the electron mass, since $\muz=m$.} \mbox{$\aem(0)=1/137.036$}.
%%%%%%%%%%%%%%%%%%%%%%%%%%%%%%%%%%%%%%%%%%%%%%%%%%%%%%%%%%%%%%%%%%%
%%%%%%%%%%%%%%%%%%%%%%%%%%%%%%%%%%%%%%%%%%%%%%%%%%%%%%%%%%%%%%%%%%%
On the scale of the plot, the differences among the various $\xi$ and 
$\hat{\xi}$ parameters are invisible, since the linear term in $t$ 
overwhelmingly dominates. 
In order to see those differences, in the right panel of fig.~\ref{fig:xi} 
we display the ratios
\beq
1-\frac{\xi_1}{\xi_0}\,,\;\;\;\;\;\;
-1+\frac{\xi_2}{\xi_1}\,,\;\;\;\;\;\;
-1+\frac{\xi_{1,r}}{\xi_1}\,,\;\;\;\;\;\;
1-\frac{\xi_{2,r}}{\xi_2}\,,\;\;\;\;\;\;
\label{xirat}
\eeq
(as well as their $\hat{\xi}$ counterparts) as blue, red, green, and black 
solid lines (dashed for $\hat{\xi}$), respectively, 
as a function of $t$. The brown bullet 
points on the $y$ axis give the values of $(\aem(0)/\pi)^k$, with $k=1,2,3$,
and show that a simple coupling-constant scaling can be used to obtain
a rough estimate of the next-best $\xi$ and $\hat{\xi}$ perturbative
results. Having said that, it should be clear that while both $\xi_2$ 
and $\xi_{1,r}$ improve upon $\xi_1$ with a relative impact of 
\mbox{$\ord\left((\aem(0)/\pi)^2\right)$}, they account for different
effects. Specifically, $\xi_2$ includes NNLL collinear contributions,
while $\xi_{1,r}$ includes NLL running-coupling contributions.

%%%%%%%%%%%%%%%%%%%%%%%%%%%%%%%%%%%%%%%%%%%%%%%%%%%%%%%%%%%%%%%%%%%
\begin{figure}[thb]
  \begin{center}
  \includegraphics[width=0.65\textwidth]{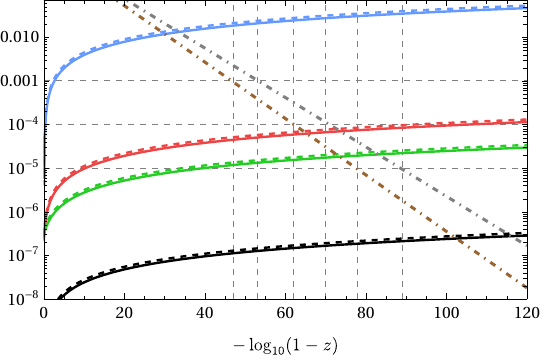}
\caption{\label{fig:xiplot} 
Equation~(\ref{Szxidef}) (diagonal gray and brown lines)
and eq.~(\ref{Sratios}) (the other coloured lines).
See the text for details.
}
  \end{center}
\end{figure}
%%%%%%%%%%%%%%%%%%%%%%%%%%%%%%%%%%%%%%%%%%%%%%%%%%%%%%%%%%%%%%%%%%%
At this point, the natural question to ask is how the perturbative
improvements on the $\xi$ parameters affect the accuracy of cross
section computations. A definite answer to that question is
process dependent. Still, one can have an idea of the general
behaviour thanks to the fact that, in the $\Delta$ scheme, the
dominant lepton PDF is essentially given by the Gribov-Lipatov
prefactor, times terms which vanish or tend to a constant as $z$ 
tends to one (see eqs.~(\ref{DelNLOlogs}) and~(\ref{Delcoeff}), and
sect.~\ref{sec:asy}). Still, even in the $\Delta$ scheme what matters
is ultimately the convolution of the PDFs with the short-distance
cross sections, i.e.~again a process-dependent result. However,
since the Gribov-Lipatov prefactor is so steeply peaked towards
one, a small shift in the position of the peak of the short-distance
cross section may lead to a visibly different prediction. In order
to estimate an effect of this kind without actually performing any
convolution, one can look at the speed at which the integral of
the Gribov-Lipatov prefactor approaches its asymptotic value;
equivalently, one can compute the following quantity:
\beq
S(z;\xi,\hat{\xi})=
\int_0^1 dx
\frac{e^{-\gE\xi}e^{\hat{\xi}}}{\Gamma(1+\xi)}\,
\xi(1-x)^{-1+\xi}
-\int_0^z dx
\frac{e^{-\gE\xi}e^{\hat{\xi}}}{\Gamma(1+\xi)}\,
\xi(1-x)^{-1+\xi}\,,
\label{Szxidef}
\eeq
for which
\beq
\lim_{z\to 1}S(z;\xi,\hat{\xi})=0\,.
\eeq
In fig.~\ref{fig:xiplot} we plot $S(z;\xi_{2,r},\hat{\xi}_{2,r})$
of eq.~(\ref{Szxidef}), as a function of \mbox{$-\log_{10}(1-z)$},
for $t=0.02840$ (gray dot-dashed line) and $t=0.03244$ (brown dot-dashed
line). The horizontal dashed lines are there to guide the eye, and
correspond to typical relative precision targets ($10^{-3}$, $10^{-4}$, 
$10^{-5}$) of some future lepton collider. Where these horizontal lines
cross the diagonal gray and brown lines a dashed vertical line is drawn,
which is thus at the value of $z$ where $S(z;\xi_{2,r},\hat{\xi}_{2,r})$
is equal to the chosen relative-precision target. For example, at $t=0.02840$
the $10^{-5}$ target is attained by integrating the running-coupling
resummed NNLO cross section (because of the use of $\xi_{2,r}$ and
$\hat{\xi}_{2,r}$) up to \mbox{$-\log_{10}(1-z)\simeq 90$}.
After selecting a relative-precision target and having determined
the value of $z$ according to the procedure above, one then wonders
what are the values assumed by the quantity in eq.~(\ref{Szxidef})
computed with $\xi$ and $\hat{\xi}$ parameters different from
$\xi_{2,r}$ and $\hat{\xi}_{2,r}$, which are associated with  
lower perturbative accuracies. If the relative differences of those
values w.r.t.~$S(z;\xi_{2,r},\hat{\xi}_{2,r})$ are smaller than
the value of $S(z;\xi_{2,r},\hat{\xi}_{2,r})$ itself, then the
corresponding perturbative accuracy will give a prediction 
compatible with the NNLO-resummed one within the target accuracy.
In view of this, in fig.~\ref{fig:xiplot} we plot the following 
relative differences:
\beqn
&&\abs{1-\frac{S(z,\xi_0,\hat{\xi}_0)}{S(z,\xi_{2,r},\hat{\xi}_{2,r})}}\,,
\;\;\;\;
\abs{1-\frac{S(z,\xi_1,\hat{\xi}_1)}{S(z,\xi_{2,r},\hat{\xi}_{2,r})}}\,,
\nonumber
\\*&&
\abs{1-\frac{S(z,\xi_{1,r}\hat{\xi}_{1,r})}{S(z,\xi_{2,r},\hat{\xi}_{2,r})}}\,,
\;\;\;\;
\abs{1-\frac{S(z,\xi_2,\hat{\xi}_2)}{S(z,\xi_{2,r},\hat{\xi}_{2,r})}}\,,
\label{Sratios}
\eeqn
as blue, red, green, and black lines, respectively; the solid (dashed)
pattern is relevant to \mbox{$t=0.02840$} ($t=0.03244$). According to the
previous discussion, if one of these coloured lines is below (above)
the relevant diagonal dashed line to the left of the point in $z$ where 
that diagonal line assumes a value equal to the chosen target accuracy, 
then the corresponding perturbative accuracy is sufficient (is not sufficient)
to meet that target in theoretical computations. We then conclude that
LO PDFs (blue lines) are inadequate even if we aim at a relative
precision of $10^{-3}$. If the target is lowered to $10^{-4}$,
we see that NLO and, to a lesser extent, NLO-resummed PDFs are borderline 
cases. Since the present exercise is meant to give one a reasonable 
idea of the accuracy one hopes to obtain, but cannot be taken literally,
one prefers to err on the safe side and deem NLO PDFs to be of
insufficient accuracy to deal with relative-precision targets of
$10^{-4}$ or smaller. Conversely, NNLO-accurate PDFs are expected
to be adequate for up to relative precisions of $10^{-6}$, 
especially if coupled with running-coupling resummation effects.
From fig.~\ref{fig:xiplot}, we observe that these conclusions are 
largely independent of the collider energy\footnote{Conversely, how to
attain the required accuracy, i.e.~up to which $z$ value the
correct handling of the integration is crucial, {\em does} depend
on the collider energy, since the two diagonal lines have markedly
different slopes -- the steeper the slope, the easier the integral.
This is what we expect: by increasing the energy, the lepton PDF is
less close to a $\delta(1-z)$ function, thus its integrable singularity
is less difficult to integrate.}.

We re-iterate that these findings must be confirmed and refined 
by results specific to given production processes in the context of
realistic simulations.

\section{Large-$z$ PDF results\label{sec:asy}}
As is discussed in sect.~\ref{sec:eopK}, in a generic factorisation
scheme $K$ the PDFs are evaluated at $t$ by means of
\beq
\Gamma^{(K)}(z,t)=\invM\left[\Eop_N^{(K)}(t)\,\Gamma_{0,N}^{(K)}\right]\,.
\label{PDFz}
\eeq
It is essentially impossible to compute analytically this inverse
Mellin transform, and one must use numerical methods instead. However,
in view of the steepness of the electron PDF for $z\to 1$, for large
values of $z$ it is the numerical Mellin inversion which fails,
and thus it is mandatory to obtain, in that region, the analytical
result.

This problem is discussed extensively in refs.~\cite{Bertone:2019hks,
Frixione:2021wzh}; here, we proceed in keeping with those papers
in order to obtain NNLO-accurate predictions in both the $\MSb$
and the $\Delta$ scheme. We mainly deal with the non-singlet case,
which is sufficient to understand the main issues, and discuss only most
briefly the photon PDF at the end of this section (as far as the large-$z$ 
behaviour is concerned, the singlet is basically identical to the non-singlet).

The sought large-$z$ PDF is found by means of eq.~(\ref{PDFz}),
in the r.h.s.~of which one uses the large-$N$ expressions of
the evolution operator and of the initial conditions.
The non-singlet $\MSb$ evolution operator in the large-$N$ limit
is given by eq.~(\ref{xihxidef}), with the $\xi$ and $\hat{\xi}$
parameters taken from eqs.~(\ref{xi0res})--(\ref{hxi2res}),
and~(\ref{xi2rres})--(\ref{hxi1rres}) depending on the perturbative
accuracy one wants to achieve. Equation~(\ref{xihxidef}) neglects
terms of $\ord(1/N)$ and higher, and therefore one is entitled (but
not obliged) to ignore analogous contributions in the initial 
conditions. In any case, in a generic factorisation scheme $K$,
eqs.~(\ref{EopNSsol}), (\ref{EopNSsolMSb}), and~(\ref{xihxidef})
lead one to:
\beqn
E_{N}^{(K)}(t)&\stackrel{N\to\infty}{\longrightarrow}&
\exp\left[\log\frac{r\!\left(K_{NS,N}^{(K)}(t)\right)}
{r\!\left(K_{NS,N}^{(K)}(0)\right)}
-\xi\log\bN+\hat{\xi}\right]
\\&&\phantom{aa}=
e^{-\gE\xi}\,e^{\hat{\xi}}\,N^{-\xi}\,
\frac{r\!\left(\widetilde{K}_{NS,N}^{(K)}(t)\right)}
{r\!\left(\widetilde{K}_{NS,N}^{(K)}(0)\right)}\,,
\eeqn
and therefore eq.~(\ref{PDFz}) implies
\beq
\Gamma_{NS}^{(K)}(z,t)
\;\stackrel{z\to 1}{\longrightarrow}\;
\invM\left[e^{-\gE\xi}\,e^{\hat{\xi}}\,N^{-\xi}\,
\frac{r\!\left(\widetilde{K}_{NS,N}^{(K)}(t)\right)}
{r\!\left(\widetilde{K}_{NS,N}^{(K)}(0)\right)}\,
\widetilde{\Gamma}_{NS,0,N}^{(K)}\right]\,,
\label{PDFzlgz}
\eeq
where, with the same notation as was employed before, 
$\widetilde{\Gamma}_{NS,0,N}^{(K)}$ is the large-$N$ 
limit of the $K$-scheme non-singlet initial conditions\footnote{The
reader must bear in mind eq.~(\ref{wKvsoK}), which implies
that the calculations that will follow will produce the same 
results regardless of whether one employs 
$\widetilde{\Kmat}_{N}^{(\Delta)[i]}$ or 
$\overline{\Kmat}_{N}^{(\Delta)[i]}+\Cmat_{N}^{[i]}$.}. This quantity, 
in the two cases we are interested in (namely $\MSb$ and $\Delta$)
reads as follows (see eq.~(\ref{iniMSb})):
\beqn
\widetilde{\Gamma}_{NS,0,N}^{(\MSb)}&=&
\sum_{m=0}^4 p_{0,m}^{(NS)}\log^m\bN\,,
\label{iniMSblgN}
\\
\widetilde{\Gamma}_{NS,0,N}^{(\Delta)}&=&1\,.
\label{iniDellgN}
\eeqn
As was anticipated, here we have retained only contributions which
are not suppressed by a power of $N$ (i.e.~$\rho=-1$ in this context). 
Equation~(\ref{iniMSblgN}) follows from eq.~(\ref{iniMSb}) where, in 
keeping with our general setup, we understand $L_0=0$ (see
footnote~\ref{ft:L0}); conversely, eq.~(\ref{iniDellgN})
is fulfilled by construction of the $\Delta$ scheme, as is
discussed in sect.~\ref{sec:DS}. Since in $\MSb$ $r(x)=1$,
from eq.~(\ref{PDFzlgz}) we finally obtain:
\beqn
\Gamma_{NS}^{(\MSb)}(z,t)
&\stackrel{z\to 1}{\longrightarrow}&
\invM\left[e^{-\gE\xi}\,e^{\hat{\xi}}\,N^{-\xi}\,
\sum_{m=0}^4 p_{0,m}^{(NS)}\log^m\bN\right]\,,
\label{PDFzlgzMSb}
\\
\Gamma_{NS}^{(\Delta)}(z,t)
&\stackrel{z\to 1}{\longrightarrow}&
\invM\left[e^{-\gE\xi}\,e^{\hat{\xi}}\,N^{-\xi}\,
\frac{r\!\left(\widetilde{K}_{NS,N}^{(\Delta)}(t)\right)}
{r\!\left(\widetilde{K}_{NS,N}^{(\Delta)}(0)\right)}\right]\,.
\label{PDFzlgzDel}
\eeqn
Thanks to the linearity of the Mellin transform, eq.~(\ref{PDFzlgzMSb})
can be computed by employing eq.~(\ref{invMlogqN}). By doing that,
and by replacing the coefficients $p_{0,j}^{(NS)}$ with their
explicit values derived from the results of appendix~\ref{sec:res},
at the NNLO we obtain:
\beqn
&&\Gamma_{NS}^{(\MSb)}(z,t)
\stackrel{z\to 1}{\longrightarrow}
\frac{e^{-\gE\xi}e^{\hat{\xi}}}{\Gamma(1+\xi)}\,
\xi(1-z)^{-1+\xi}
\label{MSbPDFevol}
\\&&\phantom{aaa}
\times\Bigg\{
1+
\frac{\aem(0)}{2\pi}
\Big[2-\frac{\pi^2}{3}+\frac{3L_0}{2}+2(1-L_0)d_1(\xi)-2d_2(\xi) 
\nonumber
\\&&\phantom{aaaaaaaaa}
-\big(2(1-L_0)-4 d_1(\xi)\big)\log(1-z) - 2\log^2(1-z)\Big] 
\nonumber
\\&&\phantom{aaa}
+\left(\frac{\aem(0)}{2\pi}\right)^2
\Bigg[\frac{241}{32}+\frac{3139\NF}{648}-\frac{11\pi^4}{180}+
\left(\frac{1}{4}-2\log 2-\frac{\NF}{3}\right)\pi^2 -
\left(\frac{3}{2}-\frac{2\NF}{3}\right)\zeta_3
\nonumber
\\&&\phantom{aaaaaaa}
+\left(\frac{27}{8}-\frac{3\NF}{2}-\pi^2+6\zeta_3\right)L_0 
+\left(\frac{9}{8}-\frac{\NF}{2}\right) L_0^2
\nonumber
\\&&\phantom{aaaaaaa}
+\left(4-\frac{56\NF}{27}-\frac{2\pi^2}{3}-
\left(1-\frac{8\NF}{9}-\frac{2\pi^2}{3}\right)L_0 -
\left(3-\frac{2\NF}{3}\right) L_0^2\right) d_1(\xi) 
\nonumber
\\&&\phantom{aaaaaaa}
-\left(2-\frac{2\pi^2}{3}+\left(7-\frac{4\NF}{3}\right)L_0-
2 L_0^2\right) d_2(\xi) 
-4 (1-L_0) d_3(\xi)+2 d_4(\xi) 
\nonumber
\\&&\phantom{aaaaaaa}
+\Bigg( 
-\left(4-\frac{56\NF}{27}-\frac{2\pi^2}{3}
-\left(1-\frac{8\NF}{9}-\frac{2\pi^2}{3}\right)L_0-
\left(3-\frac{2\NF}{3}\right)L_0^2\right) 
\nonumber
\\&&\phantom{aaaaaaa}
-2\left(-2+\frac{2\pi^2}{3}-\left(7-\frac{4\NF}{3}\right) L_0
+2 L_0^2\right) d_1(\xi) 
\nonumber
\\&&\phantom{aaaaaaaaaa}
+12\left(1-L_0\right) d_2(\xi)-8 d_3(\xi) \Bigg) \log(1-z) 
\nonumber
\\&&\phantom{aaaa}
+\left( -\frac{1}{3}\left(6-2\pi^2+(21-4 \NF)L_0-6 L_0^2\right)
-12 (1-L_0) d_1(\xi)+12 d_2(\xi) \right) \log^2(1-z) 
\nonumber
\\&&\phantom{aaaaaaa}
+\left(4(1-L_0)-8 d_1(\xi) \right) \log^3(1-z) +
2 \log^4(1-z) \Bigg]\Bigg\}\,.
\nonumber
\eeqn
Here, we can choose either $\xi=\xi_2$ and $\hat{\xi}=\hat{\xi}_2$,
or $\xi=\xi_{2,r}$ and $\hat{\xi}=\hat{\xi}_{2,r}$. It is manifest
that, up to $\ord(\aem)$, eq.~(\ref{MSbPDFevol}) coincides with the
result of ref.~\cite{Bertone:2019hks}. We note that we could obtain
a better analytical approximation to the exact large-$z$ result
by employing eq.~(\ref{invMlogqNex}) rather than eq.~(\ref{invMlogqN}) 
for the inverse Mellin transform. The result can be read directly
from eq.~(\ref{MSbPDFevol}), by means of the formal replacements:
\beqn
(1-z)^{-1+\xi}&\longrightarrow&(-\log z)^{-1+\xi}\,,
\label{frepl1}
\\
\log^{q}(1-z)&\longrightarrow&\big(\log(-\log z)\big)^q\,.
\label{frepl2}
\eeqn
The expression in terms of $\log(1-z)$ is standard, while its
alternative is, to the best of our knowledge, proposed here
for the first time. The important thing is that both forms convey
the same message, namely the proliferation of soft logarithms in
the $\MSb$ electron PDF, whose maximal power grows as the square
of the power of the coupling constant, which implies an ever
increasing amount of cancellation between the $\MSb$ PDFs and
the $\MSb$ short distance cross sections.

The computation of the inverse Mellin transform of eq.~(\ref{PDFzlgzDel})
requires that a choice be made for the characteristic function $r(x)$.
Here, we use eq.~(\ref{rfundef}); other forms will be considered 
in sect.~\ref{sec:others}. With that expression, by using the
large-$N$ limits of the $\MSb$ results presented in appendix~\ref{sec:res}
to obtain the elements of the matrices $\widetilde{\Kmat}_{N}^{(\Delta)[i]}$,
and by ignoring all of the contributions suppressed by powers of $N$ there,
at the NNLO we have:
\beq
\frac{r\!\left(\widetilde{K}_{NS,N}^{(\Delta)}(t)\right)}
{r\!\left(\widetilde{K}_{NS,N}^{(\Delta)}(0)\right)}
\;=\;
F^{(\Delta)}\left(\log\bN\right)\,,
\label{rtoFD}
\eeq
where
\beqn
F^{(\Delta)}(x)&=&
\frac{F_n^{(\Delta)}(\aem(t),x)}{F_n^{(\Delta)}(\aem(0),x)}\,,
\\*
F_n^{(\Delta)}(\aem,x)&=&
1+\frac{\aem}{\pi}\left(x^2-x+f_1\right)+
\frac{\aem^2}{4\pi^2}\left[2\left(x^2-x+f_1\right)^2+f_2x+f_3\right]\,,
\eeqn
with
\beqn
f_1&=&\frac{\pi^2}{6}-1\,,
\\
f_2&=&\frac{56\,\NF}{27}\,,
\\
f_3&=&-\frac{177}{32}-\frac{3139\NF}{648}+\frac{7\pi^4}{60}+
\left(2\log 2-\frac{11}{12}+\frac{\NF}{3}\right)\pi^2+
\left(\frac{3}{2}-\frac{2\NF}{3}\right)\zeta_3\,.
\eeqn
With the result of eq.~(\ref{rtoFD}), the inverse Mellin transform
of eq.~(\ref{PDFzlgzDel}) can be computed by using eqs.~(\ref{invMFFlogqN}),
(\ref{invMFFlogqNsum2}), (\ref{invMFFlogqNex}), or~(\ref{invMFFlogqNsum2ex}).
By means of eq.~(\ref{invMFFlogqN}) we obtain:
\beqn
\Gamma_{NS}^{(\Delta)}(z,t)
&\stackrel{z\to 1}{\longrightarrow}&
\frac{e^{-\gE\xi}e^{\hat{\xi}}}{\Gamma(1+\xi)}\,
\xi(1-z)^{-1+\xi}
\nonumber
\\&&\phantom{a}\times
\Bigg[F^{(\Delta)}\big(-\log(1-z)\big)+
d_1(\xi)F^{(\Delta)^\prime}\big(-\log(1-z)\big)
\nonumber
\\&&\phantom{aa}
+\half d_2(\xi)F^{(\Delta)^{\prime\prime}}\big(-\log(1-z)\big)+\ldots\Bigg],
\label{PDFzlgzDelfin}
\eeqn
having denotes by primes the derivatives of $F^{(\Delta)}$ w.r.t.~its
argument. Conversely, with eq.~(\ref{invMFFlogqNsum2}) we arrive at
the following resummed form:
\beqn
\Gamma_{NS}^{(\Delta)}(z,t)
&\stackrel{z\to 1}{\longrightarrow}&
\frac{ie^{-\gE\xi}e^{\hat{\xi}}}{2\pi}(1-z)^{-1+\xi}
\label{PDFzlgzDelfin2}
\\&&\phantom{aaa}\times
\int_{H^{-}} d\omega\,e^{-\omega}(-\omega)^{-\xi}
F^{(\Delta)}\Big(\log(-\omega)+\gE-\log(1-z)\Big)\,.
\nonumber
\eeqn
The usage of eqs.~(\ref{invMFFlogqNex}) and~(\ref{invMFFlogqNsum2ex})
leads to results that can be obtained from eqs.~(\ref{PDFzlgzDelfin})
and~(\ref{PDFzlgzDelfin2}), respectively, by means of the formal 
replacements of eqs.~(\ref{frepl1}) and~(\ref{frepl2}).

The expressions above allow one to show that the $\Delta$ scheme
achieves the goals for which it has been defined. In particular,
it is easy to see that, for $x\to\infty$:
\beq
\frac{d^i}{dx^i}F^{(\Delta)}(x)=\delta_{i0}R_\alpha^2
-(-)^i \frac{2\pi\Gamma(i+2)}{\aem(0)}\,R_\alpha\left(R_\alpha-1\right)
\frac{1}{x^{2+i}}+\ord\left(\frac{1}{x^{3+i}}\right)\,,
\label{fintder}
\eeq
where
\beq
R_\alpha=\frac{\aem(t)}{\aem(0)}\,.
\label{Ralpha}
\eeq
Therefore, terms stemming from higher derivatives in eq.~(\ref{PDFzlgzDelfin})
are progressively more suppressed, allowing one to truncate the series.
In fact, eq.~(\ref{fintder}) shows that, apart from the Gribov-Lipatov
prefactor, the strict $z\to 1$ limit can be taken, and reads:
\beqn
\Gamma_{NS}^{(\Delta)}(z,t)
&\stackrel{z\to 1}{\longrightarrow}&
\frac{e^{-\gE\xi}e^{\hat{\xi}}}{\Gamma(1+\xi)}\,
\xi(1-z)^{-1+\xi}
\left(\frac{\aem(t)}{\aem(0)}\right)^2\,.
\label{PDFzDelstrict}
\eeqn
More importantly, this proves that {\em all} of the soft logarithms present
in the electron PDF which are associated with a plus distribution (lest
they lead to a non-integrable divergence) are recast by the $\Delta$ scheme 
into regular functions which either vanish or tend to a constant
when $z\to 1$; this implies that they will not play any role in a
procedure which aims to resum soft-logarithms effects; such a procedure
can therefore be restricted to dealing solely with the short-distance
cross sections.

Equation~(\ref{PDFzDelstrict}) is the generalisation to NNLO of the
NLO result of eq.~(4.41) of ref.~\cite{Frixione:2021wzh}; in the present
notation, the latter result (at $L_0=0$) is identical to that of
eq.~(\ref{PDFzDelstrict}), with \mbox{$(\aem(t)/\aem(0))^2$}
replaced by \mbox{$(\aem(t)/\aem(0))$}. Needless to say, this is
also borne out directly from the procedure we have followed in this
paper, by restricting $F^{(\Delta)}(x)$ to receiving only NLO contributions.
Conversely, the organisation of the contributions vanishing with
large $-\log(1-z)$ that we have come up with in this paper is different
w.r.t.~that of ref.~\cite{Frixione:2021wzh}; while the two are mutually
compatible, at a finite number of terms the present one is more efficient,
in the sense that it guarantees a better approximation of the
complete result (and thus a faster convergence).

%%%%%%%%%%%%%%%%%%%%%%%%%%%%%%%%%%%%%%%%%%%%%%%%%%%%%%%%%%%%%%%%%%%
\begin{figure}[thb]
  \begin{center}
  \includegraphics[width=0.47\textwidth]{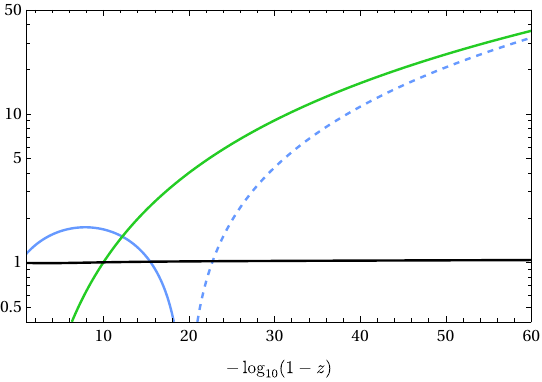}
$\phantom{a}$
  \includegraphics[width=0.47\textwidth]{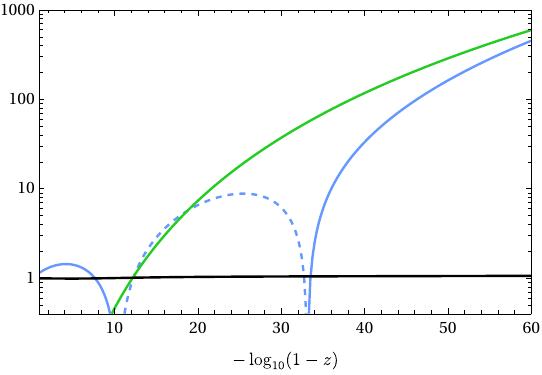}
\caption{\label{fig:PDFkfac} 
NLO (left panel) and NNLO (right panel) PDFs (bar the Gribov-Lipatov
prefactors) in the $\Delta$ scheme (black lines), and their $\MSb$
counterparts in absolute value (blue lines; dashed patterns indicate 
the regions where the PDFs are negative). The green
lines depict a suitably rescaled $\log^k(1-z)$ with $k=2$ and $k=4$
in the left and right panel, respectively; here, $t=0.02840$.
}
  \end{center}
\end{figure}
%%%%%%%%%%%%%%%%%%%%%%%%%%%%%%%%%%%%%%%%%%%%%%%%%%%%%%%%%%%%%%%%%%%
An intuitive understanding of the behaviour of the $\MSb$ and
$\Delta$-scheme PDFs can be obtained by inspecting fig.~\ref{fig:PDFkfac}.
There, the left (right) panel is relevant to NLO (NNLO) results.
We plot as a blue line the absolute value of the asymptotic form 
of the $\MSb$ NS PDF (eq.~(\ref{MSbPDFevol})) as a function of 
\mbox{$x=-\log_{10}(1-z)$}, divided by the Gribov-Lipatov 
prefactor\footnote{In other words, by choosing the same $\xi$
and $\hat{\xi}$ parameters irrespective of the perturbative order
we consider, these curves coincide with the ratio of (N)NLO PDFs over 
the LO ones.}; a dashed pattern indicates that the (N)NLO PDF is
negative. The black line is the same quantity 
computed with $\Delta$-scheme PDFs (eq.~(\ref{PDFzlgzDelfin2})). 
Finally, the green curve on the left (right) panel is proportional to 
\mbox{$\log^2(1-z)$} (\mbox{$\log^4(1-z)$}), which represents the leading 
behaviour of the corresponding $\MSb$ PDF (bar the prefactor). There
are two prominent features in this figure. Firstly, we see that, on 
the scales of the plots, the $\Delta$-scheme PDF is indistinguishable from
being equal to one, while the extremely large (absolute) values assumed by 
the $\MSb$ PDFs show the extent of the cancellations eventually taking 
place between them and the short-distance cross sections, and the dramatic 
impact of the factorisation-scheme definition. Secondly, as the dashed
patterns in the blue curves indicate, the $\MSb$ PDFs assume negative
values in non-zero measure sets; what is worse, for $z\to 1$ such PDFs
diverge to $-\infty$ at the NLO, and to $+\infty$ at the NNLO, which is
a behaviour manifestly driven by the negative (positive) coefficient of 
the $\ord(\aem)$ $\log^2(1-z)$ ($\ord(\aem^2)$ $\log^4(1-z)$) term
in eq.~(\ref{MSbPDFevol}). Clearly, a PDF need not be positive, and
the fact that it is not so does not pose any problems of principle
if one considers sufficiently inclusive quantities. For example,
we obtain the following results\footnote{Needless to say, these are
unphysical since we integrate down to $z=0$ quantities whose functional 
form is in principle valid only at large $z$'s. For the sake of the present 
argument this fact is not relevant.}:
\beqn
\int_0^1 dz\Gamma_{NS}(z,t)&=&
1.0408287({\rm LO}),\;
\\*&&
1.0430859({\rm NLO}~\MSb),\;\;\;\,
1.0407327({\rm NLO}~\Delta)\,\;\phantom{\Big(}
\\*&&
1.0430789({\rm NNLO}~\MSb),\;
1.0407333({\rm NNLO}~\Delta),\;\phantom{\Big(}
\eeqn
amply in keeping with what is expected from a good perturbative behaviour,
for both factorisation schemes. Still, fig.~\ref{fig:PDFkfac} tells one
that, as soon as observables sensitive to soft radiation are analysed,
$\MSb$ PDFs and their cross-section counterparts must be resummed for
the emerging results to be sensible.

\begin{figure}[thb]
  \begin{center}
  \includegraphics[width=0.65\textwidth]{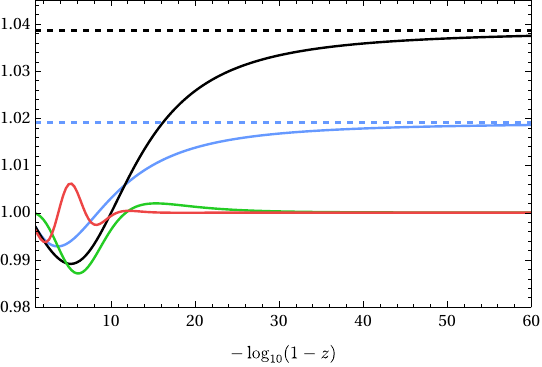}
\caption{\label{fig:Delzoom} 
Ratio of $\Delta$-scheme NLO (blue curve) and NNLO (black curve) PDFs
over LO ones, compared with the respective strict asymptotic limits
(dotted curves). The red and green curves have been obtained
with eq.~(\ref{stabrat}).
}
\end{center}
\end{figure}

Conversely, in order to appreciate the features of the $\Delta$-scheme PDFs we 
have zoomed fig.~\ref{fig:PDFkfac} around the value of one in the codomain, 
thus arriving at fig.~\ref{fig:Delzoom}. There, we plot again (at the NLO,
blue, and the NNLO, black) the asymptotic form of the $\Delta$-scheme lepton
PDFs, bar the Gribov-Lipatov prefactor, as continuous lines; the dotted lines 
are equal to $R_\alpha$ (blue) and $R_\alpha^2$ (black), with $R_\alpha$ 
defined in eq.~(\ref{Ralpha}), i.e.~the strict $z\to 1$ limits of the PDF at 
the NLO and NNLO, respectively. As one can see, the PDFs do indeed tend to 
%%%%%%%%%%%%%%%%%%%%%%%%%%%%%%%%%%%%%%%%%%%%%%%%%%%%%%%%%%%%%%%%%%%
%%%%%%%%%%%%%%%%%%%%%%%%%%%%%%%%%%%%%%%%%%%%%%%%%%%%%%%%%%%%%%%%%%%
those limits, in a way compatible with the ``large'' parameter of the problem 
growing logarithmically.

It is also interesting to see how well the truncation of the
series in eq.~(\ref{PDFzlgzDelfin}) converges to the exact result
of eq.~(\ref{PDFzlgzDelfin2}). In fig.~\ref{fig:Delzoom} we plot
\beq
1+s\left(1-\frac{{\rm eq}.~({\protect\ref{PDFzlgzDelfin}})}
{{\rm eq}.~({\protect\ref{PDFzlgzDelfin2}})}\right)
\label{stabrat}
\eeq
as red and green curves, which correspond to truncating the
series in eq.~(\ref{PDFzlgzDelfin}) to two and eight terms,
respectively. In these two cases, the parameter $s$ has been
set equal to $10^2$ and $2\mydot 10^7$, which is necessary in
order to enhance the visibility. It is clear that eq.~(\ref{PDFzlgzDelfin}) 
converges very fast to eq.~(\ref{PDFzlgzDelfin2}); in particular,
the size of the parameter $s$ relevant to the red and green curves
in fig.~\ref{fig:Delzoom} implies that only a handful of terms in 
the summation are necessary in order to obtain fairly accurate results.

\noindent
{\bf The photon-singlet sector.} 
In refs.~\cite{Bertone:2019hks,Frixione:2021wzh} a detailed description
is given of the procedure followed, in $\MSb$ and $\Delta$ respectively, 
in order to understand the large-$z$ behaviour of the singlet and of the 
photon PDFs. Analytically, it turns out to be impossible to arrive at closed
forms for such PDFs at a level comparable with that of the non-singlet.
Fortunately, this does not constitute a problem: the singlet is extremely 
similar to the non-singlet, the differences being of order $1/N^2$, and 
the photon PDF has its leading contribution at order $1/N$, so that the
numerical-to-analytical switch at large $z$ is, as a matter of fact,
unimportant in practical applications.

Since the methodology applied to the NLO PDFs is unchanged at the NNLO,
we refrain from repeating its description here. We limit ourselves to
stating that the pattern relevant at the NLO is the same as at the NNLO.
In particular, while the photon PDF is much suppressed at $z\to 1$
w.r.t.~its electron counterpart, it is still singular (and integrable)
in $\MSb$, where it features a double power of $\log(1-z)$. Conversely,
in the $\Delta$ scheme the photon does not diverge, in keeping with
what happens at the NLO. The interested reader can find more details
in appendix~B of ref.~\cite{Bertone:2019hks} ($\MSb$) and in sect.~5
of ref.~\cite{Frixione:2021wzh} ($\Delta$).

\section{Alternative choices within the $\Delta$ scheme
\label{sec:others}}
We have so far assumed that the $r(x)$ function of eq.~(\ref{rfun})
is chosen according to eq.~(\ref{rfundef}). In view of future 
developments, it is interesting to consider what happens when
other choices are made, still within the constraints dictated
by eq.~(\ref{rfun}), and the condition that $\Rop$ be invertible.

We start by observing that, for any $n\ge 1$, in Mellin space and
in the evolution basis, one has:
\beq
\Kmat_{E,N}^{(\Delta)^n}=\left(
\begin{array}{ccc}
K_{\Sigma\Sigma,N}^{(\Delta)^n} & 0 & 0 \\
K_{\gamma\Sigma,N}^{(\Delta)}K_{\Sigma\Sigma,N}^{(\Delta)^{n-1}} & 0 & 0 \\
0 & 0 & K_{NS,N}^{(\Delta)^n} \\
\end{array}
\right)
\eeq
and therefore, from eqs.~(\ref{rfun}) and~(\ref{Ropdef})
\beqn
\Rop_{E,N}&=&I+\sum_{i=1}^\infty r_i\Kmat_{e,N}^{(\Delta)^i}
=\left(
\begin{array}{ccc}
1+\sum_{i=1}^\infty r_iK_{\Sigma\Sigma,N}^{(\Delta)^i} & 0 & 0 \\
\sum_{i=1}^\infty r_iK_{\gamma\Sigma,N}^{(\Delta)}
K_{\Sigma\Sigma,N}^{(\Delta)^{i-1}} & \;\;\;1\;\;\; & 0 \\
0 & 0 & 1+\sum_{i=1}^\infty r_iK_{NS,N}^{(\Delta)^i} \\
\end{array}
\right)
\nonumber
\\*
&=&\left(
\begin{array}{ccc}
r\!\left(K_{\Sigma\Sigma,N}^{(\Delta)}\right) & 0 & 0 \\
\frac{K_{\gamma\Sigma,N}^{(\Delta)}}{K_{\Sigma\Sigma,N}^{(\Delta)}}
\left[r\!\left(K_{\Sigma\Sigma,N}^{(\Delta)}\right)-1\right] & 
\;\;\;1\;\;\; & 0 \\
0 & 0 & r\!\left(K_{NS,N}^{(\Delta)}\right) \\
\end{array}
\right),
\eeqn
or, in the physical basis
\beqn
\!\!\!\Rop_{P,N}&\!\!=\!\!&T^{-1}\,\Rop_{E,N}\,T=
\half\left(
\begin{array}{ccc}
r\!\left(K_{\Sigma\Sigma,N}^{(\Delta)}\right) +
r\!\left(K_{NS,N}^{(\Delta)}\right) & 0 & 
r\!\left(K_{\Sigma\Sigma,N}^{(\Delta)}\right) -
r\!\left(K_{NS,N}^{(\Delta)}\right) \\
2\frac{K_{\gamma\Sigma,N}^{(\Delta)}}{K_{\Sigma\Sigma,N}^{(\Delta)}}
\left[r\!\left(K_{\Sigma\Sigma,N}^{(\Delta)}\right)-1\right] & 
\;\;\;2\;\;\; & 
2\frac{K_{\gamma\Sigma,N}^{(\Delta)}}{K_{\Sigma\Sigma,N}^{(\Delta)}}
\left[r\!\left(K_{\Sigma\Sigma,N}^{(\Delta)}\right)-1\right]  \\
r\!\left(K_{\Sigma\Sigma,N}^{(\Delta)}\right) -
r\!\left(K_{NS,N}^{(\Delta)}\right) & 0 & 
r\!\left(K_{\Sigma\Sigma,N}^{(\Delta)}\right) +
r\!\left(K_{NS,N}^{(\Delta)}\right) \\
\end{array}
\right).\phantom{aa}
\label{Ropfull}
\eeqn
Also:
\beqn
\Rop_{E,N}^{-1}&=&
\left(
\begin{array}{ccc}
\frac{1}{r\!\left(K_{\Sigma\Sigma,N}^{(\Delta)}\right)} & 0 & 0 \\
-\frac{K_{\gamma\Sigma,N}^{(\Delta)}}{K_{\Sigma\Sigma,N}^{(\Delta)}}
\frac{r\!\left(K_{\Sigma\Sigma,N}^{(\Delta)}\right)-1}
{r\!\left(K_{\Sigma\Sigma,N}^{(\Delta)}\right)} & 
\;\;\;1\;\;\; & 0 \\
0 & 0 & \frac{1}{r\!\left(K_{NS,N}^{(\Delta)}\right)} \\
\end{array}
\right).
\eeqn
The coefficients $r_i$ of the expansion of $r(x)$ at $x=0$ are 
relevant only up to $i\le k$ at order N$^k$LO. Thus, the generalisation
of the NNLO conditions of eqs.~(\ref{wKaswG1})--(\ref{wKaswG6}) only
involves $r_1$ and $r_2$, and reads as follows:
\beqn
\widetilde{K}_{NS,N}^{(\Delta)[1]}&=&
-\frac{1}{r_1}\,\widetilde{\Gamma}_{NS,0,N}^{[1]}\,,
\label{rwKaswG1}
\\
\widetilde{K}_{NS,N}^{(\Delta)[2]}&=&
\frac{r_1^2-r_2}{r_1^3}
\left(\widetilde{\Gamma}_{NS,0,N}^{[1]}\right)^2
-\frac{1}{r_1}\,\widetilde{\Gamma}_{NS,0,N}^{[2]}\,,
\label{rwKaswG2}
\\
\widetilde{K}_{\Sigma\Sigma,N}^{(\Delta)[1]}&=&
-\frac{1}{r_1}\,\widetilde{\Gamma}_{\Sigma,0,N}^{[1]}\,,
\label{rwKaswG3}
\\
\widetilde{K}_{\Sigma\Sigma,N}^{(\Delta)[2]}&=&
\frac{r_1^2-r_2}{r_1^3}
\left(\widetilde{\Gamma}_{\Sigma,0,N}^{[1]}\right)^2
-\frac{1}{r_1}\,\widetilde{\Gamma}_{\Sigma,0,N}^{[2]}\,,
\label{rwKaswG4}
\\
\widetilde{K}_{\gamma\Sigma,N}^{(\Delta)[1]}&=&
-\frac{1}{r_1}\,\widetilde{\Gamma}_{\gamma,0,N}^{[1]}\,,
\label{rwKaswG5}
\\
\widetilde{K}_{\gamma\Sigma,N}^{(\Delta)[2]}&=&
\frac{r_1^2-r_2}{r_1^3}\,
\widetilde{\Gamma}_{\gamma,0,N}^{[1]}\widetilde{\Gamma}_{\Sigma,0,N}^{[1]}
-\frac{1}{r_1}\,\widetilde{\Gamma}_{\gamma,0,N}^{[2]}\,.
\label{rwKaswG6}
\eeqn
These results render it clear that the maximally-simple form of
eq.~(\ref{rfundef}) does not correspond to maximally-simple forms
for the $K$ matrices. At the NNLO, the latter characteristic (identified
with the absence of squares of the initial conditions) is achieved
by choosing $r_2=r_1^2$. While there are infinitely many functions
whose perturbative coefficients satisfy this equation, one can
iterate the exercise made above, and consider the N$^3$LO case,
where the $\ord(\aem^3)$ term of the non-singlet element of the 
$K$ matrices turns out to be:
\beqn
\widetilde{K}_{NS,N}^{(\Delta)[3]}&=&
-\frac{r_1^4-2r_1^2r_2+2r_2^2-r_1r_3}{r_1^5}\,
\left(\widetilde{\Gamma}_{NS,0,N}^{[1]}\right)^3
+2\,\frac{r_1^2-r_2}{r_1^3}\,
\widetilde{\Gamma}_{NS,0,N}^{[1]}\widetilde{\Gamma}_{NS,0,N}^{[2]}
\nonumber\\&&
-\frac{1}{r_1}\,\widetilde{\Gamma}_{NS,0,N}^{[3]}\,,
\label{rwKaswG2N3}
\eeqn
and analogous solutions in the singlet-photon sector. Therefore,
if one wants to avoid dealing with products of lower-order
initial conditions, so that only the rightmost terms on the
r.h.s.~of eq.~(\ref{rwKaswG2N3}) is non-null, one is lead to
\beq
r_2=r_1^2\,,\;\;\;\;\;\;\;\;r_3=r_1^3\,.
\eeq
By repeating this procedure at yet higher orders one finds an obvious
pattern, and thus obtains\footnote{As is shown in eq.~(\ref{Ropfull}),
for the matrices of the form given in eq.~(\ref{KmatDel}) the knowledge
of the $r(x)$ function in the non-singlet sector is sufficient.}
\beq
r(x)=1+\sum_{i=1}^\infty r_ix^i=1+\sum_{i=1}^\infty r_1^ix^i=
\frac{1}{1-r_1x}\,,
\label{rfunex1}
\eeq
i.e.~the reciprocal of eq.~(\ref{rfundef}) when $r_1=-1$.

Needless to say, the ``maximal-simplicity'' argument used before is
largely arbitrary. In particular, another obvious one stems from
the observation that the results of sect.~\ref{sec:ini} show that
\beq
\widetilde{\Gamma}_{NS,0,N}^{[2]}=
\half\left(\widetilde{\Gamma}_{NS,0,N}^{[1]}\right)^2
+\frac{56\NF}{27}\,\log\bN+\ldots\,.
\eeq
Therefore, eq.~(\ref{rwKaswG2}) implies that 
$\widetilde{K}_{NS,N}^{(\Delta)[2]}$ has a mild $\log\bN$ dependence
(growing linearly instead with the fourth power) if
\beq
\frac{r_1^2-r_2}{r_1^3}=\frac{1}{2r_1}
\;\;\;\;\Longrightarrow\;\;\;\;\
r_2=\half r_1^2\,.
\label{r2half}
\eeq
Clearly, the iteration of this argument would require, at variance
with what has been done before, the knowledge of the $\MSb$ initial
conditions at N$^3$LO and beyond. The two obvious guesses stemming
from eq.~(\ref{r2half}) are
\beq
r(x)=1+\sum_{i=1}^\infty r_ix^i=1+\sum_{i=1}^\infty \frac{r_1^i}{i}x^i=
1-\log\left(1-r_1x\right)\,,
\label{rfunex2}
\eeq
and
\beq
r(x)=1+\sum_{i=1}^\infty r_ix^i=1+\sum_{i=1}^\infty \frac{r_1^i}{i!}x^i=
\exp\left(r_1x\right)\,.
\label{rfunex3}
\eeq
In all cases, the chosen function $r(x)$ determines the asymptotic
behaviour of the $\Delta$-scheme PDFs in according to 
eq.~(\ref{PDFzlgzDelfin}) or eq.~(\ref{PDFzlgzDelfin2}), with
$F^{(\Delta)}$ constructed by means of $r(x)$ according to 
eq.~(\ref{rtoFD})\footnote{We stress that this is remarkable: the
argument of $r(x)$ is originally meant to live in the Mellin space,
and one can ultimately employ that function also in the $z$ space thanks 
to the properties of the inverse Mellin transforms presented
in appendix~\ref{sec:Mell}.}. By doing so with the functions of
eqs.~(\ref{rfunex1}), (\ref{rfunex2}), and~(\ref{rfunex3}), one
finds that the latter two cases lead to numerically challenging
behaviour; in particular, eq.~(\ref{rfunex3}) leads to an
essential singularity \mbox{$\sim\exp[(\aem-\aem(0))\log^2(1-z)]$},
which is peculiar of QED, since in that theory $\aem>\aem(0)$.
This is not terribly important in the present context, but it
helps underscore how the latter maximal-simplicity argument
introduced above does not necessarily lead to results which are
perturbatively stable. In fact, one can see that the choice
\beq
r(x)=1+\log\left(1-x\right)\,,
\label{rfunex4}
\eeq
closely related to that of eq.~(\ref{rfunex2}), is in fact much better
behaved, in spite of stemming from a more involved $N$-space structure.
%%%%%%%%%%%%%%%%%%%%%%%%%%%%%%%%%%%%%%%%%%%%%%%%%%%%%%%%%%%%%%%%%%%
\begin{figure}[thb]
  \begin{center}
  \includegraphics[width=0.65\textwidth]{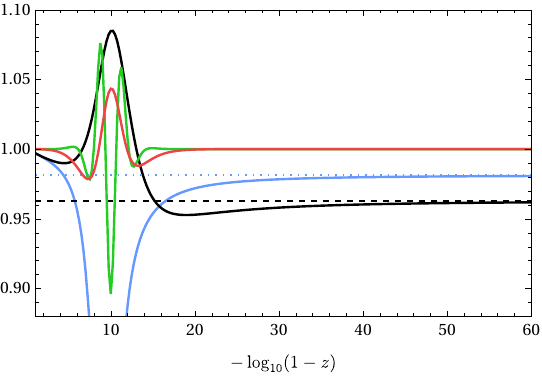}
\caption{\label{fig:Delzoom2} 
As in fig.~\ref{fig:Delzoom}, for the functional form of eq.~(\ref{rfunex1}),
with $r_1=-1$.
The factor $s$ of eq.~(\ref{stabrat}) has been set equal to $10$ and
$5\mydot 10^4$ for the red and green curve, respectively.
}
\end{center}
\end{figure}
%%%%%%%%%%%%%%%%%%%%%%%%%%%%%%%%%%%%%%%%%%%%%%%%%%%%%%%%%%%%%%%%%%%
%%%%%%%%%%%%%%%%%%%%%%%%%%%%%%%%%%%%%%%%%%%%%%%%%%%%%%%%%%%%%%%%%%%
\begin{figure}[thb]
  \begin{center}
  \includegraphics[width=0.65\textwidth]{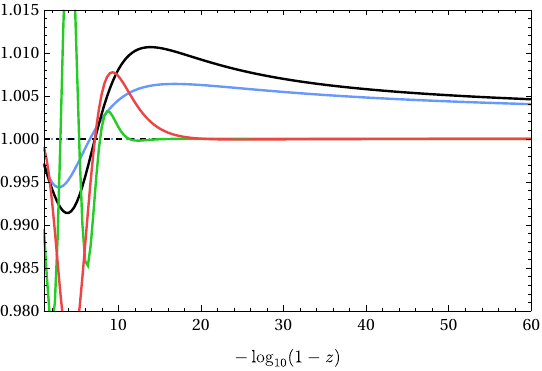}
\caption{\label{fig:Delzoom3} 
As in fig.~\ref{fig:Delzoom}, for the functional form of eq.~(\ref{rfunex4}).
The factor $s$ of eq.~(\ref{stabrat}) has been set equal to $10^2$ and
$5\mydot 10^6$ for the red and green curve, respectively.
}
\end{center}
\end{figure}
%%%%%%%%%%%%%%%%%%%%%%%%%%%%%%%%%%%%%%%%%%%%%%%%%%%%%%%%%%%%%%%%%%%
A summary of the situation is given in figs.~\ref{fig:Delzoom2}
(which is relevant to eq.~(\ref{rfunex1})) and~\ref{fig:Delzoom3}
(which is relevant to eq.~(\ref{rfunex4})). For these two choices
of $r(x)$, the strict asymptotic behaviour, i.e.~the analogue
of eq.~(\ref{PDFzDelstrict}) is presented in eq.~(\ref{PDFzDelstrict2})
and~(\ref{PDFzDelstrict3}):
\beqn
\Big.r(x)\Big|_{{\rm eq}.~({\protect\ref{rfunex1}})}
\;\;\;\;\Longrightarrow\;\;\;\;
\Gamma_{NS}^{(\Delta)}(z,t)
&\stackrel{z\to 1}{\longrightarrow}&
\frac{e^{-\gE\xi}e^{\hat{\xi}}}{\Gamma(1+\xi)}\,
\xi(1-z)^{-1+\xi}
\left(\frac{\aem(0)}{\aem(t)}\right)^2\,,
\label{PDFzDelstrict2}
\\*
\Big.r(x)\Big|_{{\rm eq}.~({\protect\ref{rfunex4}})}
\;\;\;\;\Longrightarrow\;\;\;\;
\Gamma_{NS}^{(\Delta)}(z,t)
&\stackrel{z\to 1}{\longrightarrow}&
\frac{e^{-\gE\xi}e^{\hat{\xi}}}{\Gamma(1+\xi)}\,
\xi(1-z)^{-1+\xi}\,.
\label{PDFzDelstrict3}
\eeqn
The result of eq.~(\ref{PDFzDelstrict2}) is that due to the NNLO PDF;
at the NLO, there is a single power of that ratio of $\aem$ factors.
Conversely, the strict asymptotic limit emerging from the choice
of $r(x)$ of eq.~(\ref{rfunex4}) is identical to the Gribov-Lipatov
prefactor, at both the NLO and NNLO. This should imply an extremely
stable perturbative behaviour, but an explicit implementation of the
exact RGE solutions is necessary in order to confirm or disprove
this indication.

We point out again that, also for the latter two choices of $r(x)$,
eq.~(\ref{PDFzlgzDelfin}) converges very fast to eq.~(\ref{PDFzlgzDelfin2}),
as is shown by the rescaled red and green curves in fig.~\ref{fig:Delzoom2}
and fig.~\ref{fig:Delzoom3}, and can therefore be used effectively 
in practical computations.

\section{Cross sections\label{sec:DYxsec}}
Cross sections in a factorisation scheme different from $\MSb$ can
be obtained from the latter by means of eq.~(\ref{xsecrot}). In order
to arrive at explicit results, we start by considering the case of
a single incoming leg, whereby the short-distance cross sections carry
a single flavour index, as in eq.~(\ref{hxsec}). The coefficients
of an expansion in $\aem$ of such cross sections are denoted in keeping 
with the notation of eq.~(\ref{fcoeff}):
\beq
\hsig_{\alpha,N}^{(K)}=
\sum_{i=0}^\infty \hsig_{\alpha,N}^{(K)[i]}\left(\aemotpi\right)^i\,.
\label{hScoeff}
\eeq
In the evolution basis and in Mellin space, with the general form of 
eq.~(\ref{Ropfull}) we obtain what follows:
\beqn
\hsig_{NS,N}^{(K)[0]}&=&\hsig_{NS,N}^{[0]}\,,
\label{ndegNS0}
\\
\hsig_{NS,N}^{(K)[1]}&=&\hsig_{NS,N}^{[1]}
-r_1\hsig_{NS,N}^{[0]}K_{NS,N+1}^{(K)[1]}\,,
\label{ndegNS1}
\\
\hsig_{NS,N}^{(K)[2]}&=&\hsig_{NS,N}^{[2]}
- r_1\hsig_{NS,N}^{[1]}K_{NS,N+1}^{(K)[1]}
+ \left(r_1^2-r_2\right)
\hsig_{NS,N}^{[0]}\left(K_{NS,N+1}^{(K)[1]}\right)^2
\nonumber
\\&&
- r_1\hsig_{NS,N}^{[0]}K_{NS,N+1}^{(K)[2]}\,,
\label{ndegNS2}
\\
\hsig_{\Sigma,N}^{(K)[0]}&=&\hsig_{\Sigma,N}^{[0]}\,,
\label{ndegS0}
\\
\hsig_{\Sigma,N}^{(K)[1]}&=&\hsig_{\Sigma,N}^{[1]}
-r_1\hsig_{\gamma,N}^{[0]}K_{\gamma\Sigma,N+1}^{(K)[1]}
-r_1\hsig_{\Sigma,N}^{[0]}K_{\Sigma\Sigma,N+1}^{(K)[1]}\,,
\label{ndegS1}
\\
\hsig_{\Sigma,N}^{(K)[2]}&=&\hsig_{\Sigma,N}^{[2]}
-r_1\hsig_{\gamma,N}^{[1]}K_{\gamma\Sigma,N+1}^{(K)[1]}
-r_1\hsig_{\Sigma,N}^{[1]}K_{\Sigma\Sigma,N+1}^{(K)[1]}
\nonumber
\\&&
-r_1\hsig_{\gamma,N}^{[0]}K_{\gamma\Sigma,N+1}^{(K)[2]}
-r_1\hsig_{\Sigma,N}^{[0]}K_{\Sigma\Sigma,N+1}^{(K)[2]}
\nonumber
\\&&
+\left(r_1^2-r_2\right)
\hsig_{\gamma,N}^{[0]}K_{\gamma\Sigma,N+1}^{(K)[1]}
K_{\Sigma\Sigma,N+1}^{(K)[1]}
+\left(r_1^2-r_2\right)
\hsig_{\Sigma,N}^{[0]}\left(K_{\Sigma\Sigma,N+1}^{(K)[1]}\right)^2,
\label{ndegS2}
\\
\hsig_{\gamma,N}^{(K)[0]}&=&\hsig_{\gamma,N}^{[0]}\,,
\label{ndegg0}
\\
\hsig_{\gamma,N}^{(K)[1]}&=&\hsig_{\gamma,N}^{[1]}\,,
\label{ndegg1}
\\
\hsig_{\gamma,N}^{(K)[2]}&=&\hsig_{\gamma,N}^{[2]}\,.
\label{ndegg2}
\eeqn
Equations~(\ref{ndegNS1}), (\ref{ndegS1}), and~(\ref{ndegg1})
coincide, as they should, with the scheme-change contributions
to the so-called $(n+1)$-body degenerate cross sections in the
FKS~\cite{Frixione:1995ms,Frixione:1997np} formalism. Those
cross sections are the remainders of the subtractions of
initial-state collinear singularities and, flavour-wise, they
have the structure of a reduced cross section (which, at the NLO,
is a Born-level one), times generalised branching kernels (which 
include the scheme-change functions $K$). Equations~(\ref{ndegNS2}), 
(\ref{ndegS2}), and~(\ref{ndegg2}) then suggest one how to generalise 
such scheme-change contributions to all orders. In fact, we can re-write
all of eqs.~(\ref{ndegNS0})--(\ref{ndegg2}), and their all-order
generalisations, in a single formula, which reads as follows:
\beqn
\hsig_{\alpha,N}^{(K)[j]}&=&
\sum_{m_1,\ldots m_j=0}^{\sum_p m_p\le j}
\sum_{\delta_1\cdots\delta_j}
c_{m_1,\ldots m_j}\hsig_{\delta_j,N}^{[j-\sum_p m_p]}
\prod_{p=1}^j K_{\delta_p\delta_{p-1},N+1}^{(K)[m_p]}\,,
\label{Kxsecsingle}
\eeqn
with the coefficients $c_{m_1,\ldots m_j}$ depending on the $r_i$'s
(being all equal to one when $r(x)$ is that of eq.~(\ref{rfundef})),
and where we have conventionally set
\beq
\delta_0=\alpha,\;\;\;\;
K_{ab,N+1}^{(K)[0]}=1\,.
\label{dKconv}
\eeq
The l.h.s.~of eq.~(\ref{Kxsecsingle}) is the scheme-change contribution
to an N$^j$LO cross section (i.e.~one of $\ord(\aem^j)$ relative to
the Born). The r.h.s.~features the sum over all possible branchings,
each of which at all possible perturbative orders, compatible with the
constraint that the orders of all branchings, plus that of the reduced
cross section, be equal to $j$. Schematically, we can write this
chain of branchings as follows:
\beq
\hsig_\alpha(\ord(\aem^j)):
\alpha\equiv\delta_0\stackrel{\ord(\aem^{m_1})}{\longrightarrow}\delta_1
\stackrel{\ord(\aem^{m_2})}{\longrightarrow}\delta_2\ldots
\stackrel{\ord(\aem^{m_j})}{\longrightarrow}\delta_j:
\hsig_{\delta_j}(\ord(\aem^{j-m_1-\ldots m_j}))\,.
\label{KxsecsingleB}
\eeq
In words: flavour $\alpha\equiv\delta_0$ branches at $\ord(\aem^{m_1})$
into flavour $\delta_1$ (plus whatever remnants are necessary to conserve
charge and flavour); $\delta_1$ then branches at $\ord(\aem^{m_2})$
into flavour $\delta_2$, and so forth till flavour $\delta_j$ is
obtained, which enters the reduced cross section. Note that while
we have written a chain of $j$ branchings, those with $m_p=0$ are
trivial (consistently with eq.~(\ref{dKconv})); thus, a non-trivial
scheme-change contribution occurs only when $m_p\ge 1$.

Equations~(\ref{Kxsecsingle}) and~(\ref{KxsecsingleB}) render it clear
how these results can be generalised to the physical case of two 
incoming legs; it is sufficient to associate a branching chain of
that type with each of the two legs, keeping in mind that now the
constraint on the perturbative order applies to both legs 
simultaneously. Equation~(\ref{KxsecsingleB}) leads to:
{\beq
\hsig_{\alpha\beta}(\ord(\aem^j)):\left\{
\begin{array}{c}
\alpha\equiv\delta_0\stackrel{\ord(\aem^{m_1})}{\longrightarrow}\delta_1
\stackrel{\ord(\aem^{m_2})}{\longrightarrow}\delta_2\ldots
\stackrel{\ord(\aem^{m_{j_1}})}{\longrightarrow}\delta_{j_1}\\
\beta\equiv\lambda_0\stackrel{\ord(\aem^{n_1})}{\longrightarrow}\lambda_1
\stackrel{\ord(\aem^{n_2})}{\longrightarrow}\lambda_2\ldots
\stackrel{\ord(\aem^{n_{j_2}})}{\longrightarrow}\lambda_{j_2}
\end{array}
\right.\;\;
:\hsig_{\delta_{j_1}\lambda_{j_2}}(\ord(\aem^{j-m_1-\ldots n_{j_2}}))\,,
\label{KxsecdoubleB}
\eeq
for all possible values of $j_1$ and $j_2$ with \mbox{$j_1+j_2=j$}.
In formulae:
\beqn
\hsig_{\alpha\beta,N}^{(K)[j]}&=&\sum_{j_1,j_2=0}^{j_1+j_2=j}\;
\sum_{m_1,\ldots m_{j_1}=0}^{\sum_p m_p\le {j_1}}\;
\sum_{\delta_1\cdots\delta_{j_1}}\;
\sum_{n_1,\ldots n_{j_2}=0}^{\sum_q n_q\le {j_2}}\;
\sum_{\lambda_1\cdots\lambda_{j_2}}\,
c_{m_1,\ldots m_{j_1}}
c_{n_1,\ldots n_{j_2}}
\label{Kxsecdouble}
\\&&\phantom{aaa}\times\,
\hsig_{\delta_{j_1}\lambda_{j_2},N}^{[j-\sum_p m_p-\sum_q n_q]}
\prod_{p=1}^{j_1} K_{\delta_p\delta_{p-1},N+1}^{(K)[m_p]}
\prod_{q=1}^{j_2} K_{\lambda_q\lambda_{q-1},N+1}^{(K)[n_q]}\,.
\nonumber
\eeqn
The $K$ matrices that appear in eqs.~(\ref{ndegNS0})--(\ref{ndegg2})
are given in appendix~\ref{sec:resK}, for the choice of $r(x)$ 
of eq.~(\ref{rfundef}). However, one needs also to consider the
product of two such $K$ matrices, and compute the inverse Mellin transform
of the results. These quantities are also given in  appendix~\ref{sec:resK}.

There is a subtlety related to the use of the evolution basis for
cross section computations. Specifically, we have defined the singlet 
and non-singlet components according to eq.~(\ref{snsdef}), which stems 
from eq.~(\ref{Tmatdef}). However, one must keep in mind that that 
definition assumes the particle whose partons we consider is the electron.
If one deals with a positron instead, one must make the choice of
either keeping the same definition of the non-singlet as for the electron,
or to flip its sign. The latter option implies that the non-singlet thus
defined is invariant under charge conjugation, whereas the former leads
to a positron non-singlet which is essentially negative on the whole $z$ 
range. For this reason, we adopt the charge-conjugation invariant
definition. Explicitly, by using the fully unambiguous notation where 
the PDFs have two flavour indices, the rightmost of which denotes the
incoming particle, then eq.~(\ref{snsdef}) would be re-written
as follows:
\beqn
&&\PDF{{\rm S}}{\lm}=\PDF{\lm}{\lm}+\PDF{\lp}{\lm}\,,
\;\;\;\;\;\;\;\;\;\;\;\;\;\;\,
\PDF{{\rm NS}}{\lm}=\PDF{\lm}{\lm}-\PDF{\lp}{\lm}\,,
\label{snsdef2}
\eeqn
while its positron counterpart reads thus:
\beqn
&&\PDF{{\rm S}}{\lp}=\PDF{\lp}{\lp}+\PDF{\lm}{\lp}\,,
\;\;\;\;\;\;\;\;\;\;\;\;\;\;\,
\PDF{{\rm NS}}{\lp}=\PDF{\lp}{\lp}-\PDF{\lm}{\lp}\,.
\label{snsdef3}
\eeqn
Further to that, we want to write the incoherent parton sums in the
factorisation theorem in an unique way regardless of whether we
use the physical or the evolution basis, i.e.
\beq
\sum_{\alpha\beta}\PDF{\alpha}{P}\PDF{\beta}{Q}\hat{\sigma}_{\alpha\beta}\,,
\;\;\;\;\;\;\;\;
\alpha,\,\beta\in\{\lm,\lp,\gamma\}\;\;\;{\rm or}\;\;\;
\alpha,\,\beta\in\{NS,\Sigma,\gamma\}\,,
\eeq
for any pairs of incoming particles $P$ and $Q$. By imposing this
condition we end up by defining short-distance cross sections for an 
incoming $(\lm,\lp)$ pair (note the order: the electron is coming 
from the left) in the evolution basis as follows:
\beqn
\hat{\sigma}_{NS,NS}&=&\frac{1}{4}
\left(\hat{\sigma}_{\lm\lp}-\hat{\sigma}_{\lp\lp}+
      \hat{\sigma}_{\lp\lm}-\hat{\sigma}_{\lm\lm}\right),
\label{sNSNS}
\\
\hat{\sigma}_{\Sigma,\Sigma}&=&\frac{1}{4}
\left(\hat{\sigma}_{\lm\lp}+\hat{\sigma}_{\lp\lp}+
      \hat{\sigma}_{\lp\lm}+\hat{\sigma}_{\lm\lm}\right),
\\
\hat{\sigma}_{\Sigma,NS}&=&\frac{1}{4}
\left(\hat{\sigma}_{\lm\lp}+\hat{\sigma}_{\lp\lp}-
      \hat{\sigma}_{\lp\lm}-\hat{\sigma}_{\lm\lm}\right)=0,
\label{sSNS}
\\
\hat{\sigma}_{NS,\Sigma}&=&\frac{1}{4}
\left(\hat{\sigma}_{\lm\lp}-\hat{\sigma}_{\lp\lp}-
      \hat{\sigma}_{\lp\lm}+\hat{\sigma}_{\lm\lm}\right)=0,
\\
\hat{\sigma}_{NS,\gamma}&=&\frac{1}{2}
\left(\hat{\sigma}_{\lm\gamma}-\hat{\sigma}_{\lp\gamma}\right)=0,
\\
\hat{\sigma}_{\Sigma,\gamma}&=&\frac{1}{2}
\left(\hat{\sigma}_{\lm\gamma}+\hat{\sigma}_{\lp\gamma}\right),
\\
\hat{\sigma}_{\gamma,NS}&=&\frac{1}{2}
\left(\hat{\sigma}_{\gamma\lp}-\hat{\sigma}_{\gamma\lm}\right)=0,
\label{sgNS}
\\
\hat{\sigma}_{\gamma,\Sigma}&=&\frac{1}{2}
\left(\hat{\sigma}_{\gamma\lp}+\hat{\sigma}_{\gamma\lm}\right),
\eeqn
where the cross sections equal to zero are so because of charge-conjugation
invariance. This implies that the non-singlet component in the cross 
section decouples from the singlet-photon sector. This must happen for 
a fundamental reason: non-singlet PDF evolution also decouples. Therefore, 
there could not be any factorisation-scale independence of the l.h.s.~of 
the factorisation theorem without an analogous decoupling at the level of 
short-distance cross sections.

We point out that by adopting a charge-invariant non-singlet definition
we are forced to keep track of the identity of the incoming particles
in the construction of the short-distance cross sections, which is typically
not the case. In the present situation, by adopting the alternative definition
for the positron non-singlet one would need to change the sign of the
r.h.s.~of eq.~(\ref{sNSNS}) (which is a consequence of the negative
nature of the positron non-singlet PDF stemming from this 
definition)\footnote{Also the signs of eqs.~\eqref{sSNS} and \eqref{sgNS} 
would have to be changed, however, this is irrelevant, since these 
channels vanish.}.

\subsection{Application to Drell-Yan-like cross sections\label{DYapp}}
In this section we consider the large-$z$ behaviour of Drell-Yan-like 
hard scattering cross sections in the $\Delta$ scheme. These cross sections,
stemming from the tree-level \mbox{$e^+ e^- \to V^*$} process, in
fact encompass several key reactions as far as QED ISR is concerned, 
such as \mbox{$e^+ e^- \to Z^*/\gamma^* \to \mu^+ \mu^-$} and
\mbox{$e^+ e^- \to Z^* \to Z H$}.
The partonic (QCD) cross sections at the NNLO have been first calculated 
in refs.~\cite{Hamberg:1990np,Harlander:2002wh}, and extended to NNNLO in 
refs.~\cite{Duhr:2020seh,Duhr:2021vwj}. We use the ancillary files of the 
latter paper to obtain the partonic cross sections in configuration space, and 
subsequently abelianise them so as to obtain the QED cross sections
we are interested in. In order to be fully self-contained we shall provide
such QED cross sections in a file ancillary to this publication.

For the Drell-Yan cross section we have two incoming legs and we can 
organize the cross section into a matrix form $\tsig_{\alpha\beta}$
(henceforth, in order to avoid any confusion, Drell-Yan cross sections 
will be denoted by $\tilde{\sigma}$). As is discussed above, the
non-singlet cross section does not mix in the evolution, and therefore
we abbreviate $\tsig_{NS \, NS} \to \tsig_{NS}$; note also that
$\tsig_{\Sigma \, \gamma} = \tsig_{\gamma \, \Sigma}$.

By computing explicitly the two-incoming-leg cross sections in a
generic $K$ factorisation scheme from the general formulae  given 
before (eq.~(\ref{Kxsecdouble})), we obtain what follows at the NNLO:
\beqn
\tsig_{NS,N}^{(K)[0]}&=&\tsig_{NS,N}^{[0]}\,,
\label{ndegNS0dy}
\\
\tsig_{NS,N}^{(K)[1]}&=&\tsig_{NS,N}^{[1]}
-2r_1\tsig_{NS,N}^{[0]}K_{NS,N}^{(K)[1]}\,,
\label{ndegNS1dy}
\\
\tsig_{NS,N}^{(K)[2]}&=&\tsig_{NS,N}^{[2]}
- 2r_1\tsig_{NS,N}^{[1]}K_{NS,N}^{(K)[1]}
+ \left(3r_1^2-2r_2\right)
\tsig_{NS,N}^{[0]}\left(K_{NS,N}^{(K)[1]}\right)^2
\nonumber
\\&&
- 2r_1\tsig_{NS,N}^{[0]}K_{NS,N}^{(K)[2]}\,,
\label{ndegNS2dy}
\\
\tsig_{\Sigma \Sigma,N}^{(K)[0]}&=&\tsig_{\Sigma \Sigma,N}^{[0]}\,,
\label{ndegS0dy}
\\
\tsig_{\Sigma \Sigma,N}^{(K)[1]}&=&\tsig_{\Sigma \Sigma,N}^{[1]}
-2r_1\tsig_{\gamma \Sigma,N}^{[0]}K_{\gamma\Sigma,N}^{(K)[1]}
-2r_1\tsig_{\Sigma \Sigma,N}^{[0]}K_{\Sigma\Sigma,N}^{(K)[1]}\,,
\label{ndegS1dy}
\\
\tsig_{\Sigma \Sigma,N}^{(K)[2]}&=&\tsig_{\Sigma \Sigma,N}^{[2]}
-2r_1\tsig_{\gamma \Sigma,N}^{[1]}K_{\gamma\Sigma,N}^{(K)[1]}
-2r_1\tsig_{\Sigma \Sigma,N}^{[1]}K_{\Sigma\Sigma,N}^{(K)[1]}
-2r_1\tsig_{\gamma \Sigma,N}^{[0]}K_{\gamma\Sigma,N}^{(K)[2]}
\nonumber
\\&&
-2r_1\tsig_{\Sigma \Sigma,N}^{[0]}K_{\Sigma\Sigma,N}^{(K)[2]}
+\left(4r_1^2-2r_2\right)
\tsig_{\gamma \Sigma,N}^{[0]}K_{\gamma\Sigma,N}^{(K)[1]}
K_{\Sigma\Sigma,N}^{(K)[1]}
\nonumber
\\&&
+\left(3r_1^2-2r_2\right)
\tsig_{\Sigma \Sigma,N}^{[0]}\left(K_{\Sigma\Sigma,N}^{(K)[1]}\right)^2
+ r_1^2 \tsig_{\gamma \gamma,N}^{[0]}\left(K_{\gamma\Sigma,N}^{(K)[1]}\right)^2\,,
\label{ndegS2dy}
\\
\tsig_{\gamma \Sigma,N}^{(K)[0]}&=&\tsig_{\gamma \Sigma,N}^{[0]}\,,
\label{ndeggS0dy}
\\
\tsig_{\gamma \Sigma,N}^{(K)[1]}&=&\tsig_{\gamma \Sigma,N}^{[1]}
- r_1 \tsig_{\gamma \Sigma,N}^{[0]}K_{\Sigma\Sigma,N}^{(K)[1]}
- r_1 \tsig_{\gamma \gamma,N}^{[1]}K_{\gamma\Sigma,N}^{(K)[1]} \,,
\label{ndeggS1dy}
\\
\tsig_{\gamma \Sigma,N}^{(K)[2]}&=&\tsig_{\gamma \Sigma,N}^{[2]}
-r_1\tsig_{\gamma \Sigma,N}^{[1]}K_{\Sigma\Sigma,N}^{(K)[1]}
-r_1\tsig_{\gamma \gamma,N}^{[1]}K_{\gamma\Sigma,N}^{(K)[1]}
-r_1\tsig_{\gamma \Sigma,N}^{[0]}K_{\Sigma\Sigma,N}^{(K)[2]}
\nonumber
\\&&
-r_1\tsig_{\gamma \gamma,N}^{[0]}K_{\gamma\Sigma,N}^{(K)[2]}
+\left(r_1^2-r_2\right)
\tsig_{\gamma \gamma,N}^{[0]}K_{\gamma\Sigma,N}^{(K)[1]}
K_{\Sigma\Sigma,N}^{(K)[1]}
\nonumber
\\&&
+\left(r_1^2-r_2\right)
\tsig_{\gamma \Sigma,N}^{[0]}\left(K_{\Sigma\Sigma,N}^{(K)[1]}\right)^2
\label{ndeggS2dy}
\\
\tsig_{\gamma \gamma,N}^{(K)[0]}&=&\tsig_{\gamma \gamma,N}^{[0]}\,,
\label{ndegg0dy}
\\
\tsig_{\gamma \gamma,N}^{(K)[1]}&=&\tsig_{\gamma \gamma,N}^{[1]}\,,
\label{ndegg1dy}
\\
\tsig_{\gamma \gamma,N}^{(K)[2]}&=&\tsig_{\gamma \gamma,N}^{[2]}\,.
\label{ndegg2dy}
\eeqn
Note that, at variance with the general formulae given before, here
the short-distance cross sections and the $K$-matrix elements are both 
computed at $N$. This is because, conventionally and differently w.r.t.~what 
has been assumed thus far (specifically, see footnote~\ref{ft:facth}), the 
results of refs.~\cite{Duhr:2021vwj} do {\em not} include partonic 
flux factors. It is a simple exercise to show that the consequence of 
this is to shift the Mellin moments of the scheme-change functions from 
$N+1$ to $N$. 

By using the equations above, we arrive at the large-$N$ behaviour of 
the Drell-Yan abelianised cross sections in both the $\MSb$ and the 
$\Delta$ scheme. Importantly, in the latter case the dependence on
the $r_i$ coefficients of the $r(x)$ function completely drops out.
While this is the expected behaviour, it nevertheless constitutes
a significant self-consistency check. At $\mu^2=Q^2$ we find:
\beqn
  \tsig_{\Sigma\Sigma,N}^{[0]}&=& \frac{1}{2}\,,
  \\
  \tsig_{\Sigma\Sigma,N}^{(\Delta)[0]}&=& \frac{1}{2}\,,
  \\
  \tsig_{\Sigma\Sigma,N}^{[1]}&=& 2 \log^2(\bN) + \frac{2 \pi^2}{3} - 4 
%  + \frac{2}{N} \log(\bN) 
%  - \frac{1}{N^2}
%  \left(
%    \frac{5}{2} - \frac{7}{3} \log(\bN)
%  \right)
  + \mathcal{O}(N^{-1}) \,,
  \\
  \tsig_{\Sigma\Sigma,N}^{(\Delta)[1]}&=&
  2 \log(\bN) + \frac{\pi^2}{3} - 2 
%  + \frac{3}{N} - \frac{13}{6 N^2} 
  + \mathcal{O}(N^{-1}) \,,
  \\
  \tsig_{\Sigma\Sigma,N}^{[2]}&=& 
  4 \log^4(\bN)
  - \frac{8}{9} \NF \log^3(\bN)
  + \biggl(   
    \frac{8 \pi^2}{3}-16 - \frac{20 \NF}{9}
  \biggr) \log^2(\bN)
  - \frac{56}{27} \NF \log(\bN)
  \nonumber \\ &&
  + \frac{511}{32}
  - \frac{33 \pi^2}{8}
  + \frac{23 \pi^4}{60}
  - \frac{15 \zeta_3}{2}
  + \biggl(   
    \frac{127}{24}
    - \frac{8 \pi^2}{9}
    + \frac{2 \zeta_3}{9}
  \biggr) \NF
  + \mathcal{O}(N^{-1}) \,,
  \\
  \tsig_{\Sigma\Sigma,N}^{(\Delta)[2]}&=&
  - \frac{8}{9} \NF \log^3(\bN)
  + \biggl(   
    4 - \frac{20 \NF}{9}
  \biggr) \log^2(\bN)
  + \biggl(   
    \frac{4 \pi^2}{3} - 8 - \frac{112 \NF}{27}
  \biggr) \log(\bN)
  + \frac{19}{2}
  \nonumber \\ &&
  + \frac{19 \pi^2}{24}
  - \frac{\pi^4}{15}
  - 9 \zeta_3
  - 2 \pi^2 \log(2)
  + \biggl(   
    \frac{821}{81}
    - \frac{11 \pi^2}{9}
    + \frac{8 \zeta_3}{9}
  \biggr) \NF
  + \mathcal{O}(N^{-1}) \,.
\eeqn
In configuration space this leads to the following large-$z$ behaviour:
\beqn
  \tsig_{\Sigma\Sigma}^{[0]}&=& \frac{1}{2} \delta(1-z) \,,
  \\
  \tsig_{\Sigma\Sigma}^{(\Delta)[0]}&=& \frac{1}{2} \delta(1-z) \,,
  \\
  \tsig_{\Sigma\Sigma}^{[1]}&=& \delta(1-z) \left( \frac{\pi^2}{3} - 4 \right) + 4 \left( \frac{\log(1-z)}{1-z} \right)_{+}
  + \mathcal{O}\left( (1-z)^0 \right) \,,
  \\
  \tsig_{\Sigma\Sigma}^{(\Delta)[1]}&=& \delta(1-z) \left( \frac{\pi^2}{3} - 2 \right) - 2 \left( \frac{1}{1-z} \right)_{+}
  + \mathcal{O}\left( (1-z)^0 \right) \,,
  \\
  \tsig_{\Sigma\Sigma}^{[2]}&=& 
  \delta(1-z)
  \biggl(   
      \frac{511}{32}
    - \frac{35 \pi^2}{24}
    + \frac{ \pi^4}{180}
    - \frac{15 \zeta_3}{2}
  + \biggl[   
    \frac{127}{24}
    - \frac{14 \pi^2}{27}
    + \frac{2 \zeta_3}{9}
  \biggr] \NF
  \biggr)
  \nonumber \\ &&
  + 16 \left( \frac{\log^3(1-z)}{1-z} \right)_{+}
  + \frac{8}{3} \NF \left( \frac{\log^2(1-z)}{1-z} \right)_{+}
  - \biggl(   
   32
   + \frac{8 \pi^2}{3}
   + \frac{40 \NF}{9} 
  \biggr) \nonumber \\ &&
  \times \left( \frac{\log(1-z)}{1-z} \right)_{+}
  + 
  \biggl(   
    32 \zeta_3 
    + \biggl(  
      \frac{56}{27}
      - \frac{4 \pi^2}{9}
    \biggr) \NF
  \biggr) \left( \frac{1}{1-z} \right)_{+}
  \nonumber \\ &&
  + \mathcal{O}\left( (1-z)^0 \right) \,,
  \\
  \tsig_{\Sigma\Sigma}^{(\Delta)[2]}&=&
  \delta(1-z)
  \biggl(   
     \frac{19}{2}
    + \frac{\pi^2}{8}
    - \frac{\pi^4}{15}
    - 2 \pi^2 \log(2)
    - 9 \zeta_3
  + \biggl[   
    \frac{821}{81}
    - \frac{23 \pi^2}{27}
    + \frac{8 \zeta_3}{3}
  \biggr] \NF
  \biggr)
  \nonumber \\ &&
  + \frac{8}{3} \NF \left( \frac{\log^2(1-z)}{1-z} \right)_{+}
  + \biggl(   
   8
   - \frac{40 \NF}{9} 
  \biggr) \left( \frac{\log(1-z)}{1-z} \right)_{+}
  + \biggl(    
    8 
    - \frac{4 \pi^2}{3}
    \nonumber \\ &&
    + \biggl(  
      \frac{112}{27}
      - \frac{4 \pi^2}{9}
    \biggr) \NF
  \biggr) \left( \frac{1}{1-z} \right)_{+}
  + \mathcal{O}\left( (1-z)^0 \right) \,.
\eeqn

For the mixed channel we obtain
\beqn   
\tsig_{\gamma\Sigma,N}^{[0]}&=&\tsig_{\Sigma\Sigma,N}^{(\Delta)[0]}=0
\,, \\
\tsig_{\gamma\Sigma,N}^{[1]}&=&\tsig_{\Sigma\Sigma,N}^{(\Delta)[1]}= -4 \frac{\log(\bN)}{N}
+ \mathcal{O}(N^{-2}) \,, \\
\tsig_{\gamma\Sigma,N}^{[2]}&=&\frac{1}{N}
\biggl[ 
  - \frac{35}{3} \log^3(\bN) 
  -6\log^2(\bN) 
  + \biggl(
    21
    - \frac{9 \pi^2}{9}
  \biggr) \log(\bN)
  \nonumber \\ &&
  -5 
  + \frac{44 \zeta_3}{3}
\biggr]
+ \mathcal{O}(N^{-2}) \,, \\
\tsig_{\gamma\Sigma,N}^{(\Delta)[2]}&=&\frac{1}{N}
\biggl[ 
  - \frac{11}{3} \log^3(\bN) 
  -14\log^2(\bN) 
  + \biggl(
    13
    - \frac{19 \pi^2}{6}
  \biggr) \log(\bN)
  \nonumber \\ &&
  -5 
  + \frac{44 \zeta_3}{3}
\biggr]
+ \mathcal{O}(N^{-2}) \,.
\eeqn   
In configuration space this leads to the following large-$z$ behaviour:
\beqn   
\tsig_{\gamma\Sigma}^{[0]}&=& \tsig_{\gamma\Sigma}^{(\Delta)[0]} = 0
\,, \\
\tsig_{\gamma\Sigma}^{[1]}&=& \tsig_{\gamma\Sigma}^{(\Delta)[1]} = 
4 \log(1-z)
+ \mathcal{O}\bigl((1-z)\bigr)
\,, \\
\tsig_{\gamma\Sigma}^{[2]}&=&
\frac{35}{3} \log ^3(1-z)
-6 \log ^2(1-z)
- \biggl( 
  \frac{4 \pi ^2}{3}  
  +21 
\biggr) \log (1-z)
\nonumber \\ &&
+38 \zeta_{3}
+\pi ^2
-5
+ \mathcal{O}\bigl((1-z)\bigr)
\,, \\
\tsig_{\gamma\Sigma}^{(\Delta)[2]}&=&
\frac{11}{3} \log ^3(1-z)
-14 \log ^2(1-z)
+\left(\frac{4 \pi ^2}{3}-13\right) \log (1-z)
\nonumber \\ &&
+22 \zeta_{3}
+\frac{7 \pi ^2}{3}
-5
+ \mathcal{O}\bigl((1-z)\bigr)
\,.
\eeqn
We do not present results for the $\tsig_{\gamma\gamma}$, since it is 
not affected by the scheme change.

We observe that the singlet-singlet (and therefore also the non-singlet) cross 
section in the $\MSb$ scheme at $\ord(\aem^k)$ has a leading behaviour 
proportional to $\log^{2k}(\bN)$ when $N \to \infty$. Conversely, in 
the $\Delta$ scheme, this is reduced to  $\log^{2k-1}(\bN)$, thus making 
the cancellation of the leading logarithmic contribution manifest before 
convoluting the PDFs with the partonic cross sections.
This early cancellation will lead to a better numerical stability and 
reduced computational cost.
The large-$N$ behaviour of the $\gamma$-singlet cross section is altered,
but the power supression and the logarithmic structure is not changed.

\section{Conclusions\label{sec:concl}}
The main results of this paper are the generalisation of the definition
of the $\Delta$ factorisation scheme~\cite{Frixione:2021wzh} to all
orders in QED, and the computation of the evolved electron PDFs at 
the NNLO in QED, building on the fixed-order results of 
refs.~\cite{Blumlein:2011mi,Ablinger:2020qvo}, with analytical
predictions in the asymptotic $z\to 1$ region provided for both the
$\MSb$ and the $\Delta$ scheme lepton PDFs.
Given the relevant $\MSb$ initial conditions and Altarelli-Parisi kernels,
the procedure we have defined will allow the determination of the evolved
QED lepton PDFs at N$^k$LO, with $k\ge 3$, in the context of any factorisation 
scheme; this, thanks to the fact that, in Mellin space, those kernels are
expected to be linear in $\log\bN$, while the $\MSb$ initial conditions
will have a $\log^{2k}\bN$ leading behaviour.

The $\Delta$ scheme we have defined is in fact a class of schemes,
all members of which are characterised by the fact that no $\delta(1-z)$ 
terms or plus distributions with $z=1$ endpoint are present in the resulting 
lepton initial conditions, in this way implying that the evolved PDFs are
completely free of integrable $z\to 1$ singularities, except for those
included (and resummed) in the Gribov-Lipatov prefactor. With the
exception of this underlying common feature, the individual $\Delta$
factorisation schemes can be rather freely defined through the use of 
a function $r(x)$, which is only required to have a meaningful expansion 
at $x=0$ with non-null first-order coefficient, and $r(0)=1$.

We have underscored how the definition of the $\Delta$ scheme, and
in fact of any scheme which modifies the large-$z$ behaviour of the
initial conditions, has the potential to alter the small-$z$ functional
forms of the evolved PDFs. While small-$z$ dynamics will not play a role
at any planned future $\epem$ colliders, we have still preferred to
avoid this problem altogether, by means of a procedure we have
called $\log z\to 1-z$ prescription, and which relies on exact
Mellin and inverse-Mellin transforms of elementary functions, several of 
which are presented here for the first time.

As a token demonstration of the $\Delta$ scheme at work, we have
considered the abelianised Drell-Yan process, and shown how
in such a scheme the dominant power of the soft logarithms simply
vanishes from the short-distance cross sections, order by order
in $\aem$.

Finally, we have argued how the NNLO PDFs presented here are necessary
for phenomenological predictions of relative accuracy equal to or better
than $10^{-4}$, and will most likely remain sufficient up to $10^{-6}$; 
beyond that value, N$^3$LO results will become necessary. These figures 
have been obtained without reference to specific processes, and must be 
confirmed and validated in the context of realistic simulations.

Regardless of this, what is certain is that one must integrate 
cross sections up to $z$ values {\em extremely} close to one in
order to obtain numerically accurate predictions
(see e.g.~fig.~\ref{fig:xiplot}). We stress that this statement is
meant to be understood in an inclusive sense -- in an MC implementation,
one can safely ignore final-state effects due to sub-eV photons associated
with such extreme $z$ values. With that being said, we point out that there 
is still an exclusive aspect of any MC implementation which is much easier
to deal with in the $\Delta$ scheme than in $\MSb$, since the 
initial conditions in the former scheme are regular functions (whose
handling is standard), while in the latter one they are plus distributions, 
which force one to have an awkward coupled event-counterevent shower structure 
(or any equivalent variants of that).

All of the PDFs calculated here have been obtained by working in the
context of a single fermion family. However, as has been already
shown e.g.~in ref.~\cite{Bertone:2022ktl}, this is actually all of
the substantial technical work required to consider the generic
variable-flavour number scheme case, which we postpone to a
forthcoming paper.

\section*{Acknowledgments}
SF is grateful to M.~Bonvini and G.~Stagnitto for discussions on
the small-$x$ implications of a non-canonical $\Delta$-like scheme,
and to P.~Torrielli for information. We thank P.~Monni for a few 
brainstorming sessions on topics relevant to this paper and its possible 
by-products, and M.~Mangano for helping us improve the manuscript.
KS~is supported by the European Union under the Marie Sk{\l}odowska-Curie 
Actions (MSCA) Grant 101204018.

\appendix
\section{Mellin transforms\label{sec:Mell}}
We call $z$ the configuration-space variable \mbox{$z\in [0,1]$}, 
and denote by $N$ its counterpart in the conjugate space. The Mellin 
transform of a (generalised) function $f(z)$ is defined thus:
\beq
M[f](N)\equiv f_N=\int_0^1 dz\,z^{N-1}f(z)\,.
\label{Mellconv}
\eeq
We also introduce the auxiliary quantity
\beq
\Lz=-\log\left(-\log z\right)\,,\;\;\;\;\;\;\;\;
%%%\LE=\Lz-\gE\,,
\label{LEAdef}
\eeq
and systematically understand a factor $\stepf(0\le z\le 1)$ which enforces
the support for the test functions to be equal to \mbox{$[0,1]$}.

We are interested in obtaining results for the Mellin (inverse
Mellin) transforms of function featuring $\log z$, $\Lz$, and $\log(1-z)$
($\log N$, and $\log\bN$). This can be done by starting from a few
elementary integrals, the Mellin transforms of relatively simple
functions. These integrals are:
\beqn
I_1(N;a)&=&M\left\{\left[\frac{1}{(-\log z)^a}\right]_+\right\}
=\int_0^1 dz\left[\frac{1}{(-\log z)^a}\right]_+ z^{N-1}
\nonumber
\\*&=&\phantom{\Big[}
\left(-1+N^{a-1}\right)\Gamma(1-a)\,,
\label{masterI1}
\\
I_2(N;a)&=&M\left\{(-\log z)^a\right\}
=\int_0^1 dz(-\log z)^a z^{N-1}
\nonumber
\\*&=&\phantom{\Big[}
N^{-a-1}\Gamma(1+a)\,,
\label{masterI2}
\\
I_3(N;a)&=&M\left\{\left[\frac{1}{(1-z)^{1-a}}\right]_+\right\}
=\int_0^1 dz\left[\frac{1}{(1-z)^{1-a}}\right]_+ z^{N-1}
\nonumber
\\*&=&
-\frac{1}{a}+\frac{\Gamma(a)\Gamma(N)}{\Gamma(a+N)}\,,
\label{masterI3}
\\
I_4(N;a)&=&M\left\{(1-z)^{a}\right\}
=\int_0^1 dz(1-z)^{a} z^{N-1}
\nonumber
\\*&=&
\frac{\Gamma(1+a)\Gamma(N)}{\Gamma(1+a+N)}\,.
\label{masterI4}
\eeqn
By applying the inverse Mellin transform on both sides of
eqs.~(\ref{masterI1}) and~(\ref{masterI2}) one obtains
what follows:
\beqn
J_1(z;a)=\invM\left[N^{a-1}\right]&=&\delta(1-z)+
\frac{1}{\Gamma(1-a)}\left[\frac{1}{(-\log z)^a}\right]_+\,,
\label{mastersa}
\\
J_2(z;a)=\invM\left[N^{-a-1}\right]&=&
\frac{(-\log z)^a}{\Gamma(1+a)}\,,
\label{mastersa2}
\eeqn
The sought results emerge from repeated derivations of the previous
equations w.r.t.~$a$, assuming strong convergence of the integrals, and 
subsequent expansions in $a$ around suitable (generally integer) values,
in order to exploit the following relationships:
\beqn
&&\frac{d^m}{da^m}\frac{1}{(-\log z)^a}=
\frac{\Lz^m}{(-\log z)^a}
\nonumber\\&&\phantom{aaaa}\Longrightarrow\;\;\;\;
M\left\{\left[\frac{\Lz^m}{-\log z}\right]_+\right\}=
\left.\frac{d^m}{da^m}I_1(N;a)\right|_{a=1}\,,
\\
&&\frac{d^m}{da^m}(-\log z)^a=(-\Lz)^m(-\log z)^a
\nonumber\\&&\phantom{aaaa}\Longrightarrow\;\;\;\;
M\Big\{(-\Lz)^m(-\log z)^q\Big\}=
\left.\frac{d^m}{da^m}I_2(N;a)\right|_{a=q}\,,
\eeqn
for any $m\ge 0$, $q\ge 0$, as well as:
\beqn
&&\frac{d^m}{da^m}N^{a-1}=
N^{a-1}\log^m N
\nonumber\\&&\phantom{aaaa}\Longrightarrow\;\;\;\;
M^{-1}\Big[\log^m N\Big]=
\left.\frac{d^m}{da^m}J_1(z;a)\right|_{a=1}\,,
\label{invlgmN}
\\
&&\frac{d^m}{da^m}N^{-a-1}=
N^{-a-1}(-)^m\log^m N
\nonumber\\&&\phantom{aaaa}\Longrightarrow\;\;\;\;
M^{-1}\left[\frac{\log^m N}{N^q}\right]=
(-)^m\left.\frac{d^m}{da^m}J_2(N;a)\right|_{a=q-1}\,,
\label{invlgmNoNq}
\eeqn
for any $m\ge 0$, $q\ge 1$. Explicit computations then lead to 
the following results. With the auxiliary quantities ($q\ge 0$):
\beqn
W&=&\Big(\Big\{\psi_k(1)\Big\}_{k\ge 0}\Big)\,,
\\
W_q&=&\Big(\Big\{(-1)^kk!H_q^{(k+1)}+\psi_k(1)\Big\}_{k\ge 0}\Big)\,,
\eeqn
where $H_n^{(r)}$ is the harmonic number of order $r$, one obtains:
\beqn
M\left\{\left[\frac{(-\Lz)^m}{\log z}\right]_+\right\}&=&
\log N\sum_{i=0}^m(-)^i\binomial{m}{i}B_{m-i}(W)
\frac{\log^i N}{i+1}\,,
\label{Mlgzplus}
\\
M\Big\{(-\Lz)^m\log^q z\Big\}&=&
\frac{(-)^qq!}{N^{q+1}}
\sum_{i=0}^m(-)^i\binomial{m}{i}B_{m-i}(W_q)\log^i N\,,
\label{Mlgz}
\eeqn
where $B_n$ are the complete exponential Bell polynomials,
whereas by introducing ($q\ge 1$):
\beqn
Y&=&\Big(\Big\{-(1-\delta_{k0})(-)^{k+1}\psi_k(1)\Big\}_{k\ge 0}\Big)\,,
\label{Yvec}
\\
Y_q&=&\Big(\Big\{(-1)^{k+1}k!H_{q-1}^{(k+1)}-
(1-\delta_{k0})\psi_k(1)\Big\}_{k\ge 0}\Big)\,,
\label{Yqvec}
\eeqn
one finds the related inversion formulae\footnote{By writing the result 
relevant to logarithms of $\bN$, rather than of $N$, one obtains more 
compact expressions.}:
\beqn
\invM\left[\log^m \bN\right]&\!\!=\!\!&\gE^m\delta(1-z)
-m\sum_{i=0}^{m-1}\binomial{m-1}{i}B_{m-i-1}(Y)
\left[\frac{\Lz^i}{-\log z}\right]_+\,,
\label{invMlgzplus}
\\
\invM\left[\frac{\log^m \bN}{N^q}\right]&\!\!=\!\!&
\frac{(-)^{q+m+1}}{(m+1)(q-1)!}
\nonumber\\&&\times
\sum_{i=0}^{m+1}i\binomial{m+1}{i}B_{i-1}(Y_q)(-\Lz)^{m+1-i}\log^{q-1}z\,.
\label{invMlgz}
\eeqn
As far as the Mellin transforms of functions that feature $\log(1-z)$
terms are concerned, we need to employ the following auxiliary vectors 
($q\ge 0$):
\beqn
Z&=&\Big(\Big\{-\psi_k(N)+\psi_k(1)\Big\}_{k\ge 0}\Big)\,,
\\
Z_q&=&\Big(\Big\{-\psi_k(N+q+1)+\psi_k(q+1)\Big\}_{k\ge 0}\Big)\,,
\eeqn
with which
\beqn
M\left\{\left[\frac{\log^m(1-z)}{1-z}\right]_+\right\}&=&
\frac{1}{m+1}\,B_{m+1}(Z)\,,
\label{Mlgomzplus}
\\
M\Big\{(1-z)^q\log^m(1-z)\Big\}&=&
\frac{q!\Gamma(N)}{\Gamma(N+q+1)}\,B_{m}(Z_q)\,.
\label{Mlgomz}
\eeqn
Note that all of the transforms above are exact. We can explicitly verify
that by applying eq.~(\ref{invMlgzplus}) to the terms on the r.h.s.~of
eq.~(\ref{Mlgzplus}) we obtain the argument of the Mellin transform
on the l.h.s.~of the latter equation, for any choice of $m\ge 0$. 
In formulae:
\beqn
&&M\left\{\left[\frac{(-\Lz)^m}{\log z}\right]_+\right\}
\Big(\log N\to\log\bN-\gE\Big);
\\*&&
\phantom{M\left\{\left[\frac{(-\Lz)^m}{\log z}\right]_+\right\}\;\;}
\log^k\bN\to
\invM\left[\log^k \bN\right]=
\left[\frac{(-\Lz)^m}{\log z}\right]_+\,.
\nonumber
\eeqn
The analogous property holds true in the case of eqs.~(\ref{invMlgz}) 
and~(\ref{Mlgz}), for any $m\ge 0$ and $q\ge 0$. This is due to the
fact that both the Mellin transforms and their inverse in those
equations are exact.

In view of the fact that at $z\simeq 1$ $\log z\simeq 1-z$, it is
interesting to consider the relationships between eq.~(\ref{Mlgomzplus})
and~(\ref{Mlgzplus}), and between eq.~(\ref{Mlgomz}) and~(\ref{Mlgz}).
The following series expansions are therefore relevant ($q\ge 0$):
\beqn
&&\!\!\!\!\log^q z=
\sum_{n=0}^\infty f_{n}^{(q)}(z-1)^n\,,
\label{lgzvslgom0}
\\*
&&\!\!\!\!\frac{1}{\log z}=-\frac{1}{1-z}+
\sum_{n=0}^\infty g_{n+1}(z-1)^n\,,
\label{lgzvslgom1}
\\*
&&\!\!\!\!\log(-\log z)=\log(1-z)+
\sum_{n=1}^\infty h_{n+1}(z-1)^n\,,
\label{lgzvslgom2}
\eeqn
where
\beqn
f_{n}^{(q)}&=&
\frac{q!}{n!}B_{n,q}\left(\big\{(-)^j j!\big\}_{j\ge 0}\right)=
\frac{q!}{n!}(-)^{n+q}\stirlingSo{n}{q}\,,
\label{cfffnq}
\\*
g_n&=&\frac{1}{n!}
\sum_{i=1}^n\frac{1}{i+1}B_{n,i}\left(\big\{(-)^j j!\big\}_{j\ge 0}\right)
\nonumber \\*&=&
\frac{1}{n!}\sum_{i=1}^n
\frac{(-)^{n+i}}{i+1}\stirlingSo{n}{i}\,,
\label{Gcff}
\\*
h_n&=&\frac{1}{(n-1)(n-1)!}
\sum_{i=1}^{n}
\frac{(-)^{n+i}}{i}\stirlingSo{n}{i}\,,
\label{Hcff}
\eeqn
and having denoted by \mbox{$[n\;\;k]$} the unsigned Stirling numbers
of the first kind, and by $B_{n,k}$ the incomplete exponential Bell 
polynomials. By employing the Fa\`a di Bruno formula and the one for
the derivatives of a product we arrive at:
\beqn
&&\!\!\!\!\!\!\!\!\frac{\big[\log(-\log z)\big]^{m}}{\log z}=
-\frac{\log^{m}(1-z)}{1-z}+\sum_{j=0}^m\log^j(1-z)\sum_{n=0}^\infty
\ell_{j,n}^{[m,-1]}(z-1)^n\,,\phantom{aa}
\label{lgzvslgom3}
\\
&&\!\!\!\!\!\!\!\!\big[\log(-\log z)\big]^{m}\log^q z=
\sum_{j=0}^m\log^j(1-z)\sum_{n=q}^\infty
\ell_{j,n}^{[m,q]}(z-1)^n\,,
\label{lgzvslgom4}
\eeqn
with coefficients:
\beqn
u_{j,n}^{[m]}&=&\frac{\Gamma(m+1)}{n!\Gamma(j+1)}
B_{n,m-j}\left(\big\{\Gamma(k+1)h_{k+1}\big\}_{k\ge 1}\right)\,,
\\
\ell_{j,n}^{[m,-1]}&=&u_{j,n+1}^{[m]}+\sum_{i=0}^n g_{n+1-i}u_{j,i}^{[m]}\,,
\\
\ell_{j,n}^{[m,q]}&=&\sum_{i=0}^n f_{n-i}^{(q)}u_{j,i}^{[m]}\,.
\eeqn
Since for any integrable function $f(z)$ we have
\beq
\big[f(z)\big]_+ = f(z)-\delta(1-z)\int_0^1 dxf(x)\,,
\eeq
eq.~(\ref{lgzvslgom3}) implies that
\beqn
\left[\frac{\big[\log(-\log z)\big]^{m}}{\log z}\right]_+&=&
-\left[\frac{\log^{m}(1-z)}{1-z}\right]_+ 
-\delta(1-z){\cal I}_m
\nonumber\\*&+&
\sum_{j=0}^m\log^j(1-z)\sum_{n=0}^\infty
\ell_{j,n}^{[m,-1]}(z-1)^n\,,
\label{lgzvslgom3dis}
\eeqn
where for any $m\ge 0$:
\beqn
{\cal I}_m&=&\int_0^1 dz\left(
\frac{\big[\log(-\log z)\big]^{m}}{\log z}+
\frac{\log^{m}(1-z)}{1-z}\right)
\label{calIm1}
\\*&=&
\frac{(-)^m}{m+1}B_{m+1}\left(\big\{(-)^{k+1}\psi_k(1)\big\}_{k\ge 0}\right)
\label{calIm2}
\\*&=&
\frac{(-)^m}{m+1}B_{m+1}\left(\big\{\gE,\vec{0}\big\}-Y\right),
\label{calIm3}
\eeqn
and $Y$ has been introduced in eq.~(\ref{Yvec}).
We note that for any regular function $g(z)$ we have
\beq
M\left[g(z)\right]\Melleq\frac{r_N}{N^p}\;\;\;\;\Longrightarrow\;\;\;\;
M\left[(1-z)^qg(z)\right]\Melleq\frac{r_N}{N^{p+q}}\,,
\eeq
for some $r_N$, which is useful to understand what follows. If we define
\beqn
\left[\frac{\big[\log(-\log z)\big]^{m}}{\log z}\right]_+^{(\rho)}&=&
-\left[\frac{\log^{m}(1-z)}{1-z}\right]_+ 
-\delta(1-z){\cal I}_m
\nonumber\\*&+&
\sum_{j=0}^m\log^j(1-z)\sum_{n=0}^\rho
\ell_{j,n}^{[m,-1]}(z-1)^n\,,
\label{lgzvslgom3distr}
\eeqn
i.e.~we neglect terms of $\ord((z-1)^{\rho+1})$ in the expansion on 
the r.h.s.~of eq.~(\ref{lgzvslgom3dis}), then
\beqn
&&M\left\{\left[\frac{\big[\log(-\log z)\big]^{m}}{\log z}\right]_+\right\}=
%\nonumber\\&&\phantom{aaaaaa}
M\left\{\left[
\frac{\big[\log(-\log z)\big]^{m}}{\log z}\right]_+^{(\rho)}\right\}+
\ord\left(\frac{1}{N^{\rho+2}}\right).
\eeqn
With eqs.~(\ref{Mlgomzplus}) and~(\ref{Mlgomz}) this implies:
\beqn
&&M\left\{\left[\frac{\big[\log(-\log z)\big]^{m}}{\log z}\right]_+\right\}=
-{\cal I}_m-\frac{1}{m+1}\,B_{m+1}(Z)
\nonumber\\&&\phantom{aaaa}+
\sum_{j=0}^m\sum_{n=0}^\rho
\ell_{j,n}^{[m,-1]}
\frac{(-)^n n!\Gamma(N)}{\Gamma(N+n+1)}\,B_{j}(Z_n)+
\ord\left(\frac{1}{N^{\rho+2}}\right)\,.
\label{llmolexvsapp}
\eeqn
Note that if both sides of eq.~(\ref{llmolexvsapp}) are expanded
up to $\ord(1/N^{p+1})$, with $p<\rho$, that equation is still valid;
conversely, if $p>\rho$, that equality is broken by terms of
$\ord(1/N^{\rho+2})$. In order to amend this situation, more terms
in the expansion of eq.~(\ref{lgzvslgom3distr}) are necessary, which
is equivalent to saying that one must change $\rho\to p$ there.
The bottom line is that the $\log z\to 1-z$ prescription can be 
realised at order $\rho$ by means of the replacements:
\beqn
\left[\frac{\big[\log(-\log z)\big]^{m}}{\log z}\right]_+
&\longrightarrow&
\left[\frac{\big[\log(-\log z)\big]^{m}}{\log z}\right]_+^{(\rho)}\,,
\label{lztoomz}
\\
M\left\{\left[\frac{\big[\log(-\log z)\big]^{m}}{\log z}\right]_+\right\}
&\longrightarrow&
-{\cal I}_m-\frac{1}{m+1}\,B_{m+1}(Z)
\label{lztoomzN}
\\&&+
\sum_{j=0}^m\sum_{n=0}^\rho
\ell_{j,n}^{[m,-1]}
\frac{(-)^n n!\Gamma(N)}{\Gamma(N+n+1)}\,B_{j}(Z_n)\,,
\nonumber
\eeqn
in the configuration and Mellin spaces, respectively, with the r.h.s.~of
eq.~(\ref{lztoomz}) taken from eq.~(\ref{lgzvslgom3distr}).
While these forms are closed, what is done above shows that
the Mellin-space replacement of eq.~(\ref{lztoomzN}) can also
obtained by direct {\em exact} Mellin transform of the r.h.s.~of
eq.~(\ref{lztoomz}). By Mellin-transform both sides of 
eq.~(\ref{invMlgzplus}) and by exploiting eq.~(\ref{llmolexvsapp}), 
we can also arrive at the following result:
\beqn
\log^m \bN&\!\!=\!\!&\gE^m
-m\sum_{i=0}^{m-1}(-)^{i+1}\binomial{m-1}{i}B_{m-i-1}(Y)
\label{MlgNappr}
\\&&\times\Bigg\{
-{\cal I}_i-\frac{1}{i+1}\,B_{i+1}(Z)
\nonumber\\&&\phantom{aaaa}+
\sum_{j=0}^i\sum_{n=0}^\rho
\frac{(-)^n n!\Gamma(N)}{\Gamma(N+n+1)}\,
\ell_{j,n}^{[i,-1]}B_{j}(Z_n)\Bigg\}+
\ord\left(\frac{1}{N^{\rho+2}}\right)\,,
\nonumber
\eeqn
which can be employed e.g.~in the case where $\log\bN$ terms emerge from
asymptotic expressions which need to be multiplied, and inverted back
in the configuration space, while keeping there only contributions up to and
including $\ord((1-z)^\rho)$. Explicitly:
\beqn
&&\invM\left[\log^m \bN\right]=\gE^m\delta(1-z)
-m\sum_{i=0}^{m-1}(-)^{i+1}\binomial{m-1}{i}B_{m-i-1}(Y)\phantom{aaa}
\\&&\phantom{aaa}\times\Bigg\{
-{\cal I}_i\delta(1-z)-
\left[\frac{\log^i(1-z)}{1-z}\right]_+
\nonumber\\&&\phantom{aaaaaa}+
\sum_{j=0}^i\sum_{n=0}^\rho
(-)^n\ell_{j,n}^{[i,-1]}(1-z)^n\log^j(1-z)\Bigg\}+
\ord\left((1-z)^{\rho+1}\right)\,.
\nonumber
\eeqn
Likewise, upon defining ($q\ge 0$)
\beq
\Big[\big[\log(-\log z)\big]^{m}\log^q z\Big]^{(\rho)}=
\sum_{j=0}^m\log^j(1-z)\sum_{n=q}^\rho
\ell_{j,n}^{[m,q]}(z-1)^n\,,
\label{lgzvslgom4rho}
\eeq
i.e.~we neglect terms of $\ord((z-1)^{\rho+1})$ in the expansion on 
the r.h.s.~of eq.~(\ref{lgzvslgom4}), then
\beqn
&&M\Big\{\big[\log(-\log z)\big]^{m}\log^q\Big\}=
\nonumber\\*&&\phantom{aaaaaa}
M\left\{
\Big[\big[\log(-\log z)\big]^{m}\log^q z\Big]^{(\rho)}\right\}+
\ord\left(\frac{1}{N^{\rho+2}}\right).
\eeqn
Therefore, eq.~(\ref{Mlgomz}) then implies:
\beqn
&&M\Big\{\big[\log(-\log z)\big]^{m}\log^q\Big\}=
\nonumber\\&&\phantom{aaaaaa}
\sum_{j=0}^m\sum_{n=q}^\rho
\ell_{j,n}^{[m,q]}
\frac{(-)^nn!\Gamma(N)}{\Gamma(N+n+1)}\,B_{j}(Z_n)
+\ord\left(\frac{1}{N^{\rho+2}}\right).
\eeqn
In other words, for these terms the $\ord(\rho)$ $\log z\to 1-z$ 
prescription amounts to the replacements ($q\ge 0$):
\beqn
\big[\log(-\log z)\big]^{m}\log^q
&\longrightarrow&
\Big[\big[\log(-\log z)\big]^{m}\log^q z\Big]^{(\rho)}\,,
\label{lzqtoomz}
\\
M\Big\{\big[\log(-\log z)\big]^{m}\log^q\Big\}
&\longrightarrow&
\sum_{j=0}^m\sum_{n=q}^\rho
\ell_{j,n}^{[m,q]}
\frac{(-)^nn!\Gamma(N)}{\Gamma(N+n+1)}\,B_{j}(Z_n)\,.
\label{lzqtoomzN}
\eeqn
In keeping with eq.~(\ref{MlgNappr}), we thus obtain from eq.~(\ref{invMlgz})
($q\ge 1$):
\beqn
\frac{\log^m \bN}{N^q}&\!\!=\!\!&
\frac{(-)^{q+m+1}}{(m+1)(q-1)!}
\sum_{i=0}^{m+1}i\binomial{m+1}{i}B_{i-1}(Y_q)
\nonumber\\&&\phantom{aaa}\times
\sum_{j=0}^{m+1-i}\sum_{n=q-1}^\rho
\ell_{j,n}^{[m+1-i,q-1]}
\frac{(-)^nn!\Gamma(N)}{\Gamma(N+n+1)}\,B_{j}(Z_n)+
\ord\left(\frac{1}{N^{\rho+2}}\right)\,,\phantom{aa}
\label{Mlgzqappr}
\eeqn
and:
\beqn
&&\!\!\invM\left[\frac{\log^m \bN}{N^q}\right]=
\frac{(-)^{q+m+1}}{(m+1)(q-1)!}
\sum_{i=0}^{m+1}i\binomial{m+1}{i}B_{i-1}(Y_q)
\\&&\phantom{aaaaa}\times
\sum_{j=0}^{m+1-i}\sum_{n=q-1}^\rho(-)^n\ell_{j,n}^{[m+1-i,q-1]}
\log^j(1-z)(1-z)^n+
\ord\left((1-z)^{\rho+1}\right)\,.
\nonumber
\eeqn
In summary, given a function $\widetilde{G}_N$ in Mellin space
\beq
\widetilde{G}_N=
\sum_{m=0}^{m_{\max}}\sum_{q=0}^{q_{\max}}G_{q,m}\frac{\log^m\bN}{N^q}
\label{GNfun}
\eeq
the $\log z\to 1-z$ prescription amounts to the replacement
\beqn
\widetilde{G}_N&\longrightarrow&
\overline{G}_N=\sum_{m=0}^{m_{\max}}G_{0,m}\,
\Bigg\{\gE^m
-m\sum_{i=0}^{m-1}(-)^{i+1}\binomial{m-1}{i}B_{m-i-1}(Y)
\label{cNMlgNappr}
\\&&\phantom{aaaa}\times\Bigg[
-{\cal I}_i-\frac{1}{i+1}\,B_{i+1}(Z)
\nonumber\\&&\phantom{aaaaaaa}+
\sum_{j=0}^i\sum_{n=0}^{q_{\max}-1}
\frac{(-)^n n!\Gamma(N)}{\Gamma(N+n+1)}\,
\ell_{j,n}^{[i,-1]}B_{j}(Z_n)\Bigg]\Bigg\}
\nonumber
\\&&
+\sum_{m=0}^{m_{\max}}\sum_{q=1}^{q_{\max}}G_{q,m}\,
\frac{(-)^{q+m+1}}{(m+1)(q-1)!}
\sum_{i=0}^{m+1}i\binomial{m+1}{i}B_{i-1}(Y_q)
\nonumber\\&&\phantom{aaa}\times
\sum_{j=0}^{m+1-i}\sum_{n=q-1}^{q_{\max}-1}
\ell_{j,n}^{[m+1-i,q-1]}
\frac{(-)^nn!\Gamma(N)}{\Gamma(N+n+1)}\,B_{j}(Z_n)\,,\phantom{aa}
\nonumber
\eeqn
and its related exact inverse Mellin transform
\beqn
\invM\left[\overline{G}_N\right]&=&
\sum_{m=0}^{m_{\max}}G_{0,m}\,
\Bigg\{
\gE^m\delta(1-z)
-m\sum_{i=0}^{m-1}(-)^{i+1}\binomial{m-1}{i}B_{m-i-1}(Y)\phantom{aaa}
\label{czMlgNappr}
\\&&\phantom{aaa}\times\Bigg[
-{\cal I}_i\delta(1-z)-
\left[\frac{\log^i(1-z)}{1-z}\right]_+
\nonumber\\&&\phantom{aaaaaa}+
\sum_{j=0}^i\sum_{n=0}^{q_{\max}-1}
(-)^n\ell_{j,n}^{[i,-1]}(1-z)^n\log^j(1-z)\Bigg]\Bigg\}
\nonumber
\\&&
+\sum_{m=0}^{m_{\max}}\sum_{q=1}^{q_{\max}}G_{q,m}\,
\frac{(-)^{q+m+1}}{(m+1)(q-1)!}
\sum_{i=0}^{m+1}i\binomial{m+1}{i}B_{i-1}(Y_q)
\nonumber
\\&&\phantom{aaaaa}\times
\sum_{j=0}^{m+1-i}\sum_{n=q-1}^{q_{\max}-1}(-)^n\ell_{j,n}^{[m+1-i,q-1]}
\log^j(1-z)(1-z)^n\,,\phantom{aa}
\nonumber
\eeqn
in the Mellin and $z$ spaces, respectively, having set $\rho=q_{\max}-1$
in order to have the same subleading powers of $1/N$ as the original
function $\widetilde{G}_N$ of eq.~(\ref{GNfun}).

These results are useful in the context of fixed-order expansions.
However, when considering the large-$z$ behaviour of the
evolved PDFs, one needs results such as eq.~(A.18) of
ref.~\cite{Frixione:2021wzh}, namely:
\beq
M^{-1}\!\left[N^{-\xi}\log^q\bN\right]\,\stackrel{z\to 1}{\longrightarrow}\,
\frac{\xi(1-z)^{-1+\xi}}{\Gamma(1+\xi)}
\sum_{i=0}^q (-)^{q-i}\binomial{q}{i}d_i(\xi)\,\log^{q-i}(1-z)\,,
\label{invMlogqN}
\eeq
with $\xi$ a (small and positive) real number, and where
\beq
d_k(\xi)=B_k\big(\Psi_k(\xi)\big)\,,
\label{dkBell}
\eeq
with
\beq
\Psi_k(\xi)=\left(\gE+\psi_0(\xi),-\psi_1(\xi),\ldots,
(-)^{k-1}\psi_{k-1}(\xi)\right)\,,
\label{Psivec}
\eeq
or, equivalenty
\beq
d_k(\xi)=\frac{1}{{\cal G}_d(\xi,0)}
\left.\frac{\partial^k{\cal G}_d(\xi,\delta)}{\partial\delta^k}
\right|_{\delta=0}\,,
\label{dffbygen}
\eeq
with
\beq
{\cal G}_d(\xi,\delta)=
\frac{e^{-\gE(\xi-\delta)}}{\Gamma(\xi-\delta)}\,.
\label{dffgenfun}
\eeq
We point out that, as the notation indicates,
eq.~(\ref{invMlogqN}) is valid in the $z\to 1$ limit. By exploiting
again eq.~(\ref{mastersa2}), one can obtain an improved version of
eq.~(\ref{invMlogqN}), where power-suppressed terms are also retained.
Specifically, after including in eq.~(\ref{mastersa2}) a factor
\mbox{$\exp[(-a-1+\xi)\gE]$} to account for the difference between
$N$ and $\bN$, and deriving $q$ times w.r.t.~$a$, one expands the result 
around $a=-1+\xi$, thereby arriving at:
\beq
M^{-1}\!\left[N^{-\xi}\log^q\bN\right]=
\frac{\xi}{\Gamma(1+\xi)}
\sum_{i=0}^q (-)^{q-i}\binomial{q}{i}d_i(\xi)\,
\frac{[\log(-\log z)]^{q-i}}{(-\log z)^{1-\xi}}\,.
\label{invMlogqNex}
\eeq
At variance with eq.~(\ref{invMlogqN}), eq.~(\ref{invMlogqNex}) is
exact; we point out that the former equation can be derived from
the latter by exploiting eqs.~(\ref{lgzvslgom1}) and~(\ref{lgzvslgom3}),
and by dropping all of the non-leading terms there.

Equation~(\ref{invMlogqN}) constitutes the basic building block for
a more general formula which we need to use in the context of the
$\Delta$ scheme. Let $F(x)$ be a regular function which can be
expanded around $x=0$:
\beq
F(x)=\sum_{k=0}^\infty\frac{F^{(k)}(0)}{k!}x^k\,.
\label{Fser}
\eeq
Our goal is the computation of
\beq
f(z)=\invM\left[N^{-\xi}\,F\left(\log\bN\right)\right]
\eeq
and of its large-$z$ asymptotic form
\beq
f_\infty(z)\;\stackrel{z\to 1}{\longleftarrow}\;f(z)\,.
\eeq
The idea is that of using eq.~(\ref{invMlogqN}) in the case of $f_\infty(z)$, 
or eq.~(\ref{invMlogqNex}) in the case of $f(z)$, after expanding in
series $F(\log\bN)$ and (assuming strong convergence) commuting the
summation with the inverse Mellin transform. Since eqs.~(\ref{invMlogqN}) 
and~(\ref{invMlogqNex}) have identical structures, we shall work with
the former in order to be definite, and consider the latter only at the
very end of the procedure. We have
\beqn
f_\infty(z)&=&\invM\left[N^{-\xi}
\sum_{k=0}^\infty\frac{F^{(k)}(0)}{k!}\log^k\bN\right]
=\sum_{q=0}^\infty\frac{F^{(q)}(0)}{q!}
\invM\left[N^{-\xi}\log^q\bN\right]
\nonumber \\*&=&
\frac{\xi(1-z)^{-1+\xi}}{\Gamma(1+\xi)}
\sum_{q=0}^\infty\frac{F^{(q)}(0)}{q!}
\sum_{i=0}^q (-)^{q-i}\binomial{q}{i}d_i(\xi)\,\log^{q-i}(1-z)\,.
\label{fzinf}
\eeqn
We now observe that
\beq
(-)^{q-i}\binomial{q}{i}\log^{q-i}(1-z)=
\frac{1}{i!}\left.\frac{d^i}{dx^i}\,x^q\right|_{x=-\log(1-z)}\,,
\label{iqiden}
\eeq
and that this identity holds true also for $i>q$, when both sides
of the equation are identically equal to zero. This implies that,
when eq.~(\ref{iqiden}) is used in eq.~(\ref{fzinf}), the upper
limit of the summation over $i$ can be set equal to infinity.
By doing so, one is also able to exchange the order of the summations,
making that over $i$ the outer one. Thus:
\beqn
f_\infty(z)&=&
\frac{\xi(1-z)^{-1+\xi}}{\Gamma(1+\xi)}
\sum_{i=0}^\infty\frac{d_i(\xi)}{i!}\left.\frac{d^i}{dx^i}
\sum_{q=0}^\infty\frac{F^{(q)}(0)}{q!}x^q\right|_{x=-\log(1-z)}
\nonumber \\*&=&
\frac{\xi(1-z)^{-1+\xi}}{\Gamma(1+\xi)}
\sum_{i=0}^\infty\frac{d_i(\xi)}{i!}\left.\frac{d^i}{dx^i}
F(x)\right|_{x=-\log(1-z)}\,,
\label{invMFFlogqN}
\eeqn
having used eq.~(\ref{Fser}). By treating $d^i/dx^i$ as an
expansion parameter, and by employing eq.~(\ref{dffbygen}), the
summation in eq.~(\ref{invMFFlogqN}) has a structure of a formal
series expansion, and therefore can be re-written more compactly
as follows:
\beqn
f_\infty(z)&=&
\frac{\xi(1-z)^{-1+\xi}}{\Gamma(1+\xi)}
\left.\frac{1}{{\cal G}_d(\xi,0)}\,{\cal G}_d\left(\xi,\frac{d}{dx}\right)
F(x)\right|_{x=-\log(1-z)}\,.
\label{invMFFlogqNsum}
\eeqn
It is now easy to see that the whole procedure could be repeated 
identically by starting from eq.~(\ref{invMlogqNex}) rather than from
eq.~(\ref{invMlogqNex}). By doing that, the analogues of 
eqs.~(\ref{invMFFlogqN}) and~(\ref{invMFFlogqNsum}) read as follows:
\beqn
f(z)&=&
\frac{\xi(-\log z)^{-1+\xi}}{\Gamma(1+\xi)}
\sum_{i=0}^\infty\frac{d_i(\xi)}{i!}\left.\frac{d^i}{dx^i}
F(x)\right|_{x=-\log(-\log z)}
\label{invMFFlogqNex}
\\*&=&
\frac{\xi(-\log z)^{-1+\xi}}{\Gamma(1+\xi)}
\left.\frac{1}{{\cal G}_d(\xi,0)}\,{\cal G}_d\left(\xi,\frac{d}{dx}\right)
F(x)\right|_{x=-\log(-\log z)}\,.
\label{invMFFlogqNsumex}
\eeqn
When applying these results to the $\Delta$ scheme, luckily the
terms of the series in eq.~(\ref{invMFFlogqN}) are increasingly
suppressed at large $z$ with increasing index $i$, and the series
can thus be safely truncated. When this is not the case, one must
use the resummed expressions of eqs.~(\ref{invMFFlogqNsum}) 
or~(\ref{invMFFlogqNsumex}); these, however, are only formal
quantities, owing to the presence of the derivative operator.
While we are unable to find closed expressions which do not
feature the derivative operator, we do find an integral representation
with excellent numerical convergence properties, which we now proceed
to construct.

The starting point is the following integral representation of the
reciprocal of the $\Gamma$ function:
\beq
\frac{1}{\Gamma(\xi-\delta)}=
\frac{i}{2\pi}\int_{H^{-}} d\omega\,e^{-\omega}(-\omega)^{-\xi+\delta}\,,
\label{ooGamhank2}
\eeq
where the path $H^-$ is counterclockwise, starts and ends at 
\mbox{$(+\infty\pm i\ep)$}, and crosses the real axis at some negative
value (Hankel path). The representation is valid for any integer, real, 
and complex argument, and the path can be freely deformed, provided the 
branch cut of the integrand is not crossed. Then, from eq.~(\ref{dffgenfun}),
\beqn
{\cal G}_d(\xi,\delta)&=&
e^{-\gE(\xi-\delta)}
\frac{i}{2\pi}\int_{H^{-}} d\omega\,e^{-\omega}(-\omega)^{-\xi+\delta}
\nonumber
\\*&=&
\frac{ie^{-\gE\xi}}{2\pi}
\int_{H^{-}} d\omega\,
\,e^{-\omega}(-\omega)^{-\xi}
\exp\Big[\delta\log(-\omega)+\gE\delta\Big].
%\label{dffgenfun}
\eeqn
Therefore
\beqn
&&\left.{\cal G}_d\left(\xi,\frac{d}{dx}\right)F(x)\right|_{x=-\log(1-z)}=
\\&&\phantom{aaa}
=\frac{ie^{-\gE\xi}}{2\pi}
\int_{H^{-}} d\omega\,
\,e^{-\omega}(-\omega)^{-\xi}
\left.\exp\left[\Big(\log(-\omega)+\gE\Big)\frac{d}{dx}\right]
F(x)\right|_{x=-\log(1-z)}\,.
\nonumber
\\&&\phantom{aaa}
=\frac{ie^{-\gE\xi}}{2\pi}
\int_{H^{-}} d\omega\,
\,e^{-\omega}(-\omega)^{-\xi}
F\Big(\log(-\omega)+\gE-\log(1-z)\Big)\,,
\nonumber
\eeqn
where in the last step we have exploited the fact that the
exponential of the derivative operator is the generator of the argument 
shifts. By replacing this result in eq.~(\ref{invMFFlogqNsum})
we obtain, after some trivial simplifications which involve $\Gamma$
functions:
\beq
f_\infty(z)=
\frac{i}{2\pi}\,(1-z)^{-1+\xi}\,
\int_{H^{-}} d\omega\,
\,e^{-\omega}(-\omega)^{-\xi}
F\Big(\log(-\omega)+\gE-\log(1-z)\Big)\,.
\label{invMFFlogqNsum2}
\eeq
Analogously, strting from eq.~(\ref{invMFFlogqNsumex}):
\beq
f(z)=
\frac{i}{2\pi}\,(-\log z)^{-1+\xi}\,
\int_{H^{-}} d\omega\,
\,e^{-\omega}(-\omega)^{-\xi}
F\Big(\log(-\omega)+\gE-\log(-\log z)\Big)\,.
\label{invMFFlogqNsum2ex}
\eeq

\section{Proof of an identity\label{sec:proof}}
Assuming strong convergence, we expect
\beq
M^{-1}\!\left[N^{-\xi}\log^q\bN\right]=\invM\left[\log^q \bN\right]+
\ord(\xi)\,,
\eeq
that is, from eqs.~(\ref{invMlgzplus}), (\ref{invMlogqN}),
and~(\ref{invMlogqNex}):
\beqn
&&\!\!\!\gE^q\delta(1-z)
-q\sum_{i=0}^{q-1}\binomial{q-1}{i}B_{q-i-1}(Y)
\left[\frac{\Lz^i}{-\log z}\right]_+ 
\label{prstart}
\\&&\phantom{aaaa}
=\frac{\xi}{\Gamma(1+\xi)}
\sum_{i=0}^q (-)^{q-i}\binomial{q}{i}d_i(\xi)\,
\frac{[\log(-\log z)]^{q-i}}{(-\log z)^{1-\xi}}
+\ord(\xi)
\nonumber
\\&&\phantom{aaaa}
=\frac{\xi(1-z)^{-1+\xi}}{\Gamma(1+\xi)}
\sum_{i=0}^q (-)^{q-i}\binomial{q}{i}d_i(\xi)\,\log^{q-i}(1-z)
\nonumber
\\&&\phantom{aaaaaaaa}
+\ord(\xi,(1-z)^0)\equiv X\,.
\nonumber
\eeqn
Let us start from the rightmost side of eq.~(\ref{prstart}), where
we use eq.~(C.2) of ref.~\cite{Bertone:2019hks}, after correcting
a typo on the r.h.s.~there \mbox{$\Gamma(1+\kappa)\to\Gamma(1+i)$}:
\beq
\frac{\log^p(1-z)}{(1-z)^{1-\kappa}}=
\frac{(-1)^p\,\Gamma(1+p)}{\kappa^{1+p}}\,\delta(1-z)+
\sum_{i=0}^\infty \frac{\kappa^i}{\Gamma(1+i)}\,
\left[\frac{\log^{i+p}(1-z)}{1-z}\right]_+\,,
\label{lpomzexp1}
\eeq
which is valid for any integer $p\ge 0$ and real $\kappa>0$.
With the identity
\beq
\binomial{q}{i}\Gamma(1+q-i)=\frac{\Gamma(q+1)}{\Gamma(i+1)}
=\frac{q!}{i!}\,,
\eeq
we obtain
\beqn
X&=&\frac{\xi}{\Gamma(1+\xi)}\sum_{i=0}^q\Bigg\{
\frac{q!}{i!}\,d_i(\xi)\frac{\xi^i}{\xi^{q+1}}\,\delta(1-z)
\\&&\phantom{aaaa}
+(-)^{q-i}\binomial{q}{i}d_i(\xi)
\sum_{k=0}^\infty \frac{\xi^k}{\Gamma(1+k)}\,
\left[\frac{\log^{k+q-i}(1-z)}{1-z}\right]_+\Bigg\}.
\nonumber
\eeqn
We now manipulate the first term on the r.h.s.~of this equation,
by employing eq.~(\ref{dffbygen}), so that:
\beqn
&&\sum_{i=0}^q\frac{\xi^i}{i!}\,d_i(\xi)=
\sum_{i=0}^q\frac{\xi^i}{i!}\,
\frac{1}{{\cal G}_d(\xi,0)}
\left.\frac{\partial^i{\cal G}_d(\xi,\delta)}{\partial\delta^i}
\right|_{\delta=0}
=\sum_{i=0}^\infty-\sum_{i=q+1}^\infty\Big(\ldots\Big)
\nonumber
\\&&\phantom{aaaa}
=\frac{{\cal G}_d(\xi,\xi)}{{\cal G}_d(\xi,0)}-
\frac{\xi^{q+1}}{(q+1)!}\,d_{q+1}(\xi)-
\sum_{i=q+2}^\infty\frac{\xi^i}{i!}\,d_i(\xi)\,.
\eeqn
From eq.~(\ref{dffgenfun})
\beq
{\cal G}_d(\xi,\xi)=0\,,
\eeq
while eq.~(A.37) of ref.~\cite{Frixione:2021wzh} is ($k\ge 1$):
\beq
d_k(\xi)=-\frac{k}{\xi}\,
B_{k-1}\Big(\big\{0,\,-1!\,\zeta_2,\,-2!\,\zeta_3,
\ldots,\, -k!\,\zeta_{k+1}\big\}\Big)+\ord(\xi^0)\,.
\label{dkasy2}
\eeq
Since
\beq
\big\{0,\,-1!\,\zeta_2,\,-2!\,\zeta_3,\ldots,\, -k!\,\zeta_{k+1}\big\}=Y
\eeq
we obtain
\beqn
&&\sum_{i=0}^q\frac{\xi^i}{i!}\,d_i(\xi)=
\frac{1}{(q+1)!}\frac{q+1}{\xi}B_q(Y)\xi^{q+1}+\ord(\xi^{q+2}),
\eeqn
and therefore:
\beqn
X&=&\frac{\xi}{\Gamma(1+\xi)}\Bigg\{
\left[\frac{q!}{\xi^{q+1}}
\frac{1}{(q+1)!}\frac{q+1}{\xi}B_q(Y)\xi^{q+1}+\ord(\xi^{q+2})\right]
\delta(1-z)\phantom{aa}
\label{X1}
\\&&\phantom{aaa}
+\sum_{i=0}^q(-)^{q-i}\binomial{q}{i}d_i(\xi)
\sum_{k=0}^\infty \frac{\xi^k}{\Gamma(1+k)}\,
\left[\frac{\log^{k+q-i}(1-z)}{1-z}\right]_+\Bigg\}.
\nonumber
\eeqn
Now we use
\beq
\frac{\xi}{\Gamma(1+\xi)}=\xi+\ord(\xi^2)\,,
\eeq
and again eq.~(\ref{dkasy2}) to replace $d_i(\xi)$ in eq.~(\ref{X1}),
so that (the $i=0$ term in the sum gives a null contribution at this
order in $\xi$, since $d_0(\xi)=1$):
\beqn
&&\!\!\!\!\!X=B_q(Y)\delta(1-z)
-\sum_{i=1}^q(-)^{q-i}\binomial{q}{i}iB_{i-1}(Y)
\left[\frac{\log^{q-i}(1-z)}{1-z}\right]_+
\\&&\phantom{;}=
B_q(Y)\delta(1-z)
-\sum_{j=0}^{q-1}(-)^{j}\binomial{q}{q-j}(q-j)B_{q-j-1}(Y)
\left[\frac{\log^{j}(1-z)}{1-z}\right]_+\phantom{a}
\nonumber
\\&&\phantom{;}=
B_q(Y)\delta(1-z)
-q\sum_{j=0}^{q-1}(-)^{j}\binomial{q-1}{j}B_{q-j-1}(Y)
\left[\frac{\log^{j}(1-z)}{1-z}\right]_+
\nonumber
\\&&\phantom{aaa}
+\ord(\xi)\,,
\eeqn
having exploited
\beq
\binomial{q}{q-j}(q-j)=q\binomial{q-1}{j}\,.
\eeq
With eq.~(\ref{lgzvslgom3dis}), by ignoring all regular-function
contributions, we then obtain:
\beqn
&&\!\!\!\!\!X=B_q(Y)\delta(1-z)
-q\sum_{j=0}^{q-1}\binomial{q-1}{j}B_{q-j-1}(Y)
\label{X2}
\\&&\phantom{aaaa}\times
\Bigg\{
-\left[\frac{\big[-\log(-\log z)\big]^{j}}{\log z}\right]_+
-(-)^j{\cal I}_j\delta(1-z)\Bigg\}
\nonumber\\&&\phantom{\Big(aaa}
+\ord(\xi,(1-z)^0)\,.
\nonumber
\eeqn
Equation~(\ref{calIm3}) allows us to write
\beq
{\cal I}_j=
\frac{(-)^j}{j+1}B_{j+1}\left(\big\{\gE,\vec{0}\big\}-Y\right),
\eeq
so that the coefficient of the $\delta(1-z)$ term in eq.~(\ref{X2})
becomes
\beqn
&&B_q(Y)-q\sum_{j=0}^{q-1}\binomial{q-1}{j}B_{q-j-1}(Y)
(-)(-)^j{\cal I}_j
\\&&\phantom{aa}=
B_q(Y)+q\sum_{j=0}^{q-1}\binomial{q-1}{j}B_{q-j-1}(Y)
(-)^j\frac{(-)^j}{j+1}B_{j+1}\left(\big\{\gE,\vec{0}\big\}-Y\right).
\nonumber
\eeqn
With
\beq
\frac{q}{j+1}\binomial{q-1}{j}=\binomial{q}{j+1}
\eeq
we can exploit the binomial-like sum property of the Bell polynomials,
compensating for a single missing summand:
\beqn
&&\sum_{j=0}^{q-1}\binomial{q}{j+1}B_{q-j-1}(Y)
B_{j+1}\left(\big\{\gE,\vec{0}\big\}-Y\right)
\\&&\phantom{aa}
=\sum_{k=0}^{q}\binomial{q}{k}B_{q-k}(Y)
B_{k}\left(\big\{\gE,\vec{0}\big\}-Y\right)
-B_q(Y)B_0\left(\big\{\gE,\vec{0}\big\}-Y\right)
\nonumber
\\&&\phantom{aa\Big(}
=B_{q}\left(Y+\big\{\gE,\vec{0}\big\}-Y\right)
-B_q(Y)B_0\left(\big\{\gE,\vec{0}\big\}-Y\right)
\nonumber
\\*&&\phantom{aa\Big(}
=\gE^q-B_q(Y)\,.
\nonumber
\eeqn
With this result and by recalling eq.~(\ref{LEAdef}), eq.~(\ref{X2})
becomes
\beq
X=\gE^q\delta(1-z)
-q\sum_{j=0}^{q-1}\binomial{q-1}{j}B_{q-j-1}(Y)
\left[\frac{\Lz^{j}}{-\log z}\right]_+
+\ord(\xi,(1-z)^0)\,.
\label{X3}
\eeq
This result coincides with the leftmost side of eq.~(\ref{prstart}), 
which is what we have set out to prove. From the physical viewpoint,
this proof shows the technical side of the direct connection between
the large-$z$ behaviour of the PDFs and their underlying initial conditions.

\section{Analytical results\label{sec:res}}
In this appendix we shall report results both in the physical and in
the evolution basis, by typically choosing the shortest of the two
expressions. 
We point out that the photon is invariant under this basis transformation,
and that eq.~(\ref{Tmatdef}) implies:
\beqn
&&\ePDFs=\ePDF{\lm}+\ePDF{\lp}\,,\;\;\;\;\;\;\;\;\;\;\;\;\;\;\,
\ePDFns=\ePDF{\lm}-\ePDF{\lp}\,,
\label{snsdef}
\\*
&&\ePDF{\lm}=\half\left(\ePDFs+\ePDFns\right)\,,\;\;\;\;\;\;\;\;
\ePDF{\lp}=\half\left(\ePDFs-\ePDFns\right)\,.
\label{snsinv}
\eeqn
In order to work with initial conditions at the NNLO, and especially 
to perform (inverse) Mellin transforms and asymptotic expansions in both
Mellin and configuration space, we employ the packages 
\texttt{Sigma}~\cite{sigmaI,sigmaII} and 
\texttt{HarmonicSums}~\cite{Ablinger:2009ovq,Ablinger:2012ufz,Ablinger:2013cf,Ablinger:2014rba}.  

%outside of the scope:Ablinger:2011te,Ablinger:2014bra,Ablinger:2015gdg,Ablinger:2018cja

In configuration space we can express all of the results in terms of 
widely known polylogarithms. In Mellin space, we use 
generalized harmonic sums~\cite{Vermaseren:1998uu,Ablinger:2013eba}, 
defined recursively by:
\beqn
  S_\emptyset&=&1\,,
\\
   S_{a,\vec{b}} &\equiv& S_{a,\vec{b}}(N) 
   = \sum\limits_{i=1}^{N} \frac{(\text{sign}(a))^i}{i^{|a|}} S_{\vec{b}}(i)
   \,, 
  \\ 
S_{a,\vec{b}}\big(\{w_a,\vec{w}_{\vec{b}}\}\big) 
&\equiv& 
  S_{a,\vec{b}}\big(\{w_a,\vec{w}_{\vec{b}}\}\big)(N) 
  = \sum\limits_{i=1}^{N} \frac{(\text{sign}(a)w_a)^i}{i^{|a|}} 
S_{\vec{b}}(\{\vec{w}_{\vec{b}}\})(i)
  \,,
\eeqn
where the $w_a$, $\vec{w}_{\vec{b}}$ are sets of positive rational numbers.
Simple harmonic sums are related to polygamma and polylogarithm functions, 
as follows (see e.g.~refs.~\cite{Blumlein:1998if,Blumlein:2009ta}):
\beqn   
  \psi_{0}(N+1) &=& S_{1} - \gE \,, \\
  \psi_{j}(N+1) &=& (-)^j \big( j!S_{j+1} + \psi_j(1) \big)
  \label{eq:psi}
  \\
  &=&
  (-)^j j!\big( S_{j+1} - \text{Li}_{j+1}(1) \big) 
  \,, 
\eeqn
for any integer $j>0$ (eq.~\eqref{eq:psi} also for $j=0$).

\subsection{$\MSb$ initial conditions\label{sec:resMSbb}}

\subsubsection{Mellin space\label{sec:iniN}}
The initial conditions in the $\MSb$ scheme read as follows in Mellin space:
\begin{small}
\beqn 
  \Gamma_{e^{-},N} &=& 1 
  + \frac{\aem(0)}{2\pi} 
  \Biggl[
    \biggl(
        \frac{2+3 N+3 N^2}{2 N (1+N)}
        -2 S_1
    \biggr) L_0 
    -\frac{1+N-4 N^2-2 N^3}{N (1+N)^2}
    +\frac{2 \big(1+N+N^2\big) S_1}{N (1+N)}
    \nonumber \\ &&
    -2 S_1^2
    -2 S_2
  \Biggr]
  + \left( \frac{\aem(0)}{2\pi} \right)^2
  \Biggl[
    L_0^2 
    \biggl(
        \frac{\big(2+3 N+3 N^2\big)^2}{8 N^2 (1+N)^2}
        -\frac{\big(2+3 N+3 N^2\big) S_1}{N (1+N)}
        +2 S_1^2
    \biggr)
    \nonumber \\ &&
    +L_0 
    \biggl(
        \frac{-8-8 N-12 N^2+35 N^3+61 N^4+81 N^5+27 N^6}{8 N^3 (1+N)^3}
        +\pi ^2 \biggl[
                \frac{1}{3 N (1+N)}
                -\frac{2 S_1}{3}
        \biggr]
        \nonumber \\ &&
        +\biggl[
                 \frac{3+10 N-2 N^2-N^3}{N (1+N)^2}
                +8 S_2
        \biggr] S_1
        -\frac{\big(6+7 N+7 N^2\big) S_1^2}{N (1+N)}
        -\frac{2 \big(2+3 N+3 N^2\big) S_2}{N (1+N)}
        \nonumber \\ &&
        +4 S_1^3
        +4 S_3
        +2 \zeta_{3}
    \biggr)
    -\frac{7 \pi ^4}{180}
    +\zeta_{3} 
    \biggl(
        \frac{9+9 N-11 N^2-24 N^3-4 N^4}{3 N (1+N)}
        -6 S_1
    \biggr)
    \nonumber \\ &&
    +\frac{48+48 N+48 N^2+1008 N^3+4051 N^4+2476 N^5+834 N^6+76 N^7+19 N^8}{96 N^4 (1+N)^4}
    \nonumber \\ &&
    +\pi ^2 
    \biggl(
        \frac{-42+17 N+242 N^2+195 N^3+48 N^4}{36 N (1+N)^2}
        +\frac{5 \big(1+N+N^2\big) S_1}{3 N (1+N)}
        -\frac{5}{3} S_1^2
        -S_2
        \nonumber \\ &&
        -\frac{4}{3} S_{-1}
        +\frac{2}{3} S_{-2}
        -6 \log (2)
    \biggr)
    +\biggl(
        \frac{4+4 N+4 N^2+47 N^3+32 N^4+27 N^5+8 N^6}{2 N^3 (1+N)^3}
        \nonumber \\ &&
        -\frac{14 \big(1+N+N^2\big) S_2}{N (1+N)}
        +6 S_3
    \biggr) S_1
    +\biggl(
        -\frac{3+2 N-20 N^2+4 N^3+4 N^4}{2 N^2 (1+N)^2}
        +14 S_2
    \biggr) S_1^2
    \nonumber \\ &&
    -\frac{4 \big(1+N+N^2\big) S_1^3}{N (1+N)}
    +2 S_1^4
    -\frac{\big(9-36 N-80 N^2+172 N^3+186 N^4+48 N^5\big) S_2}{6 N^2 (1+N)^2}
    \nonumber \\ &&
    -\frac{\big(18+51 N+11 N^2-48 N^3-8 N^4\big) S_3}{6 N (1+N)}
    +6 S_4
    -16 S_{-2} S_{-1}
    +4 S_{-2}^2
    +7 S_{2,1}
    \nonumber \\ &&
    +6 S_{3,1}
    +16 S_{-2,-1}
    -6 S_{2,1,1}
    +8 S_2^2
    \nonumber \\ &&
    - (-1)^N
    \biggl(
         \frac{\big(1+N^2+4 N^3+3 N^4\big) \pi ^2}{3 N^2 (1+N)^2}
        +\frac{4 \big(1+N^2+4 N^3+3 N^4\big) S_{-2}}{N^2 (1+N)^2}
    \biggr)
    \nonumber \\ &&
    +\NF 
    \biggl(
        L_0^2 
        \biggl[
                -\frac{-12-16 N-23 N^2-11 N^3+2 N^4+9 N^5+3 N^6}{6 (N-1) N^2 (1+N)^2 (2+N)}
                +\frac{2}{3} S_1
        \biggr]
        \nonumber \\ &&
        +L_0 
        \biggl[
                -\frac{1}{18 (N-1)^2 N^3 (1+N)^3 (2+N)^2} \big(432+456 N-796 N^2-910 N^3-1268 N^4
                \nonumber \\ &&
                -1099 N^5-527 N^6-94 N^7+188 N^8+135 N^9+27 N^{10}\big)
                +\frac{4 \big(-3+2 N+2 N^2\big) S_1}{9 N (1+N)}
                \nonumber \\ &&
                +\frac{4}{3} S_1^2
        \biggr]
        -\frac{2 \pi ^2}{81 (N-1) N^2 (1+N)^2 (2+N)} \big(54+27 N+3 N^2-2 N^3+57 N^4+6 N^5
        \nonumber \\ &&
        -42 N^6-6 N^7+9 N^8+2 N^9\big)
        +\frac{1}{216 (N-1) N^4 (1+N)^4 (2+N)^3} \big(-1728-8064 N
        \nonumber \\ &&
        -16512 N^2-16208 N^3-18988 N^4-67880 N^5-59294 N^6+18305 N^7+47573 N^8
        \nonumber \\ &&
        +29174 N^9+13900 N^{10}+7097 N^{11}+2801 N^{12}+640 N^{13}+64 N^{14}\big)
        -\frac{56}{27} S_1
        \nonumber \\ &&
        +\frac{2\big(108+54 N+60 N^2+77 N^3+87 N^4-69 N^5-111 N^6-12 N^7+18 N^8+4 N^9\big) S_2}{27 (N-1) N^2 (1+N)^2 (2+N)}
        \nonumber \\ && 
        +\frac{2}{3} S_3
        + (-1)^N 
        \biggl[
                \frac{2 \big(6+3 N-15 N^2+3 N^3-13 N^5-N^6+4 N^7+N^8\big) \pi ^2}{9 (N-1) N^2 (1+N)^2 (2+N)}
                \nonumber \\ &&
                +\frac{8 \big(6+3 N-15 N^2+3 N^3-13 N^5-N^6+4 N^7+N^8\big) S_{-2}}{3 (N-1) N^2 (1+N)^2 (2+N)}
        \biggr]
      \biggr)
  \Biggr]
  \,, \\
  \Gamma_{\gamma,N} &=&
  \frac{\aem(0)}{2\pi} 
  \Biggl[
     \frac{\big(2+N+N^2\big) L_0}{(N-1) N (1+N)}
    -\frac{4-2 N-15 N^2-3 N^3-N^4+N^5}{(N-1)^2 N^2 (1+N)^2}
  \Biggr]
  \nonumber \\ && 
  + \left( \frac{\aem(0)}{2\pi} \right)^2
  \Biggl[
    +L_0^2 
    \biggl(
        \frac{\big(2+N+N^2\big)\big(2+3 N+3 N^2\big)}{4 (N-1) N^2 (1+N)^2}
        -\frac{\big(2+N+N^2\big) S_1}{(N-1) N (1+N)}
    \biggr)
    \nonumber \\ && 
    +L_0 \biggl(
         \frac{4+8 N-5 N^2-16 N^3-29 N^4-18 N^5-8 N^6}{2 (N-1) N^3 (1+N)^3}
         -\frac{3 \big(2+N+N^2\big) S_2}{(N-1) N (1+N)}
         \nonumber \\ && 
        +\frac{\big(4+16 N+25 N^2+12 N^3+7 N^4\big) S_1}{(N-1) N^2 (1+N)^2}
        -\frac{3 \big(2+N+N^2\big) S_1^2}{(N-1) N (1+N)}
    \biggr)
    \nonumber \\ && 
    - \frac{8+12 N-14 N^2-15 N^3+119 N^4+127 N^5+167 N^6+51 N^7+42 N^8+15 N^9}{4 (N-2) (N-1) N^4 (1+N)^4} 
    \nonumber \\ && 
    +\pi ^2 
    \biggl(
        \frac{8
                +12 N
                +14 N^2
                +25 N^3
                +4 N^4
                +N^5
                + 24 \cdot 2^N N^2 \big( 1 - N \big)
        }{6 (N-2) (N-1) N^2 (1+N)^2} 
        \nonumber \\ && 
        +\frac{2 \big(2+N+N^2\big) S_1(\{2\})}{3 (N-1) N (1+N)}
    \biggr)
    +\frac{\big(-76-96 N-61 N^2-22 N^3+7 N^4\big) S_1^2}{4 (N-2) (N-1) N (1+N)^2}
    \nonumber \\ && 
    +\frac{\big(2+N+N^2\big) S_1^3}{6 (N-1) N (1+N)}
    -\biggl(
         \frac{32+64 N-60 N^2-174 N^3-182 N^4-47 N^5-32 N^6-N^7}{2 (N-2) (N-1) N^3 (1+N)^3}
        \nonumber \\ && 
        -\frac{\big(2+N+N^2\big) S_2}{2 (N-1) N (1+N)}
    \biggr) S_1
    +\frac{\big(-16-100 N-124 N^2-111 N^3-30 N^4+5 N^5\big) S_2}{4 (N-2) (N-1) N^2 (1+N)^2}
    \nonumber \\ && 
    +\frac{\big(2+N+N^2\big) S_3}{3 (N-1) N (1+N)}
    -\frac{4 \big(2+N+N^2\big) S_{2,1}}{(N-1) N (1+N)}
    +\frac{3 2^{4+N} S_{1,1}\big(\{\frac{1}{2},1\}\big)}{(N-2) (1+N)^2}
    \nonumber \\ && 
    -\frac{8 \big(2+N+N^2\big) S_{1,1,1}\big(\{2,\frac{1}{2},1\}\big)}{(N-1) N (1+N)}
    \nonumber \\ && 
    +\NF 
    \biggl(
        -\frac{2 \big(2+N+N^2\big) L_0^2}{3 (N-1) N (1+N)}
        +L_0 
        \biggl[
                -\frac{4 \big(-12-10 N+34 N^2+23 N^3+8 N^4+5 N^5\big)}{9 (N-1)^2 N^2 (1+N)^2}
                \nonumber \\ &&
                +\frac{4 \big(2+N+N^2\big) S_1}{3 (N-1) N (1+N)}
        \biggr]
        +\frac{2 \big(576-136 N-198 N^2+105 N^3-56 N^4-5 N^5+2 N^6\big) S_1}{9 (-4+N) (N-3) (N-2) (N-1) N (1+N)^2}
        \nonumber \\ &&
        -\frac{2\big(
                2064+6292 N-494 N^2-3084 N^3+626 N^4-41 N^5-96 N^6+13 N^7\big)}{27 (-4+N) (N-3) (N-2) (N-1) N (1+N)^3} 
        \nonumber \\ && 
        -\frac{\big(2+N+N^2\big) S_1^2}{3 (N-1) N (1+N)}
        -\frac{\big(2+N+N^2\big) S_2}{3 (N-1) N (1+N)}
    \biggr)
  \Biggr]
  \,, \\
  \Gamma_{e^{+},N} &=&
  \left( \frac{\aem(0)}{2\pi} \right)^2
  \Biggl[ 
     L_0 
    \biggl(
        \frac{2 \big(1+2 N+2 N^2\big)}{N^3 (1+N)^3}
        + (-1)^N 
        \biggl[
                -\pi ^2 
                \bigg(
                         \frac{1}{3 N (1+N)}
                        -\frac{2 S_1}{3}
                \bigg)
                \nonumber \\ && 
                -\bigg(
                         \frac{4}{N (1+N)}
                        -8 S_1
                \bigg) S_{-2}
                +4 S_{-3}
                -8 S_{-2,1}
                -2 \zeta_{3}
        \biggr]
    \biggr)
    \nonumber \\ && 
    -\frac{\big(3+N^2+12 N^3+13 N^4+2 N^5\big) \pi ^2}{9 N^2 (1+N)^2}
    +\frac{2 \big(3+N^2+12 N^3+13 N^4+2 N^5\big) S_2}{3 N^2 (1+N)^2}
    \nonumber \\ && 
    +\frac{-6-24 N-27 N^2-33 N^3-172 N^4-121 N^5-3 N^6+32 N^7+8 N^8}{6 N^4 (1+N)^4}
    \nonumber \\ && 
    + (-1)^N 
    \biggl(
        \frac{7 \pi ^4}{180}
        +\zeta_{3}
        \bigg(
                \frac{-9+9 N+14 N^2+6 N^3+N^4}{3 N (1+N)}
                +6 S_1
        \big)
        +\pi ^2 
        \bigg(
                 S_1^2
                 +\frac{1}{3} S_2
                 \nonumber \\ && 
                -\frac{\big(3+2 N+2 N^2\big) S_1}{3 N (1+N)}
                +\frac{4}{9} \big(3+5 N+N^2\big) S_{-1}
                +\frac{2}{3} S_{-2}
                +\frac{2}{3} \big(3+5 N+N^2\big) \log (2)
                \nonumber \\ && 
                -\frac{-6-15 N-10 N^2+34 N^3+35 N^4+16 N^5+4 N^6}{18 N^2 (1+N)^2}
        \bigg)
        -\frac{8}{3} \big(3+5 N+N^2\big) S_{-1} S_2
        \nonumber \\ && 
        +\bigg(
                -\frac{2 \big(-6-15 N-10 N^2+34 N^3+35 N^4+16 N^5+4 N^6\big)}{3 N^2 (1+N)^2}
                -\frac{4 \big(3+2 N+2 N^2\big) S_1}{N (1+N)}
                \nonumber \\ && 
                +12 S_1^2
                +4 S_2
        \bigg) S_{-2}
        +\bigg(
                \frac{2 \big(-21+10 N^2+12 N^3+2 N^4\big)}{3 N (1+N)}
                +28 S_1
        \bigg) S_{-3}
        +22 S_{-4}
        \nonumber \\ && 
        +\frac{8}{3} \big(3+5 N+N^2\big) S_{2,-1}
        +\frac{4 \big(3+2 N+2 N^2\big) S_{-2,1}}{N (1+N)}
        -24 S_1 S_{-2,1}
        -20 S_{-2,2}
        \nonumber \\ && 
        -28 S_{-3,1}
        +24 S_{-2,1,1}
    \biggr)
    +\NF 
    \biggl(
        \frac{\big(2+N+N^2\big)^2 L_0^2}{2 (N-1) N^2 (1+N)^2 (2+N)}
        \nonumber \\ && 
        -\frac{2 \big(2+N+N^2\big) \pi ^2}{3 (N-1) N^2 (1+N)^2 (2+N)}
        +\frac{4 \big(2+N+N^2\big) S_2}{(N-1) N^2 (1+N)^2 (2+N)}
        \nonumber \\ && 
        -\frac{L_0\big(24+28 N-46 N^2-63 N^3-80 N^4-51 N^5-7 N^6+2 N^7+N^8\big)}{(N-1)^2 N^3 (1+N)^3 (2+N)^2} 
        \nonumber \\ && 
        -\frac{1}{6 (N-1) N^4 (1+N)^4 (2+N)^3} \big(48+240 N+504 N^2
        +480 N^3+243 N^4+993 N^5
        \nonumber \\ && 
        +633 N^6-779 N^7-741 N^8+20 N^9+199 N^{10}+72 N^{11}
        +8 N^{12}\big)
        \nonumber \\ && 
        +(-1)^N
        \bigg[
                \frac{2 \big(6+3 N-15 N^2+3 N^3-13 N^5-N^6+4 N^7+N^8\big) \pi ^2}{9 (N-1) N^2 (1+N)^2 (2+N)}
                \nonumber \\ && 
                +\frac{8 \big(6+3 N-15 N^2+3 N^3-13 N^5-N^6+4 N^7+N^8\big) S_{-2}}{3 (N-1) N^2 (1+N)^2 (2+N)}
        \bigg]  
    \biggr)
  \Biggr] \,.
\eeqn

\end{small}

\vskip 0.2truecm
\noindent 
In keeping with eq.~(\ref{wtFNdef}) and our choice $\rho=1$ for
the asymptotic $N\to\infty$ expansion, we write
\beqn
\ePDF{\alpha,N}\;\stackrel{N\to\infty}{=}\;
\widetilde{\Gamma}_{\alpha,N}+\ord\left(\frac{1}{N^3}\right)\,,
\eeqn
and in view of eq.~(\ref{GNfun}) we parametrise the asymptotic 
Mellin-space PDFs thus:
\beqn
\widetilde{\Gamma}_{\alpha,N}=
\sum_{m=0}^4\sum_{q=0}^2 p_{q,m}^{(\alpha)}
\frac{\log^m\bN}{N^q}\,.
\label{iniMSb}
\eeqn
From the results given above we obtain:
\beqn
p^{(e^-)}_{0,4} &=& 2 \left(\frac{\aem(0)}{2\pi}\right)^2 \,,
\\
 p^{(e^-)}_{0,3} &=& \left(\frac{\aem(0)}{2\pi}\right)^2 (4 L_0-4) \,,
\\
 p^{(e^-)}_{0,2} &=& -2 \frac{\aem(0)}{2\pi} + \left(\frac{\aem(0)}{2\pi}\right)^2 \left(2 L_0^2-L_0 
\left(7-\frac{4 \NF}{3}\right)+\frac{2 \pi^2}{3}-2\right)
\,,
\\
 p^{(e^-)}_{0,1} &=& \left(\frac{\aem(0)}{2\pi}\right)^2 
\left[-L_0^2\left(3-\frac{2 \NF}{3}\right)-
L_0 \left(1-\frac{8 \NF}{9}\right)-\frac{2}{3} \pi^2 (1-L_0)-
\frac{56 \NF}{27}+4\right]
\nonumber \\*&&
+\frac{\aem(0)}{2\pi} (2-2 L_0) \,,
\\
 p^{(e^-)}_{0,0} &=& \left(\frac{\aem(0)}{2\pi}\right)^2 
\Bigg[L_0^2 \left(\frac{9}{8}-\frac{\NF}{2}\right)+
L_0 \left(\frac{27}{8}-\frac{3 \NF}{2}\right)+
6 L_0 \zeta_3-\pi^2 L_0
-\left(\frac{3}{2}-\frac{2 \NF}{3}\right) \zeta_3
\nonumber \\*&&
+\frac{3139 \NF}{648}+\pi^2 \left(\frac{1}{4}-\frac{\NF}{3}\right)-
\frac{11 \pi^4}{180}+\frac{241}{32}-2 \pi^2 \log (2)\Bigg]
\nonumber \\*&&
+\frac{\aem(0)}{2\pi} \left(\frac{3 L_0}{2}-\frac{\pi^2}{3}+2\right)+1 \,,
\\
 p^{(e^-)}_{1,3} &=& 4 \left(\frac{\aem(0)}{2\pi}\right)^2 \,,
\\
 p^{(e^-)}_{1,2} &=& \left(\frac{\aem(0)}{2\pi}\right)^2 (6 L_0-15) \,,
\\
 p^{(e^-)}_{1,1} &=& \left(\frac{\aem(0)}{2\pi}\right)^2 \left[2 L_0^2-
L_0 \left(15-\frac{4 \NF}{3}\right)+\frac{2 \pi^2}{3}+\frac{25}{2}\right]
-2 \frac{\aem(0)}{2\pi} \,,
\\
 p^{(e^-)}_{1,0} &=& \left(\frac{\aem(0)}{2\pi}\right)^2 
\Bigg[L_0^2 \left(-\left(\frac{3}{2}-\frac{\NF}{3}\right)\right)+
L_0 \left(\frac{4 \NF}{9}+\frac{11}{2}\right)-
\pi^2 \left(\frac{1}{18}-\frac{L_0}{3}\right)
\nonumber \\*&&
-\frac{58 \NF}{27}+\frac{28}{3}\Bigg]+\frac{\aem(0)}{2\pi} (3-L_0) \,,
\\
 p^{(e^-)}_{2,3} &=& -\frac{14}{3} \left(\frac{\aem(0)}{2\pi}\right)^2 \,,
\\
 p^{(e^-)}_{2,2} &=& \left(\frac{\aem(0)}{2\pi}\right)^2 
\left(\frac{35}{2}-7 L_0\right) \,,
\\
 p^{(e^-)}_{2,1} &=& \left(\frac{\aem(0)}{2\pi}\right)^2 
\left[-\frac{7 L_0^2}{3}+L_0 \left(\frac{115}{6}-\frac{14 \NF}{9}\right)-
\frac{7 \pi^2}{9}-\frac{139}{6}\right]+\frac{7}{3}\frac{\aem(0)}{2\pi} \,,
\\
 p^{(e^-)}_{2,0} &=& \left(\frac{\aem(0)}{2\pi}\right)^2 
\Bigg[L_0^2 \left(\frac{\NF}{9}+\frac{9}{4}\right)-
L_0 \left(\frac{32 \NF}{27}+\frac{79}{6}\right)+
\pi^2 \left(\frac{13}{36}-\frac{7 L_0}{18}\right)
\nonumber \\*&&
+\frac{3653 \NF}{810}+\frac{1}{6}\Bigg]
+\frac{\aem(0)}{2\pi} \left(\frac{7 L_0}{6}-\frac{14}{3}\right) \,,
\\
 p^{(\gamma)}_{1,3} &=& \frac{\left(\frac{\aem(0)}{2\pi}\right)^2}{6} \,,
\\
 p^{(\gamma)}_{1,2} &=& \left(\frac{\aem(0)}{2\pi}\right)^2 \left(-3 L_0-\frac{\NF}{3}+\frac{7}{4}\right),
\\
 p^{(\gamma)}_{1,1} &=& \left(\frac{\aem(0)}{2\pi}\right)^2 
\left[-L_0^2+L_0 \left(\frac{4 \NF}{3}+7\right)+\frac{4 \NF}{9}+
\frac{\pi^2}{12}+\frac{1}{2}\right] \,,
\\
 p^{(\gamma)}_{1,0} &=& \left(\frac{\aem(0)}{2\pi}\right)^2 
\Bigg[L_0^2 \left(\frac{3}{4}-\frac{2 \NF}{3}\right)-
L_0 \left(\frac{20\NF}{9}+4\right)+\pi^2
 \left(-\frac{L_0}{2}-\frac{\NF}{18}+\frac{3}{8}\right)
\nonumber \\*&&
-\frac{26\NF}{27}+\frac{13 \zeta_3}{3}-\frac{15}{4}\Bigg]+
\frac{\aem(0)}{2\pi} (L_0-1) \,,
\\
 p^{(\gamma)}_{2,3} &=& \frac{1}{6}\left(\frac{\aem(0)}{2\pi}\right)^2 \,,
\\
 p^{(\gamma)}_{2,2} &=& \left(\frac{\aem(0)}{2\pi}\right)^2 \left(-3 L_0
 -\frac{\NF}{3}-\frac{7}{2}\right) \,,
\\
 p^{(\gamma)}_{2,1} &=& \left(\frac{\aem(0)}{2\pi}\right)^2 
\left(-L_0^2+L_0 \left(\frac{4 \NF}{3}+2\right)+\frac{19 \NF}{9}+
\frac{\pi^2}{12}+\frac{53}{4}\right) \,,
\\
 p^{(\gamma)}_{2,0} &=& \left(\frac{\aem(0)}{2\pi}\right)^2 
\Bigg[L_0^2 \left(\frac{1}{4}-\frac{2 \NF}{3}\right)+L_0
 \left(\frac{11}{2}-\frac{26 \NF}{9}\right)-\pi^2
 \left(\frac{L_0}{2}+\frac{\NF}{18}+\frac{1}{6}\right)
\nonumber \\*&&
+\frac{25\NF}{27}+
\frac{13 \zeta_3}{3}-\frac{47}{4}\Bigg]+\frac{\aem(0)}{2\pi}  (L_0+1) \,,
\\ 
 p^{(e^+)}_{2,0} &=& \left(\frac{\aem(0)}{2\pi}\right)^2 
\left(\frac{L_0^2 \NF}{2}-L_0 \NF+\frac{3 \NF}{2}-\frac{1}{2}\right) \,.
\eeqn
All $p_{q,m}^{(\alpha)}$ coefficients not explicitly given above vanish.

\subsubsection{Configuration space\label{sec:iniZ}}
The initial conditions in the $\MSb$ scheme read as follows in 
configuration space:
\beqn 
  \lefteqn{\invM\left[\Gamma_{e^{-},N}\right](z) = \delta(1-z) } \nonumber &&
  \nonumber \\ &&
  + \frac{\aem(0)}{2\pi} 
  \biggl[
    \delta(1-z) 
    \biggl(
      2+\frac{3 L_0}{2}
    \biggr)
    + \biggl(
      \frac{1}{1-z}
      \biggl[
        2 L_0
        -2
        -4 \log (1-z)
      \biggr]
    \biggr)_{+}
    \nonumber \\ &&
    + (1+z) 
    \biggl(
      1
      +2 \log (1-z)
      - L_0
    \biggr)
  \biggr]
  \nonumber \\ &&
  + \left( \frac{\aem(0)}{2\pi} \right)^2
  \biggl[
    \delta(1-z) 
    \biggl(
      \frac{27-8 \pi ^2}{24} L_0^2
      +\bigg(
        \frac{27}{8}
        +\frac{\pi ^2}{6}
        -2 \zeta_3
      \bigg) L_0
      +\frac{241}{32}
      +\frac{13}{2} \zeta_3
      \nonumber \\ &&
      +\frac{1}{12} \pi ^2 \big(7-24 \log (2)\big)
      -\frac{5 \pi ^4}{36}
      +\NF 
      \biggl[
        -\frac{L_0^2}{2}
        -\frac{27+4 \pi ^2}{18} L_0
        +\frac{3139}{648}
        -\frac{\pi ^2}{3}
        +\frac{2}{3} \zeta_3      
      \biggr]
    \biggr)
    \nonumber \\ &&
    + \biggl(
      \frac{1}{1-z}
      \biggl[
         \big(3+4 \log (1-z)\big) L_0^2
        + L_0 
        \biggl(
        1
        + \frac{4 \pi ^2}{3}
        -14 \log (1-z)
        -12 \log ^2(1-z)
        \biggr)
        \nonumber \\ &&
        -\frac{4}{3} \big(3+\pi ^2-12 \zeta_3\big)
        -\frac{4}{3} \big(3+2 \pi ^2\big) \log (1-z)
        +12 \log ^2(1-z)
        +8 \log ^3(1-z)
        \nonumber \\ &&
        +\NF \biggl(
          -\frac{2 L_0^2}{3}
          -\bigg(
                \frac{8}{9}-\frac{8}{3} \log (1-z)\bigg) L_0
          +\frac{56}{27}
        \biggr)
      \biggr]
    \biggr)_{+}
    -L_0^2 
    \biggl(
        \frac{1}{2} (5+z)
        \nonumber \\ &&
        +2 (1+z) \log (1-z)
        +\frac{\big(1+3 z^2\big) \log (z)}{2 (1-z)}
    \biggr)
    +L_0 
    \biggl(
        -\pi ^2 (1+z)
        -\frac{1}{2} (11-9 z)
        \nonumber \\ &&
        +\bigg(
          11+3 z+\frac{2 \big(1+z^2\big) \log (z)}{1-z}
        \bigg) \log (1-z)
        +6 (1+z) \log ^2(1-z)
        \nonumber \\ &&
        -\frac{\big(2-3 z^2\big) \log (z)}{1-z}
        -\frac{1}{2} (1+z) \log ^2(z)
        +2 (1+z) \text{Li}_2(z)
    \biggr)
    + \frac{45-26 z+21 z^2}{6 (1-z)}
    \nonumber \\ &&
    +\pi ^2 
    \biggl(
        \frac{3+10 z-10 z^2}{18 (1-z)}
        +\frac{\big(3-5 z^2\big) \log (1-z)}{2 (1-z)}
    \biggr)
    \nonumber \\ &&
    +\log (z) 
    \biggl(
        -\frac{21-45 z+49 z^2-131 z^3+42 z^4}{6 (1-z)^2 (1+z)}
        +\frac{8 (1+z) \log (1+z)}{1-z}
        \nonumber \\ &&
        -\frac{4 \big(1+z^2\big) \text{Li}_2(-z)}{1-z}
        -\frac{7 \big(1+z^2\big) \text{Li}_2(z)}{1-z}
    \biggr)
    +\log (1-z) 
    \biggl(
        -\frac{1}{2} (8-13 z)
        \nonumber \\ &&
        +\frac{\big(21-14 z-z^2\big) \log (z)}{3 (1-z)}
        -\frac{7 \big(1+z^2\big) \log ^2(z)}{2 (1-z)}
        +\frac{\big(-1+7 z^2\big) \text{Li}_2(z)}{1-z}
    \biggr)
    \nonumber \\ &&
    -\frac{\big(11-z^2\big) \zeta_3}{1-z}
    +\biggl(
        -10
        +\frac{\big(1+9 z^2\big) \log (z)}{2 (1-z)}
    \biggr) \log ^2(1-z)
    -4 (1+z) \log ^3(1-z)
    \nonumber \\ &&
    +\frac{\big(9+24 z-69 z^2+30 z^3+22 z^4\big) \log ^2(z)}{12 (-1+z)^3}
    -\frac{1}{12} (1+z) \log ^3(z)
    +\frac{8 (1+z) \text{Li}_2(-z)}{1-z}
    \nonumber \\ &&
    +\frac{\big(33-2 z-10 z^2\big) \text{Li}_2(z)}{3 (1-z)}
    -\frac{\big(1-7 z^2\big) \text{Li}_3(1-z)}{1-z}
    +\frac{8 \big(1+z^2\big) \text{Li}_3(-z)}{1-z}
    \nonumber \\ &&
    +\frac{\big(9+13 z^2\big) \text{Li}_3(z)}{1-z}
    +\NF 
    \biggl(
        L_0^2 \biggl[
                -\frac{-4-5 z+z^2+4 z^3}{6 z}
                +(1+z) \log (z)
        \biggr] 
        \nonumber \\ &&
        +L_0 \biggl[
                -\frac{18+35 z-79 z^2+18 z^3}{9 z}
                -\frac{4}{3} (1+z) \log (1-z)
                -3 (1+z) \log ^2(z)
                \nonumber \\ &&
                -\frac{\big(8+9 z-18 z^2-3 z^3+8 z^4\big) \log (z)}{3 (1-z) z}              
        \biggr]
        +\frac{4 (1-z) \big(1+4 z+z^2\big) \log (1-z) \log (z)}{3 z}
        \nonumber \\ &&
        +\frac{112+47 z-134 z^2-1215 z^3+1676 z^4+849 z^5-710 z^6-641 z^7+400 z^8}{54 (-1+z)^3 z (1+z)^2} 
        \nonumber \\ &&
        +\log (z) \biggl[
                 \frac{8 (1-z) \big(1+4 z+z^2\big) \log (1+z)}{3 z}
                -8 (1+z) \text{Li}_2(-z)
                -4 (1+z) \text{Li}_2(z)
                \nonumber \\ &&
                -\frac{26+9 z-59 z^2-87 z^3-123 z^4+839 z^5-321 z^6-333 z^7+157 z^8+20 z^9}{9 (1-z)^4 (1+z)^3} 
        \biggr]
        \nonumber \\ &&
        +4 (1+z) \zeta_3        
        +\frac{\big(1+84 z-81 z^2-8 z^3\big) \log ^2(z)}{12 (1-z)}
        -\frac{1}{6} (1+z) \log ^3(z)
        \nonumber \\ &&
        +\frac{8 (1-z) \big(1+4 z+z^2\big) \text{Li}_2(-z)}{3 z}
        +\frac{4 (1-z) \big(1+4 z+z^2\big) \text{Li}_2(z)}{3 z}
        \nonumber \\ &&
        +16 (1+z) \text{Li}_3(-z)
        +8 (1+z) \text{Li}_3(z)  
    \biggr)
  \biggr]
  \,, \\
  \lefteqn{\invM\left[\Gamma_{\gamma,N}\right](z) =} \nonumber \\ &&
  \frac{\aem(0)}{2\pi} 
  \biggl[
    \frac{\big(2-2 z+z^2\big) L_0}{z}
    -\frac{2-2 z+z^2}{z}
    -\frac{2 \big(2-2 z+z^2\big) \log (z)}{z}
  \biggr]
  \nonumber \\ &&
  + \left( \frac{\aem(0)}{2\pi} \right)^2
  \biggl[
     L_0^2 \biggl(
        \frac{4-z}{4}
        +\frac{\big(2-2 z+z^2\big) \log (1-z)}{z}
        +\frac{1}{2} (2-z) \log (z)
    \biggr)
    \nonumber \\ &&
    +L_0 \biggl(
        \frac{1}{2} (-3-5 z)
        +\frac{1}{3} \pi ^2 (-2+z)
        -\frac{\big(4+3 z^2\big) \log (1-z)}{z}
        +\frac{5}{2} z \log (z)
        \nonumber \\ &&
        -\frac{3 \big(2-2 z+z^2\big) \log ^2(1-z)}{z}
        +\frac{1}{2} (-2+z) \log ^2(z)
        -2 (-2+z) \text{Li}_2(z)
    \biggr)
    \nonumber \\ &&
    -\frac{32-20 z+33 z^2}{12 z}
    +\pi ^2 
    \biggl(
             \frac{32-48 z+24 z^2-3 z^3}{18 z^2}
            -\frac{2 \big(2-2 z+z^2\big) \log (1-z)}{3 z}
            \nonumber \\ &&
            +\frac{1}{3} (2-z) \log (z)
    \biggr)
    -\log ^2(1-z) 
    \biggl(
             \frac{32-66 z+42 z^2-15 z^3}{4 z^2}
             \nonumber \\ &&
            -\frac{4 \big(2-2 z+z^2\big) \log (z)}{z}
    \biggr)
    -\log (1-z) 
    \biggl(
             \frac{48-64 z+56 z^2-37 z^3}{6 z^2}
             \nonumber \\ &&
            -\frac{8 (2-z)^3 \log (2-z)}{3 z^2}
            +\frac{\big(32-48 z+12 z^2+7 z^3\big) \log (z)}{6 z^2}
            \nonumber \\ &&
            +\frac{8 \big(2-2 z+z^2\big) \text{Li}_2(z-1)}{z}
            -\frac{4 \big(2-2 z+z^2\big) \text{Li}_2(z)}{z}
    \biggr)
    +\NF \biggl(
        -\frac{2 \big(2-2 z+z^2\big) L_0^2}{3 z}
        \nonumber \\ &&
        -L_0 \biggl[
                 \frac{4 \big(4-4 z+5 z^2\big)}{9 z}
                +\frac{4 \big(2-2 z+z^2\big) \log (1-z)}{3 z}
                -\frac{8 \big(2-2 z+z^2\big) \log (z)}{3 z}
        \biggr]
        \nonumber \\ &&
        +\frac{2 \big(192-384 z+344 z^2-152 z^3-65 z^4\big)}{135 z^3}
        -\frac{\big(2-2 z+z^2\big) \log ^2(1-z)}{3 z}
        \nonumber \\ &&
        +\frac{4 \big(32-80 z+40 z^2+45 z^3-75 z^4+33 z^5\big) \log (1-z)}{45 z^4}
        +\frac{16}{15} z \log (z)   
    \biggr)
    \nonumber \\ &&
    +\frac{2 \big(4-2 z+z^2\big) \zeta_3}{z}
    -\frac{\big(2-2 z+z^2\big) \log ^3(1-z)}{6 z}
    +\frac{1}{12} (2-z) \log ^3(z)
    \nonumber \\ &&
    +\biggl[
            -\frac{64-67 z+59 z^2}{12 z}
            +(-2+z) \text{Li}_2(z)
    \biggr] \log (z)
    +\frac{1}{24} (12+23 z) \log ^2(z)
    \nonumber \\ &&
    +\frac{8 (2-z)^3 \text{Li}_2(z-1)}{3 z^2}
    +\frac{\big(-32+48 z-36 z^2+13 z^3\big) \text{Li}_2(z)}{6 z^2}
    \nonumber \\ &&
    +\frac{8 \big(2-2 z+z^2\big) \text{Li}_3(1-z)}{z}
    +\frac{16 \big(2-2 z+z^2\big) \text{Li}_3(z-1)}{z}
    -2 (2-z) \text{Li}_3(z)
  \biggr]
  \,, \\
  \lefteqn{\invM\left[\Gamma_{e^{+},N}\right](z) = } \nonumber \\ &&
  \left( \frac{\aem(0)}{2\pi} \right)^2
  \biggl[
    L_0 
    \biggl(
        4 (1-z)
        -\frac{\pi ^2 \big(1+z^2\big)}{3 (1+z)}
        +\frac{\big(1+z^2\big) \log ^2(z)}{1+z}
        -\frac{4 \big(1+z^2\big) \text{Li}_2(-z)}{1+z}
        \nonumber \\ &&
        +\bigg(
                2 (1+z)
                -\frac{4 \big(1+z^2\big) \log (1+z)}{1+z}
        \bigg) \log (z)      
    \biggr)
    -\frac{(1-z) \big(45+74 z+45 z^2\big)}{6 (1+z)^2}
    \nonumber \\ &&
    +\log ^2(z) 
    \biggl(
        -\frac{z \big(1-z-5 z^2+z^3\big)}{(1+z)^3}
        +\frac{\big(1+z^2\big) \log (1+z)}{1+z}
    \biggr)
    \nonumber \\ &&
    +\pi ^2 
    \biggl(
        \frac{15+28 z+30 z^2+36 z^3+3 z^4}{18 (1+z)^3}
        +\frac{\big(1+z^2\big) \log (1+z)}{1+z}
    \biggr)
    -\frac{5 \big(1+z^2\big) \zeta_3}{1+z}
    \nonumber \\ &&
    +\log (z) 
    \biggl(
            -\frac{9+12 z+30 z^2-20 z^3-15 z^4}{6 (1+z)^3}
            +\frac{6 \big(1+z^2\big) \log ^2(1+z)}{1+z}
            \nonumber \\ &&
            -\frac{4 \big(3+8 z+12 z^2+12 z^3+z^4\big) \log (1-z)}{3 (1+z)^3}
            -\frac{2 \big(-3+10 z+z^2\big) \log (1+z)}{3 (1+z)}    
            \nonumber \\ &&
            +\frac{6 \big(1+z^2\big) \text{Li}_2(-z)}{1+z}
            +\frac{2 \big(1+z^2\big) \text{Li}_2(z)}{1+z}
    \biggr)
    +\frac{\big(1+z^2\big) \log ^3(z)}{6 (1+z)}
    +\frac{2 \big(1+z^2\big) \text{Li}_3(-z)}{1+z}
  \nonumber \\ &&
  -\frac{2 \big(1+z^2\big) \log ^3(1+z)}{1+z}
  +\frac{2 \big(3-10 z-z^2\big) \text{Li}_2(-z)}{3 (1+z)}
  -\frac{4 \big(1+z^2\big) \text{Li}_3(z)}{1+z}
  \nonumber \\ &&
  -\frac{4 \big(3+8 z+12 z^2+12 z^3+z^4\big) \text{Li}_2(z)}{3 (1+z)^3}
  +\frac{12 \big(1+z^2\big) \text{Li}_3\big(\frac{z}{1+z}\big)}{1+z}
  \nonumber \\ &&
  +\NF 
  \biggl(
        +\biggl[
                \frac{(1-z) \big(4+7 z+4 z^2\big)}{6 z}
                +(1+z) \log (z)
        \biggr] L_0^2
        -L_0 \biggl[
                 \frac{(1-z) \big(2+5 z-2 z^2\big)}{z}
                 \nonumber \\ &&
                +\frac{\big(8+15 z-3 z^2-8 z^3\big) \log (z)}{3 z}
                +3 (1+z) \log ^2(z)
        \biggr]
        +4 (1+z) \zeta_3
        \nonumber \\ &&
        -\frac{(1-z) \big(112+345 z+898 z^2+921 z^3+400 z^4\big)}{54 z (1+z)^2}
        -\frac{1}{6} (1+z) \log ^3(z)
        \nonumber \\ &&
        +\log (z) \biggl[
                -\frac{21+93 z+245 z^2+411 z^3+210 z^4+20 z^5}{9 (1+z)^3}
                -8 (1+z) \text{Li}_2(-z)
                \nonumber \\ &&
                +\frac{4 (1-z) \big(1+4 z+z^2\big) \log (1-z)}{3 z}
                +\frac{8 (1-z) \big(1+4 z+z^2\big) \log (1+z)}{3 z}    
                \nonumber \\ &&        
                -4 (1+z) \text{Li}_2(z)
        \biggl]
        +\frac{1}{12} \big(3+87 z+8 z^2\big) \log ^2(z)
        +\frac{8 (1-z) \big(1+4 z+z^2\big) \text{Li}_2(-z)}{3 z}
        \nonumber \\ &&
        +\frac{4 (1-z) \big(1+4 z+z^2\big) \text{Li}_2(z)}{3 z}
        +16 (1+z) \text{Li}_3(-z)
        +8 (1+z) \text{Li}_3(z)
  \biggr)
  \biggr] \,.
\eeqn

The large-$z$ asymptotic expressions of the $\MSb$ initial conditions 
read as follows:
\beqn
&&\invM\left[\widetilde{\Gamma}_{e^-,N}\right](z)=
\Domz 
\nonumber \\&&\phantom{aaa}\;
+ \frac{\aem(0)}{2\pi}  \Bigg[\Domz\left(2 + \frac{3 L_0}{2}\right) -2 (1 - L_0) \plusdz 
-4 \plusdo
\nonumber
\\&&\phantom{aaaa}\;
+2(1-L_0+2\log(1-z))-(1-z)(1-L_0+2\log(1-z))
\Bigg] 
\nonumber
\\&&\phantom{aa}
+ \left(\frac{\aem(0)}{2\pi}\right)^2 \Bigg\{\left(-4 + 4 L_0^2 + L_0\left(-14 + \frac{8\NF}{3}\right) 
-\frac{8\pi^2}{3}\right) \plusdo 
\nonumber
\\&&\phantom{aa}
+ 12(1-L_0) \plusd{2} + 8 \plusd{3} 
\nonumber
\\&&\phantom{aa}
+\plusdz \left( L_0^2\left(3 - \frac{2\NF}{3}\right) -4
+ \frac{56\NF}{27} -\frac{4\pi^2}{3} + L_0 \left(1 - \frac{8\NF}{9} 
+ \frac{4\pi^2}{3}\right) + 16 \zeta_3\right) 
\nonumber
\\&&\phantom{aa}
+\Domz\Bigg(\frac{241}{32} +\frac{3139\NF}{648} 
+\frac{7\pi^2}{12} - \frac{\NF\pi^2}{3} 
- \frac{5\pi^4}{36} + L_0^2\left(9/8 - \frac{\NF}{2} 
- \frac{\pi^2}{3}\right) 
\nonumber
\\&&\phantom{aa}
-2\pi^2\log(2) + L_0\left(\frac{27}{8} - \frac{3\NF}{2} +\frac{\pi^2}{6} 
-\frac{2\NF\pi^2}{9} -  2 \zeta_3\right)
+ \frac{13\zeta_3}{2} + \frac{2}{3}\NF\zeta_3\Bigg)
\nonumber
\\&&\phantom{aa}
-\frac{86 \NF}{27}-8\log^3(1-z)-9\log^2(1-z)+\frac{8}{3}
\pi^2\log(1-z)+\frac{3}{2}\log(1-z)
\nonumber
\\&&\phantom{aa}
-16 \zeta (3)+\frac{16\pi^2}{9}+\frac{28}{3}
+L_0^2 \left(\frac{2 \NF}{3}-4\log(1-z)-1\right)
\nonumber
\\&&\phantom{aa}
+L_0 \left(-\frac{8}{3} \NF\log(1-z)+\frac{20 \NF}{9}+12\log^2(1-z)
+10\log(1-z)-\frac{4 \pi^2}{3}-2\right)
\nonumber
\\&&\phantom{aa}
-(1-z)\Bigg[L_0^2 \left(-\frac{\NF}{6}-2\log(1-z)+\frac{3}{2}\right)
\nonumber
\\&&\phantom{aa}
+L_0 \left(-\frac{4}{3} \NF\log(1-z)+\frac{31 \NF}{9}+6\log^2(1-z)
-3\log(1-z)-\frac{2 \pi^2}{3}+3\right)
\nonumber
\\&&\phantom{aa}
-\frac{1171 \NF}{270}-4\log^3(1-z)+\frac{9}{2}\log^2(1-z)
+\frac{4}{3} \pi^2\log(1-z)
\nonumber
\\&&\phantom{aa}
-\frac{13}{2}\log(1-z)-8 \zeta (3)+\frac{\pi^2}{9}+\frac{75}{4}\Bigg]
\Bigg\} \,,
\\
&&\invM\left[\widetilde{\Gamma}_{e^+,N}\right](z)=
\left(\frac{\aem(0)}{2\pi}\right)^2 (1-z) \left[\half-\frac{3\NF}{2}+L_0\NF-L_0^2\frac{\NF}{2}\right]
\,, \\
&&\invM\left[\widetilde{\Gamma}_{\gamma,N}\right](z)=
\frac{\aem(0)}{2\pi} \Big[-1+L_0+(1-z)(1+L_0)\Big]
\nonumber \\&&\phantom{aa}
+\left(\frac{\aem(0)}{2\pi}\right)^2 \Bigg\{
L_0^2 \left(-\frac{2 \NF}{3}+\log(1-z)+\frac{3}{4}\right)
\nonumber
\\&&\phantom{aa}
+L_0 \left(-\frac{4}{3} \NF\log(1-z)-\frac{20\NF}{9}
-3\log^2(1-z)-7\log(1-z)-4\right)
\nonumber
\\&&\phantom{aa}
-\frac{1}{3} \NF\log^2(1-z)-\frac{4}{9} \NF\log(1-z)
-\frac{26 \NF}{27}-\frac{1}{6}\log^3(1-z)+\frac{7}{4}\log^2(1-z)
\nonumber
\\&&\phantom{aa}
-\frac{1}{2}\log(1-z)+4 \zeta (3)+\frac{\pi^2}{12}-\frac{15}{4}
-(1-z) 
\Bigg[L_0^2 \left(\frac{2\NF}{3}-\log(1-z)+\frac{1}{4}\right)
\nonumber
\\&&\phantom{aa}
+L_0 \left(\frac{4}{3} \NF\log(1-z)+\frac{20 \NF}{9}+3\log^2(1-z)
-\log(1-z)+2\right)
\nonumber
\\&&\phantom{aa}
+\frac{1}{3} \NF\log^2(1-z)+\frac{16}{9} \NF\log(1-z)-\frac{58 \NF}{27}
+\frac{1}{6}\log^3(1-z)+\frac{13}{4}\log^2(1-z)
\nonumber
\\&&\phantom{aa}
+\frac{11}{2}\log(1-z)-4 \zeta (3)-\frac{5\pi^2}{12}+\frac{19}{4}\Bigg]
\Bigg\} \,.
\eeqn

\subsection{$\Delta$-scheme initial conditions\label{sec:resMSb}}
Having already presented the $\MSb$ initial conditions in 
appendix~\ref{sec:resMSbb}, here we limit ourselves to giving the differences
between those and the $\Delta$-scheme ones. In other words, by adding the
quantities below to the $\MSb$ initial conditions one can obtain those
in the $\Delta$ scheme. By doing so, in the $z$ space the cancellation 
of the $\delta(1-z)$ and plus-distribution contributions will become manifest. 

\subsubsection{Mellin space\label{sec:DiniN}}
%%%%%%%%%%%NONSINGLET%%%%%%%%%%%%%%%%%%%%%%%%
\beqn
%2
&&\Gamma_{NS,0,N}^{(\Delta)[1]}-\Gamma_{NS,0,N}^{[1]}=
2\harm1{1}{N}^2-\frac{2\left(N^2+N+1\right)\harm1{1}{N}}{N(N+1)}
+2\harm1{2}{N}
\nonumber \\&&\phantom{aa}
-\frac{2 N^3+4 N^2-N-1}{N (N+1)^2}\,,
\\
&&\Gamma_{NS,0,N}^{(\Delta)[2]}-\Gamma_{NS,0,N}^{[2]}=
-2 \harm1{1}{N}^4+\frac{4 \left(N^2+N+1\right)
\harm1{1}{N}^3}{N (N+1)}-12 \harm1{1}{N}^2\harm1{2}{N}
\nonumber
\\&&\phantom{aa}
+\frac{\left(12 N^3+6N^2-69 N+9\right) \harm1{1}{N}^2}{6 N (N+1)^2}
+\frac{4}{3} \pi^2\harm1{1}{N}^2+\frac{12 \left(N^2+N+1\right) 
\harm1{1}{N} \harm1{2}{N}}{N(N+1)}
\nonumber
\\&&\phantom{aa}
-16 \harm1{1}{N} \harm1{3}{N}
-\frac{4 \pi^2 \left(N^2+N+1\right)\harm1{1}{N}}{3 N (N+1)}
\nonumber
\\&&\phantom{aa}
-\frac{\left(8 N^4+29 N^3+44 N^2+63 N-8\right)
\harm1{1}{N}}{2 N (N+1)^3}+\frac{56}{27} \NF \harm1{1}{N}+16 \zeta_3
\harm1{1}{N}
\nonumber
\\&&\phantom{aa}
-6 \harm1{2}{N}^2+\frac{\left(12 N^3+6 N^2-69 N+9\right)
\harm1{2}{N}}{6 N (N+1)^2}+\frac{4}{3} \pi^2 \harm1{2}{N}
\nonumber
\\&&\phantom{aa}
+\frac{8\left(N^2+N+1\right) \harm1{3}{N}}{N (N+1)}-
12 \harm1{4}{N}+\zeta_3
\left(-\frac{13 N^2+13 N+16}{2 N^2+2 N}-\frac{2\NF}{3}\right)
\nonumber
\\&&\phantom{aa}
-\frac{\left(15695 N^2+12095 N+5592\right) \NF}{3240 N
(N+1)}+\pi^2 \left(\frac{\NF}{3}-\frac{21 N^3+58 N^2+N+12}{36 N
(N+1)^2}\right)
\nonumber
\\&&\phantom{aa}
-\frac{723 N^5+3404 N^4+5258 N^3+2820 N^2+1307 N-1240}{96 N
(N+1)^4}+\frac{5 \pi^4}{36}+2 \pi^2 \log (2)
\nonumber
\\&&\phantom{aa}
%6
+\frac{6 \alpha_{3,2} \, f_{3,2}}{{N (N+1) (N+2) (N+3)}} \,,
\\
%%%%%%%%%%%SINGLET%%%%%%%%%%%%%%%%%%%%%%%%
%8
&&\Gamma_{\Sigma,0,N}^{(\Delta)[1]}-\Gamma_{\Sigma,0,N}^{[1]}=
\Gamma_{NS,0,N}^{(\Delta)[1]}-\Gamma_{NS,0,N}^{[1]}
%10
+\frac{6\alpha_{1,1}}{N (N+1) (N+2) (N+3)}\,,
\\
%16
&&\Gamma_{\Sigma,0,N}^{(\Delta)[2]}-\Gamma_{\Sigma,0,N}^{[2]}=
\left[\Gamma_{NS,0,N}^{(\Delta)[2]}-
\Gamma_{NS,0,N}^{[2]}\right]_{\alpha_{3,2}=0}
+\frac{1-3\NF}{N(N+1)}
\nonumber 
\\&&\phantom{aa}
%18
+\frac{\alpha_{1,1}}{N (N+1) (N+2) (N+3)}\Bigg\{
-12\harm1{1}{N}^2
+\frac{12\left(N^2+N+1\right)\harm1{1}{N}}{N(N+1)}
\nonumber
\\&&\phantom{aaaa}
-12 \harm1{2}{N}
+\frac{6\left(2 N^3+4 N^2-N-1\right)}{N (N+1)^2}
\Bigg\}
%20
+\frac{6\alpha_{1,2}}{N (N+1) (N+2) (N+3)}\,,
%%%%%%%%%%%PHOTON%%%%%%%%%%%%%%%%%%%%%%%%
\\
%12
&&\Gamma_{\gamma,0,N}^{(\Delta)[1]}-\Gamma_{\gamma,0,N}^{[1]}=
\frac{1}{N+1}
%14
-\frac{409}{30 (N-1) N (N+1)}\alpha_{2,1}\,,
\\
%22
&&\Gamma_{\gamma,0,N}^{(\Delta)[2]}-\Gamma_{\gamma,0,N}^{[2]}=
-\frac{(N+2) \harm1{1}{N}^3}{6 N (N+1)}
+\frac{\left(N^2+3 N+2\right) \NF \harm1{1}{N}^2}{3 N (N+1)^2}
\nonumber 
\\&&\phantom{aaa}
+\frac{\left(-7 N^2+N+6\right) \harm1{1}{N}^2}{4 N (N+1)^2}
-\frac{(N+2) \harm1{1}{N} \harm1{2}{N}}{2 N (N+1)}
\nonumber
\\&&\phantom{aaa}
-\frac{\left(8 N^3+68 N^2+100 N+40\right) \NF \harm1{1}{N}}{18 N (N+1)^3}
-\frac{\left(N^3+27 N^2+36 N+12\right) \harm1{1}{N}}{2 N (N+1)^3}
\nonumber
\\&&\phantom{aaa}
+\frac{\left(N^2+3 N+2\right) \NF \harm1{2}{N}}{3 N (N+1)^2}
+\frac{\left(-7 N^2+N+6\right) \harm1{2}{N}}{4 N (N+1)^2}
-\frac{(N+2) \harm1{3}{N}}{3 N (N+1)}
\nonumber
\\&&\phantom{aaa}
+\frac{2 \left(13 N^4+47 N^3+30 N^2-20 N-16\right) \NF}{27 N (N+1)^4}
\nonumber
\\&&\phantom{aaa}
+\frac{15 N^4+101 N^3+165 N^2+117 N+34}{4 N (N+1)^4}
-\frac{4 (N+2) \zeta_3}{N (N+1)}-\frac{\pi^2 (N+6)}{12 N (N+1)}
\nonumber
\\&&\phantom{aaa}
%24
+\frac{409\alpha_{2,1}}{15 (N-1) N (N+1)}\Bigg\{
\harm1{1}{N}^2
-\frac{\left(N^2+N+1\right)\harm1{1}{N}}{N (N+1)}
+\harm1{2}{N}
\nonumber
\\&&\phantom{aaa}
-\frac{\left(2 N^3+4 N^2-N-1\right)}{2 N (N+1)^2}
\Bigg\}
%\nonumber
%\\&&\phantom{aaa}
%26
+\frac{2 \alpha_{2,2}}{N (N+1) (N+2)}f_{2,2} \,.
\eeqn

\subsubsection{Configuration space\label{sec:DiniZ}}
%%%%%%%%%%%NONSINGLET%%%%%%%%%%%%%%%%%%%%%%%%
\beqn
%1
&&\Gamma_{NS,0}^{(\Delta)[1]}(z)-\Gamma_{NS,0}^{[1]}(z)=
-2\Domz-2\plusdz-4\plusdo
\nonumber
\\&&\phantom{aaa}
-2-4\log(1-z)+(1-z)\big(2\log(1-z)+1\big)\,,
\\
%3
&&\Gamma_{NS,0}^{(\Delta)[2]}(z)-\Gamma_{NS,0}^{[2]}(z)=
\nonumber 
\\&&\phantom{aaa}
\Domz\left(-\frac{3139\NF}{648}-\frac{241}{32}
-\frac{7 \pi^2}{12}
+\frac{\pi^2 \NF}{3}
+2 \pi^2 \log (2)
+\frac{5 \pi^4}{36}
-\frac{13 \zeta_3}{2}-\frac{2 \NF \zeta_3}{3}
\right)
\nonumber
\\&&\phantom{a}
+\plusdz \left(\frac{56 \NF}{27}+16 \zeta_3-\frac{4\pi^2}{3}-4\right)-
\frac{4}{3} \left(3+2 \pi^2\right) \plusdo
\nonumber
\\&&\phantom{a}
+12\plusd{2}+8\plusd{3}
+\frac{2}{27} \left(43 \NF-6 \left(-36 \zeta_3+21+4 \pi^2\right)\right)
\nonumber
\\&&\phantom{a}
+(1-z) \Bigg(\frac{73}{4}-\frac{383 \NF}{135}-8 \zeta_3+\frac{\pi^2}{9}
+\frac{1}{6} \left(8 \pi^2-39\right)\log(1-z)
\nonumber
\\&&\phantom{aaaaaa}
+\frac{9}{2} \log^2(1-z)-4 \log^3(1-z)
\Bigg)+\left(-\frac{3}{2}-\frac{8 \pi^2}{3}\right) \log(1-z)
\nonumber
\\&&\phantom{a}
+8\log^3(1-z)+9 \log^2(1-z)
%5
+\alpha_{3,2}(1-z)^3 f_{3,2} \,,
%%%%%%%%%%%SINGLET%%%%%%%%%%%%%%%%%%%%%%%%
\\
%7 CHECKKKKK
&&\Gamma_{\Sigma,0}^{(\Delta)[1]}(z)-\Gamma_{\Sigma,0}^{[1]}(z)=
\Gamma_{NS,0}^{(\Delta)[1]}(z)-\Gamma_{NS,0}^{[1]}(z)
%9
+\alpha_{1,1}(1-z)^3\,,
\\
%15
&&\Gamma_{\Sigma,0}^{(\Delta)[2]}(z)-\Gamma_{\Sigma,0}^{[2]}(z)=
\left[\Gamma_{NS,0}^{(\Delta)[2]}(z)-
\Gamma_{NS,0}^{[2]}(z)\right]_{\alpha_{3,2}=0}
+(1-z)(1-3\NF)
\nonumber 
\\&&\phantom{aa}
%17
+\alpha_{1,1}\Bigg\{
\frac{2}{3} \left(\pi^2-6 \dilog_2(z)\right)
+\frac{14 \log (z)}{3}
\nonumber
\\&&\phantom{aaaa}
+(1-z) \left(6 \dilog_2(z)+4\log(1-z)-8 \log (z)-\pi^2
+\frac{2}{3}\right)
\nonumber
\\&&\phantom{aaaa}
+(1-z)^2 \left(-4 \log(1-z)+8\log (z)
-\frac{2}{3}\right)
\nonumber
\\&&\phantom{aaaa}
+(1-z)^3
\Bigg(-4 \dilog_2(z)-2 \log^2(1-z)+\frac{11}{3} \log(1-z)
\nonumber
\\&&\phantom{aaaaaaaaaaa}
-\frac{11\log(z)}{3}+\frac{2 \pi^2}{3}+\frac{17}{6}\Bigg)
\Bigg\}
%19
+\alpha_{1,2}(1-z)^3\,,
%%%%%%%%%%%PHOTON%%%%%%%%%%%%%%%%%%%%%%%%
\\
%11
&&\Gamma_{\gamma,0}^{(\Delta)[1]}(z)-\Gamma_{\gamma,0}^{[1]}(z)=z
%13
-\frac{409(1-z)^2}{60 z}\alpha_{2,1}\,,
\\
%21
&&\Gamma_{\gamma,0}^{(\Delta)[2]}(z)-\Gamma_{\gamma,0}^{[2]}(z)=
\\&&\phantom{aaa}
\frac{26\NF}{27}
-4\zeta_3
-\frac{\pi^2}{12}+\frac{15}{4}
+\left(\frac{4\NF}{9}+\frac{1}{2}\right) \log(1-z)
+\left(\frac{\NF}{3}-\frac{7}{4}\right) \log^2(1-z)
\nonumber
\\&&\phantom{aaa}
+\frac{1}{6} \log^3(1-z)
+(1-z)\Bigg[-4\zeta_3-\frac{5 \pi^2}{12} -\frac{58\NF}{27}
+\frac{19}{4}
\nonumber
\\&&\phantom{aaa}
+\left(\frac{16 \NF}{9}+\frac{11}{2}\right) \log(1-z)
+\left(\frac{\NF}{3}+\frac{13}{4}\right)\log^2(1-z)
+\frac{1}{6} \log^3(1-z)\Bigg]
\nonumber
\\&&\phantom{aaa}
%23
-\frac{409\alpha_{2,1}}{180z}
\Bigg\{6 \dilog_2(z)
+(1-z) (-6 \log(1-z)+12 \log (z)-3)-9 \log (z)-\pi^2
\nonumber
\\&&\phantom{aaaaaa}
+(1-z)^2 \Bigg(-6 \dilog_2(z)-6 \log^2(1-z)
+9\log(1-z)-3 \log (z)+\pi^2+3\Bigg)
\Bigg\}
\nonumber
\\&&\phantom{aaa}
%25
+ \alpha_{2,2} f_{2,2} (1-z)^2 \,,
\nonumber
\eeqn

\subsection{$K$ matrices\label{sec:resK}}
The results of sect.~\ref{sec:DS} for the implementation of the $\Delta$
scheme, as well as those necessary for the subsequent redefinitions of
the short-distance cross sections (sect.~\ref{sec:DYxsec}), need the
explicit forms of the $K$-matrix elements. These are given in this
appendix in a way which should allow one to implement any specific
choice within the $\Delta$-scheme class that stems from a given
$r(x)$ function (eq.~(\ref{rfun})), according to the guidelines
given in sect.~\ref{sec:others}.

The matrix-element results are as follows.

\subsubsection{Mellin space\label{sec:KN}}
%%%%%%%%%%%NONSINGLET%%%%%%%%%%%%%%%%%%%%%%%%
\beqn
%2
&&K_{NS,N}^{(\Delta)[1]}=
\Gamma_{NS,0,N}^{(\Delta)[1]}-\Gamma_{NS,0,N}^{[1]}\,,
\\
%\eeqn
%\beqn
%4
&&K_{NS,N}^{(\Delta)[2]}=
\nonumber \\&&\phantom{aa}
2 \harm1{1}{N}^4-\frac{4 \left(N^2+N+1\right)
\harm1{1}{N}^3}{N (N+1)}-4 \harm1{1}{N}^2\harm1{2}{N}
+\frac{4}{3} \pi^2\harm1{1}{N}^2
\nonumber
\\&&\phantom{aa}
+\frac{\left(-12 N^3-30N^2+57 N+24\right) \harm1{1}{N}^2}{6 N (N+1)^2}
+\frac{4 \left(N^2+N+1\right) 
\harm1{1}{N} \harm1{2}{N}}{N(N+1)}
\nonumber
\\&&\phantom{aa}
-16 \harm1{1}{N} \harm1{3}{N}
-\frac{4 \pi^2 \left(N^2+N+1\right)\harm1{1}{N}}{3 N (N+1)}
+\frac{56}{27} \NF \harm1{1}{N}+16 \zeta_3 \harm1{1}{N}
\nonumber
\\&&\phantom{aa}
+\frac{\left(8 N^5+19 N^4-4 N^3-47 N^2-8 N-8\right) 
\harm1{1}{N}}{2 N^2 (N+1)^3}-2 \harm1{2}{N}^2
+\frac{4}{3} \pi^2 \harm1{2}{N}
\nonumber
\\&&\phantom{aa}
+\frac{\left(-36 N^3-90 N^2-45 N+33\right) \harm1{2}{N}}{6 N(N+1)^2}
+\frac{8 \left(N^2+N+1\right) \harm1{3}{N}}{N (N+1)}
\nonumber
\\&&\phantom{aa}
-12 \harm1{4}{N}+\zeta_3 \left(-\frac{13 N^2+13 N+16}{2 N^2+2 N}
-\frac{2 \NF}{3}\right)
+\frac{5 \pi^4}{36}+2 \pi^2 \log (2)
\nonumber
\\&&\phantom{aa}
-\frac{\left(15695 N^2+12095 N+5592\right) \NF}{3240 N (N+1)}
+\pi^2 \left(\frac{\NF}{3}-\frac{21 N^3+58 N^2+N+12}{36 N (N+1)^2}\right)
\nonumber
\\&&\phantom{aa}
-\frac{339 N^6+1868 N^5+4106 N^4+3972 N^3+1979 N^2-1432 N-96}
{96 N^2 (N+1)^4}
\nonumber
\\&&\phantom{aa}
%6
+\frac{6 \alpha_{3,2} f_{3,2}}{{N (N+1) (N+2) (N+3)}} \,,
\\
%\eeqn
%%%%%%%%%%%SINGLET%%%%%%%%%%%%%%%%%%%%%%%%
%\beqn
%8
&&K_{\Sigma\Sigma,0,N}^{(\Delta)[1]}=
K_{NS,0,N}^{(\Delta)[1]}
%10
+\frac{6\alpha_{1,1}}{N (N+1) (N+2) (N+3)}\,,
\\
%\eeqn
%\beqn
%12
&&K_{\Sigma\Sigma,0,N}^{(\Delta)[2]}=
\left[K_{NS,0,N}^{(\Delta)[2]}
\right]_{\alpha_{3,2}=0}
+\frac{1-3\NF}{N(N+1)}
%14
+\frac{6\alpha_{1,2}}{N (N+1) (N+2) (N+3)}\,,
\\
%\eeqn
%%%%%%%%%%%PHOTON%%%%%%%%%%%%%%%%%%%%%%%%
%\beqn
%16
&&K_{\gamma\Sigma,0,N}^{(\Delta)[1]}=
\Gamma_{\gamma,0,N}^{(\Delta)[1]}-\Gamma_{\gamma,0,N}^{[1]}\,,
\\
%\eeqn
%\beqn
%20
&&K_{\gamma\Sigma,0,N}^{(\Delta)[2]}=
\left[\Gamma_{\gamma,0,N}^{(\Delta)[2]}
-\Gamma_{\gamma,0+N}^{[2]}\right]_{\alpha_{2,1}=0}
+\frac{2\harm1{1}{N}^2}{N+1}
-\frac{2\left(N^2+N+1\right)\harm1{1}{N}}{N (N+1)^2}
\nonumber \\&&\phantom{aaaaaaaaaaaaaaaaaaaaaaaa}
+\frac{2\harm1{2}{N}}{N+1}
-\frac{2N^3+4N^2-N-1}{N(N+1)^3}\,.
\eeqn
The combination of two of these matrix elements amounts, in the $N$-space,
to simply multiplying them. While this operation is trivial, it possibly
entails some simplifications, and it constitutes the starting point
for arriving at the corresponding $z$-space terms, which will be
given in appendix~\ref{sec:KZ}.
\beqn
%2
&&\left(K_{NS,N}^{(\Delta)[1]}\right)^2=
4 \harm1{1}{N}^4-\frac{8 \left(N^2+N+1\right) \harm1{1}{N}^3}{N (N+1)}
+8 \harm1{1}{N}^2 \harm1{2}{N}
\nonumber \\&&\phantom{aa}
-\frac{4 \left(N^4+2 N^3-4 N^2-3 N-1\right) \harm1{1}{N}^2}{N^2 (N+1)^2}
-\frac{8 \left(N^2+N+1\right) \harm1{1}{N} \harm1{2}{N}}{N (N+1)}
\nonumber
\\&&\phantom{aa}
+\frac{4 \left(N^2+N+1\right) \left(2 N^3+4 N^2-N-1\right) 
\harm1{1}{N}}{N^2 (N+1)^3}+4 \harm1{2}{N}^2
\nonumber
\\&&\phantom{aa}
-\frac{4 \left(2 N^3+4 N^2-N-1\right) \harm1{2}{N}}{N (N+1)^2}
+\frac{\left(2 N^3+4 N^2-N-1\right)^2}{N^2 (N+1)^4}\,,
\\
%\eeqn
%\beqn
%4
&&\left(K_{\Sigma\Sigma,N}^{(\Delta)[1]}\right)^2=
\left(K_{NS,N}^{(\Delta)[1]}\right)^2
\nonumber \\&&\phantom{aaa}
%6
+\frac{24\alpha_{1,1}}{N (N+1) (N+2) (N+3)}\Bigg\{
\harm1{1}{N}^2
-\frac{\left(N^2+N+1\right) \harm1{1}{N}}{N (N+1)}
\nonumber
\\&&\phantom{aaaaaaaaaaaaaaaaaaaaa}
+\harm1{2}{N}
-\frac{\left(2 N^3+4 N^2-N-1\right)}{2N (N+1)^2}
\Bigg\}
\nonumber
\\&&\phantom{aaa}
%8
+\frac{36\alpha_{1,1}^2}{N^2 (N+1)^2 (N+2)^2 (N+3)^2} \,,
\\
%\eeqn
%\beqn
&&K_{\gamma\Sigma,N}^{(\Delta)[1]}K_{\Sigma\Sigma,N}^{(\Delta)[1]}=
%10
\frac{2 \harm1{1}{N}^2}{N+1}
-\frac{2 \left(N^2+N+1\right) \harm1{1}{N}}{N (N+1)^2}
+\frac{2 \harm1{2}{N}}{N+1}
\nonumber \\&&\phantom{aa}
-\frac{2 N^3+4 N^2-N-1}{N (N+1)^3}
%12
-\frac{409\alpha_{1,1}\alpha_{2,1}}{5 (N-1) N^2 (N+1)^2 (N+2) (N+3)}
%14
\nonumber
\\&&\phantom{aa}
+\frac{6\alpha_{1,1}}{N (N+1)^2 (N+2) (N+3)}
%16
+\frac{409\alpha_{2,1}}{(N-1)N(N+1)}
\nonumber
\biggl(
-\harm1{1}{N}^2-\harm1{2}{N}
\\&&\phantom{aaaaaaa}
+\frac{\left(N^2+N+1\right) \harm1{1}{N}}{N (N+1)}
+\frac{\left(2 N^3+4 N^2-N-1\right)}{2N (N+1)^2}\biggr) \,.
\eeqn

\subsubsection{Configuration space\label{sec:KZ}}
%%%%%%%%%%%NONSINGLET%%%%%%%%%%%%%%%%%%%%%%%%
\beqn
%1
&&K_{NS}^{(\Delta)[1]}(z)=
\Gamma_{NS,0}^{(\Delta)[1]}(z)-\Gamma_{NS,0}^{[1]}(z)\,,
\\
%\eeqn
%\beqn
%3
&&K_{NS}^{(\Delta)[2]}(z)=
\nonumber \\&&\phantom{aaa}
\Domz\left(-\frac{2 \NF \zeta_3}{3}+\frac{\pi^2 \NF}{3}-\frac{3139\NF}{648}-
\frac{13 \zeta_3}{2}+\frac{5 \pi^4}{36}-\frac{7 \pi^2}{12}-
\frac{113}{32}+2 \pi^2 \log (2)\right)
\nonumber
\\&&\phantom{aa}
+\plusdz \left(\frac{56 \NF}{27}-16 \zeta_3+\frac{4\pi^2}{3}+4\right)
+\frac{4}{3} \left(3+2 \pi^2\right) \plusdo
\nonumber
\\&&\phantom{aa}
-12\plusd{2}-8\plusd{3}+\frac{86\NF}{27}
+32 \zeta_3+\frac{8\pi^2}{9}-\frac{4}{3}
\nonumber
\\&&\phantom{aa}
+(1-z) \Bigg[-\frac{383 \NF}{135}-4 \dilog_2(1-z)
+\left(12 \dilog_2(1-z)-2\pi^2-7\right) \log (z)
+8\dilog_3(1-z)
\nonumber
\\&&\phantom{aa}
+24 \dilog_3(z)
+\log(1-z) \left(-8 \dilog_2(1-z)+12 \log^2(z)-4
 \log (z)-\frac{4 \pi^2}{3}-\frac{21}{2}\right)
\nonumber
\\&&\phantom{aa}
+4 \log^3(1-z)+\left(\frac{17}{2}-12 \log (z)\right)\log^2(1-z)
-16 \zeta_3+\frac{\pi^2}{9}+\frac{65}{4}\Bigg]
\nonumber
\\&&\phantom{aa}
+\frac{1}{1-z}\Bigg[32 (\dilog_3(z)-\zeta_3)+\left(16 \dilog_2(1-z)
-\frac{8\pi^2}{3}-4\right) \log (z)
\nonumber
\\&&\phantom{aaa}
-16 \log (z) \log^2(1-z)
+\left(16 \log^2(z)-16 \log (z)\right) \log(1-z)\Bigg]
\nonumber
\\&&\phantom{aaa}
+\log(1-z)\left(16 \dilog_2(1-z)-24 \log^2(z)+16 \log (z)+\frac{8
 \pi^2}{3}+\frac{13}{2}\right)
\nonumber
\\&&\phantom{aaa}
+\left(-24 \dilog_2(1-z)+4 \pi^2+10\right) 
\log(z)-8 \log^3(1-z)
\nonumber
\\&&\phantom{aaa}
+(24 \log (z)-15) \log^2(1-z)
-16 \dilog_3(1-z)-48 \dilog_3(z)
\nonumber
\\&&\phantom{aaa}
%5
+\alpha_{3,2} f_{3,2} (1-z)^3 \,,
\\
%\eeqn
%%%%%%%%%%%SINGLET%%%%%%%%%%%%%%%%%%%%%%%%
%\beqn
%7
&&K_{\Sigma\Sigma,0}^{(\Delta)[1]}(z)=
K_{NS,0}^{(\Delta)[1]}(z)
%9
+\alpha_{1,1}(1-z)^3\,,
\\
%\eeqn
%\beqn
%11
&&K_{\Sigma\Sigma,0}^{(\Delta)[2]}(z)=
\left[K_{NS,0}^{(\Delta)[2]}(z)\right]_{\alpha_{3,2}=0}
+(1-z)(1-3\NF)
%13
+\alpha_{1,2}(1-z)^3\,,
\\
%\eeqn
%%%%%%%%%%%PHOTON%%%%%%%%%%%%%%%%%%%%%%%%
%\beqn
%15
&&K_{\gamma\Sigma,0}^{(\Delta)[1]}(z)=
\Gamma_{\gamma,0}^{(\Delta)[1]}(z)-\Gamma_{\gamma,0}^{[1]}(z)\,,
\\
%\eeqn
%\beqn
%19
&&K_{\gamma\Sigma,0}^{(\Delta)[2]}(z)=
\left[\Gamma_{\gamma,0}^{(\Delta)[2]}(z)
-\Gamma_{\gamma,0}^{[2]}(z)\right]_{\alpha_{2,1}=0}
+1-3z-\frac{\pi^2}{3}z+2\log(1-z)
\nonumber \\&&\phantom{aaaaaa}
+2z\log^2(1-z)+z\log(z)+2z\dilog_2(z)\,,
\eeqn
The convolutions of two of the above matrix elements have been computed
by inverse-Mellin transform of their corresponding quantities in
the $N$-space, given in appendix~\ref{sec:KN}. They read as follows:
\beqn
%1
&&\left(K_{NS}^{(\Delta)[1]}(z)\right)^2=
\nonumber \\&&\phantom{aa}
8 \left[(\pi^2 \plusdo-4\plusdz\zeta_3-2\plusd{3}\right]
-16 (\dilog_3(z)+\zeta_3)
\nonumber
\\&&\phantom{aa}
+\left(8 \dilog_2(z)-\frac{8 \pi^2}{3}\right) \log (z)
+\frac{1}{1-z}\Bigg(16(\dilog_3(z)-\zeta_3)
+\frac{8}{3} \left(\pi^2-3 \dilog_2(z)\right) \log(z)
\nonumber
\\&&\phantom{aaaaa}
-\frac{2}{3} \log^3(z)
+8 \log(1-z) \log^2(z)-24 \log^2(1-z) \log(z)\Bigg)
-16\log^3(1-z)
\nonumber
\\&&\phantom{aa}
+\frac{2 \log^3(z)}{3}+24 \log (z) \log^2(1-z)
+\left(8 \pi^2-8\log^2(z)\right) \log(1-z)\,,
\\
%\eeqn
%\beqn
%3
&&\left(K_{\Sigma\Sigma}^{(\Delta)[1]}(z)\right)^2=
\left(K_{NS}^{(\Delta)[1]}(z)\right)^2
\nonumber \\&&\phantom{aa}
%5
+\alpha_{1,1}\Bigg\{\frac{4}{3} \left(\pi^2-6 \dilog_2(z)\right)+(1-z)^3
\Bigg(4 \dilog_2(z)+4 \log^2(1-z)+2 \log^2(z)
\nonumber
\\&&\phantom{aaaaa}
+(-4 \log (z)-12) \log(1-z)+12\log (z)-\frac{4 \pi^2}{3}+9\Bigg)
-2 \log^2(z)
\nonumber
\\&&\phantom{aaaaa}
+(1-z)^2 (4 \log(1-z)-8 \log(z)-7)
\nonumber
\\&&\phantom{aaaaa}
+(1-z) (8 \log(1-z)-10 \log (z)-2)+6 \log (z)
\Bigg\}
\nonumber
\\&&\phantom{aaa}
%7
+\alpha_{1,1}^2\Bigg\{
(1-z)^3 \left(\log (z)-\frac{11}{3}\right)
+(1-z)^2 (20-12 \log (z))
\nonumber
\\&&\phantom{aaaaa}
+(1-z) (30 \log (z)-20)-20 \log (z)
\Bigg\}\,,
\\
%\eeqn
%\beqn
%9
&&K_{\gamma\Sigma}^{(\Delta)[1]}(z)K_{\Sigma\Sigma}^{(\Delta)[1]}(z)=
\nonumber \\&&\phantom{aa}
\frac{z}{3}\Bigg(-2\pi^2+6\log^2(1-z)-6\log(1-z)\log(z)
+3\log(z)^2+6 \dilog_2(z)+(1-z)\Bigg)
\nonumber
\\&&\phantom{aa}
%11
-\frac{409\alpha_{1,1}\alpha_{2,1}}{720z} 
\left(3+44z-36z^2-12z^3+z^4+12z(2+3z)\log(z)\right)
\nonumber
\\*&&\phantom{aa}
%13
+\frac{\alpha_{1,1}}{2}\left(
2+3z-6z^2+z^3+6z\log(z)\right)
\nonumber
\\&&\phantom{aa}
%15
+\frac{409\alpha_{2,1}}{180 z} \Bigg(-12 \dilog_2(z)
-3 \log^2(z)+6 \log (z)+2 \pi^2
\nonumber
\\&&\phantom{aaaa}
+(1-z)^2 \left(-6
\dilog_2(z)-6 \log^2(1-z)-3 \log^2(z)+6 \log (z) \log(1-z)+2 \pi^2\right)
\nonumber
\\&&\phantom{aaaa}
+(1-z) \left(24 \dilog_2(z)+6 \log^2(z)-6 \log (z)
+12 \log(1-z)-4 \pi^2-6\right)\Bigg)\,,
\eeqn

\section{An aside\label{sec:gE}}
The coefficients $g_n$ of eq.~(\ref{Gcff}) are known as Gregory
coefficients (or constants); they admit the following integral
representation (known as Schr\"oder integral formula):
\beq
g_n=(-)^{n+1}\int_0^\infty dx\,\frac{1}{(1+x)^n\left(\pi^2+\log^2x\right)}\,.
\label{schroder}
\eeq
The rightmost factor in the denominator of the integrand reminds one
of the reciprocal of the $\Gamma$ function, related in turn to the
computation of the Fransen-Robinson constant. By replacing the
Schr\"oder formula into eq.~(\ref{lgzvslgom1}), by using
\beq
\sum_{n=0}^\infty\frac{(-)^n(z-1)^n}{(1+x)^{n+1}}=\frac{1}{x+z}\,,
\eeq
and by knowing that
\beq
\gE=\int_0^1 dz\left(\frac{1}{\log z}+\frac{1}{1-z}\right),
\eeq
we obtain, after integrating the resulting expression over $z$,
the following representation of $\gE$:
\beq
\gE=\int_0^\infty dx\,\frac{\log(1+x)-\log x}{\pi^2+\log^2x}\,.
\eeq

\phantomsection
\addcontentsline{toc}{section}{References}
\bibliographystyle{JHEP}
\bibliography{eepdfnnlo}

\providecommand{\href}[2]{#2}\begingroup\raggedright\begin{thebibliography}{10}

\bibitem{Yennie:1961ad}
D.~R. Yennie, S.~C. Frautschi and H.~Suura, \emph{{The infrared divergence phenomena and high-energy processes}}, \href{http://dx.doi.org/10.1016/0003-4916(61)90151-8}{\emph{Annals Phys.} {\bf 13} (1961) 379--452}.

\bibitem{Gribov:1972ri}
V.~N. Gribov and L.~N. Lipatov, \emph{{Deep inelastic e p scattering in perturbation theory}}, {\emph{Sov. J. Nucl. Phys.} {\bf 15} (1972) 438--450}.

\bibitem{Lipatov:1974qm}
L.~N. Lipatov, \emph{{The parton model and perturbation theory}}, {\emph{Sov. J. Nucl. Phys.} {\bf 20} (1975) 94--102}.

\bibitem{Altarelli:1977zs}
G.~Altarelli and G.~Parisi, \emph{{Asymptotic Freedom in Parton Language}}, \href{http://dx.doi.org/10.1016/0550-3213(77)90384-4}{\emph{Nucl. Phys.} {\bf B126} (1977) 298--318}.

\bibitem{Dokshitzer:1977sg}
Y.~L. Dokshitzer, \emph{{Calculation of the Structure Functions for Deep Inelastic Scattering and e+ e- Annihilation by Perturbation Theory in Quantum Chromodynamics.}}, {\emph{Sov. Phys. JETP} {\bf 46} (1977) 641--653}.

\bibitem{Skrzypek:1990qs}
M.~Skrzypek and S.~Jadach, \emph{{Exact and approximate solutions for the electron nonsinglet structure function in QED}}, \href{http://dx.doi.org/10.1007/BF01483573}{\emph{Z. Phys.} {\bf C49} (1991) 577--584}.

\bibitem{Skrzypek:1992vk}
M.~Skrzypek, \emph{{Leading logarithmic calculations of QED corrections at LEP}}, {\emph{Acta Phys. Polon.} {\bf B23} (1992) 135--172}.

\bibitem{Cacciari:1992pz}
M.~Cacciari, A.~Deandrea, G.~Montagna and O.~Nicrosini, \emph{{QED structure functions: A Systematic approach}}, \href{http://dx.doi.org/10.1209/0295-5075/17/2/007}{\emph{Europhys. Lett.} {\bf 17} (1992) 123--128}.

\bibitem{Blumlein:2011mi}
J.~Blumlein, A.~De~Freitas and W.~van Neerven, \emph{{Two-loop QED Operator Matrix Elements with Massive External Fermion Lines}}, \href{http://dx.doi.org/10.1016/j.nuclphysb.2011.10.009}{\emph{Nucl. Phys.} {\bf B855} (2012) 508--569}, [\href{http://arxiv.org/abs/1107.4638}{{\tt 1107.4638}}].

\bibitem{Bertone:2019hks}
V.~Bertone, M.~Cacciari, S.~Frixione and G.~Stagnitto, \emph{{The partonic structure of the electron at the next-to-leading logarithmic accuracy in QED}}, \href{http://dx.doi.org/10.1007/JHEP03(2020)135}{\emph{JHEP} {\bf 03} (2020) 135}, [\href{http://arxiv.org/abs/1911.12040}{{\tt 1911.12040}}].

\bibitem{Ablinger:2020qvo}
J.~Ablinger, J.~Bl{\"u}mlein, A.~De~Freitas and K.~Sch{\"o}nwald, \emph{{Subleading Logarithmic QED Initial State Corrections to $e^+e^- \rightarrow \gamma^*/{Z^{0}}^*$ to $O(\alpha^6 L^5)$}}, \href{http://dx.doi.org/10.1016/j.nuclphysb.2020.115045}{\emph{Nucl. Phys. B} {\bf 955} (2020) 115045}, [\href{http://arxiv.org/abs/2004.04287}{{\tt 2004.04287}}].

\bibitem{Frixione:2021wzh}
S.~Frixione, \emph{{On factorisation schemes for the electron parton distribution functions in QED}}, \href{http://dx.doi.org/10.1007/JHEP07(2021)180}{\emph{JHEP} {\bf 07} (2021) 180}, [\href{http://arxiv.org/abs/2105.06688}{{\tt 2105.06688}}].

\bibitem{Delorme:2026vln}
S.~Delorme, A.~Kusina, A.~Si{\'o}dmok and J.~Whitehead, \emph{{PDF evolution in alternative factorisation schemes}},  \href{http://arxiv.org/abs/2606.23813}{{\tt 2606.23813}}.

\bibitem{Bertone:2022ktl}
V.~Bertone, M.~Cacciari, S.~Frixione, G.~Stagnitto, M.~Zaro and X.~Zhao, \emph{{Improving methods and predictions at high-energy e$^{+}$e$^{-}$ colliders within collinear factorisation}}, \href{http://dx.doi.org/10.1007/JHEP10(2022)089}{\emph{JHEP} {\bf 10} (2022) 089}, [\href{http://arxiv.org/abs/2207.03265}{{\tt 2207.03265}}].

\bibitem{Tarasov:1980au}
O.~V. Tarasov, A.~A. Vladimirov and A.~Y. Zharkov, \emph{{The Gell-Mann-Low Function of QCD in the Three Loop Approximation}}, \href{http://dx.doi.org/10.1016/0370-2693(80)90358-5}{\emph{Phys. Lett. B} {\bf 93} (1980) 429--432}.

\bibitem{Larin:1993tp}
S.~A. Larin and J.~A.~M. Vermaseren, \emph{{The Three loop QCD Beta function and anomalous dimensions}}, \href{http://dx.doi.org/10.1016/0370-2693(93)91441-O}{\emph{Phys. Lett. B} {\bf 303} (1993) 334--336}, [\href{http://arxiv.org/abs/hep-ph/9302208}{{\tt hep-ph/9302208}}].

\bibitem{Stahlhofen:2025hqd}
M.~Stahlhofen, \emph{{NNLO electron structure functions (PDFs) from SCET}}, \href{http://dx.doi.org/10.1007/JHEP11(2025)171}{\emph{JHEP} {\bf 11} (2025) 171}, [\href{http://arxiv.org/abs/2508.16964}{{\tt 2508.16964}}].

\bibitem{Schnubel:2025ejl}
M.~Schnubel and R.~Szafron, \emph{{Electron and photon structure functions at two loops}}, \href{http://dx.doi.org/10.1007/JHEP12(2025)167}{\emph{JHEP} {\bf 12} (2025) 167}, [\href{http://arxiv.org/abs/2509.09618}{{\tt 2509.09618}}].

\bibitem{Moch:2004pa}
S.~Moch, J.~A.~M. Vermaseren and A.~Vogt, \emph{{The Three loop splitting functions in QCD: The Nonsinglet case}}, \href{http://dx.doi.org/10.1016/j.nuclphysb.2004.03.030}{\emph{Nucl. Phys. B} {\bf 688} (2004) 101--134}, [\href{http://arxiv.org/abs/hep-ph/0403192}{{\tt hep-ph/0403192}}].

\bibitem{Blumlein:2021enk}
J.~Bl{\"u}mlein, P.~Marquard, C.~Schneider and K.~Sch{\"o}nwald, \emph{{The three-loop unpolarized and polarized non-singlet anomalous dimensions from off shell operator matrix elements}}, \href{http://dx.doi.org/10.1016/j.nuclphysb.2021.115542}{\emph{Nucl. Phys. B} {\bf 971} (2021) 115542}, [\href{http://arxiv.org/abs/2107.06267}{{\tt 2107.06267}}].

\bibitem{Frixione:1995ms}
S.~Frixione, Z.~Kunszt and A.~Signer, \emph{{Three jet cross-sections to next-to-leading order}}, \href{http://dx.doi.org/10.1016/0550-3213(96)00110-1}{\emph{Nucl. Phys.} {\bf B467} (1996) 399--442}, [\href{http://arxiv.org/abs/hep-ph/9512328}{{\tt hep-ph/9512328}}].

\bibitem{Frixione:1997np}
S.~Frixione, \emph{{A General approach to jet cross-sections in QCD}}, \href{http://dx.doi.org/10.1016/S0550-3213(97)00574-9}{\emph{Nucl. Phys.} {\bf B507} (1997) 295--314}, [\href{http://arxiv.org/abs/hep-ph/9706545}{{\tt hep-ph/9706545}}].

\bibitem{Hamberg:1990np}
R.~Hamberg, W.~L. van Neerven and T.~Matsuura, \emph{{A complete calculation of the order $\alpha-s^{2}$ correction to the Drell-Yan $K$ factor}}, \href{http://dx.doi.org/10.1016/0550-3213(91)90064-5}{\emph{Nucl. Phys. B} {\bf 359} (1991) 343--405}.

\bibitem{Harlander:2002wh}
R.~V. Harlander and W.~B. Kilgore, \emph{{Next-to-next-to-leading order Higgs production at hadron colliders}}, \href{http://dx.doi.org/10.1103/PhysRevLett.88.201801}{\emph{Phys. Rev. Lett.} {\bf 88} (2002) 201801}, [\href{http://arxiv.org/abs/hep-ph/0201206}{{\tt hep-ph/0201206}}].

\bibitem{Duhr:2020seh}
C.~Duhr, F.~Dulat and B.~Mistlberger, \emph{{Drell-Yan Cross Section to Third Order in the Strong Coupling Constant}}, \href{http://dx.doi.org/10.1103/PhysRevLett.125.172001}{\emph{Phys. Rev. Lett.} {\bf 125} (2020) 172001}, [\href{http://arxiv.org/abs/2001.07717}{{\tt 2001.07717}}].

\bibitem{Duhr:2021vwj}
C.~Duhr and B.~Mistlberger, \emph{{Lepton-pair production at hadron colliders at N$^{3}$LO in QCD}}, \href{http://dx.doi.org/10.1007/JHEP03(2022)116}{\emph{JHEP} {\bf 03} (2022) 116}, [\href{http://arxiv.org/abs/2111.10379}{{\tt 2111.10379}}].

\bibitem{sigmaI}
C.~Schneider, \emph{{Symbolic summation assists combinatorics}}, {\emph{Seminaire Lotharingien de Combinatoire} {\bf 56} (01, 2007) 1--36}.

\bibitem{sigmaII}
C.~Schneider, \emph{{Term Algebras, Canonical Representations and Difference Ring Theory for Symbolic Summation}},  \href{http://arxiv.org/abs/2102.01471}{{\tt 2102.01471}}.

\bibitem{Ablinger:2009ovq}
J.~Ablinger, \emph{{A Computer Algebra Toolbox for Harmonic Sums Related to Particle Physics}},  Master's thesis, Linz U., 2009.

\bibitem{Ablinger:2012ufz}
J.~Ablinger, \emph{{Computer Algebra Algorithms for Special Functions in Particle Physics}}.
\newblock PhD thesis, Linz U., 2012.
\newblock \href{http://arxiv.org/abs/1305.0687}{{\tt 1305.0687}}.

\bibitem{Ablinger:2013cf}
J.~Ablinger, J.~Bl\"umlein and C.~Schneider, \emph{{Analytic and algorithmic aspects of generalized harmonic sums and polylogarithms}}, \href{http://dx.doi.org/10.1063/1.4811117}{\emph{J. Math. Phys.} {\bf 54} (2013) 082301}, [\href{http://arxiv.org/abs/1302.0378}{{\tt 1302.0378}}].

\bibitem{Ablinger:2014rba}
J.~Ablinger, \emph{{The package HarmonicSums: Computer Algebra and Analytic Aspects of Nested Sums}}, \href{http://dx.doi.org/10.22323/1.211.0019}{\emph{PoS} {\bf LL2014} (2014) 019}, [\href{http://arxiv.org/abs/1407.6180}{{\tt 1407.6180}}].

\bibitem{Vermaseren:1998uu}
J.~A.~M. Vermaseren, \emph{{Harmonic sums, Mellin transforms and integrals}}, \href{http://dx.doi.org/10.1142/S0217751X99001032}{\emph{Int. J. Mod. Phys. A} {\bf 14} (1999) 2037--2076}, [\href{http://arxiv.org/abs/hep-ph/9806280}{{\tt hep-ph/9806280}}].

\bibitem{Ablinger:2013eba}
J.~Ablinger, J.~Bl\"umlein and C.~Schneider, \emph{{Generalized Harmonic, Cyclotomic, and Binomial Sums, their Polylogarithms and Special Numbers}}, \href{http://dx.doi.org/10.1088/1742-6596/523/1/012060}{\emph{J. Phys. Conf. Ser.} {\bf 523} (2014) 012060}, [\href{http://arxiv.org/abs/1310.5645}{{\tt 1310.5645}}].

\bibitem{Blumlein:1998if}
J.~Blumlein and S.~Kurth, \emph{{Harmonic sums and Mellin transforms up to two loop order}}, \href{http://dx.doi.org/10.1103/PhysRevD.60.014018}{\emph{Phys. Rev.} {\bf D60} (1999) 014018}, [\href{http://arxiv.org/abs/hep-ph/9810241}{{\tt hep-ph/9810241}}].

\bibitem{Blumlein:2009ta}
J.~Bl\"umlein, \emph{{Structural relations of harmonic sums and Mellin transforms up to weight $w = 5$}}, \href{http://dx.doi.org/10.1016/j.cpc.2009.07.004}{\emph{Comput. Phys. Commun.} {\bf 180} (2009) 2218--2249}, [\href{http://arxiv.org/abs/0901.3106}{{\tt 0901.3106}}].

\end{thebibliography}\endgroup

\end{document}